\documentclass[prc, twocolumn, superscriptaddress, preprintnumbers,
amsmath, amssymb]{revtex4}

\allowdisplaybreaks
\allowdisplaybreaks[4]
\usepackage{dcolumn}
\usepackage[
            pdfstartview=FitH,
            CJKbookmarks=true,
            bookmarksnumbered=true,
            bookmarksopen=true,
            colorlinks=true,
            linkcolor=blue,
            anchorcolor=blue,
            citecolor=blue
            ]{hyperref}
\usepackage{bm}
\usepackage{graphicx}
\usepackage{color}

\newcommand{\beq}{\begin{equation}}
\newcommand{\eeq}{\end{equation}}
\newcommand{\bea}{\begin{eqnarray}}
\newcommand{\eea}{\end{eqnarray}}
\def\am{angular momentum }
\def\amd{angular momentum}
\def\momi{moment of inertia }
\def\momis{moments of inertia }

\def\momid{moment of inertia}
\def\momisd{moments of inertia}

\newcommand{\vth}{\vartheta}

\newcommand{\De}{\Delta}
\newcommand{\ga}{\gamma}

\newcommand{\f}{\varphi}

\begin{document}

\title{Interpretation of the quasiparticle  plus triaxial rotor model}

\author{Q. B. Chen}
\affiliation{Physik-Department, Technische Universit\"{a}t
M\"{u}nchen, D-85747 Garching, Germany}

\author{S. Frauendorf}\email{sfrauend@nd.edu}
\affiliation{Physics Department, University of Notre Dame, Notre
Dame, IN 46556, USA}


\date{\today}

\begin{abstract}

We discuss in depth the application of the classical concepts for 
interpreting the quantal results from  the triaxial 
rotor core without and with odd-particle. The corresponding limitations  
caused by the discreteness and finiteness of the \am Hilbert space and 
the extraction of the relevant features from the complex wave function 
and distributions of various \am components are discussed in detail. 
New methods based on spin coherent states and spin squeezed states are 
introduced. It is demonstrated that the spin coherent state map is a 
powerful tool to visualize the angular momentum geometry of rotating 
nuclei. The topological nature of the concepts of transverse and 
longitudinal wobbling is clarified and the transitional axis-flip
regime is analysed for the first time.

\end{abstract}

\maketitle

\graphicspath{{./figs/}}

\section{Introduction}
\label{sec:intro}

The description of the structure of rotating nuclei in terms of classical 
\am vectors is a very illuminating tool, which has been and is being widely 
used. Rotational alignment, magnetic rotation, chiral doubling, and transverse 
wobbling are examples (see e.g., the recent review~\cite{Frauendorf2018PS}).
For the quantitative comparison with experiment one uses  theoretical approaches
which treat the \am (or parts of) quantum mechanically. The interpretation 
of the quantal results in classical terms is often only qualitative. There 
is a need to establish more direct connections between the quantal information 
about the various kinds \am operators encountered in rotating nuclei and their 
classical counterparts. 
 
The model of several quasiparticles coupled to a triaxial rotor has been 
very successful in describing the results of experimental studies of 
rotating nuclei. The model divides the total \am into the contributions 
of the quasiparticles and a rotor part, which accounts for the rest.
The various types of \am are treated quantum mechanically. In this work 
we discuss in depth the application of the classical concepts for 
interpreting the quantal results from the simplest case, the one 
particle-plus triaxial rotor model. The limitations of the correspondence
caused by the discreteness and finiteness of the \am Hilbert space will be 
addressed, as well as, the question how to extract the relevant features of 
various \am components from the complex wave function. New methods based 
on Spin Coherent States (SCS) and Spin Squeezed States (SSS) will be 
introduced. Much of the insight gained from the present study will apply 
to more complex theoretical approaches to rotating nuclei. 

Angular momentum is a vector operator, so it is appropriate to represent 
it as a classical vector, which becomes quite accurate for large magnitude.  
Diagrams of the various \am vectors that determine the rotational behaviour 
have turned out to be a very illuminating tool, which is widely used. 
Of course, in a dynamical system like the nucleus the \am 
components of the constituent elements are not rigidly arranged. 
Their orientations change with time in a classical system, 
which corresponds to distributions in a quantum state. Vector diagrams 
represent only the average or most likely values of such distributions, 
which contain more information about the dynamics. In this paper 
we will investigate the detailed \am structure of selected 
examples by means of various visualization techniques 
in order to demonstrate their capabilities and limitations. 

A central aspect is the non-commutativity of the three components of
the \am operator. The ensuing uncertainty implies that drawing a classical 
vector with three fixed components may lead to an over simplified picture. 
One time-proven way to illustrate the uncertainty is to draw precession 
cones of the \am vectors, which are widely used for qualitative illustrations. 
However, their construction becomes tedious if not problematic when 
illustrating a quantal results in a quantitative way. In this paper 
we will discuss in detail the ``most classical'' representation of 
the \am operators, which are the SCS of the SU(2) group~\cite{Atkins1971PRSLA, 
Radcliffe1971JPA} and for the $\textrm{SU(2)} \times \textrm{SU(2)}$ group 
of the triaxial rotor~\cite{Janssen1977SJNP}. We will focus on the main players 
in the dynamics of rotating nuclei: the total angular momentum, for which 
$\hat{J}^2$ is exactly conserved, and the \am of the high-$j$ orbitals $f_{7/2}$, 
$g_{9/2}$, $h_{11/2}$, $i_{13/2}$, $j_{15/2}$, for which $\hat{j}^2$ is conserved 
to good approximation. The uncertainty is restricted to the orientation. As the 
SCS constitutes a complete, though non-orthogonal, set it is straight forward 
to project the quantal results on to this basis, which generates 
quantitative illustrations of the \am structure of the rotating nuclei. 

The other aspect addressed is how to extract the properties of the
various \am components from complex wave functions. For this
purpose the appropriate density matrices are introduced.

The paper is organized as follows. Sec.~\ref{s:j-space} reviews the 
properties of the discrete Hilbert space of fixed total \am and introduces 
the SCS basis. Sec.~\ref{s:TR} studies the triaxial rotor model (TRM).  
It applies the commonly used ways to illustrate the \am structure 
and demonstrates the new perspectives provided by the SCS representation. 
Sec.~\ref{s:PTR} studies the particle-plus-triaxial rotor model (PTR). 
The approaches exposed in Sec.~\ref{s:TR} will be generalized by 
introducing the appropriate density matrices. The partial loss of phase 
coherence will be addressed. As example, transverse wobbling and its 
transition to longitudinal wobbling at large spin~\cite{Frauendorf2014PRC} 
will be discussed in detail. Recent approximate treatments of the PTR  
model~\cite{Tanabe2017PRC} and \cite{Raduta2020PRC} (and earlier work 
cited therein) will be compared with the complete PTR solutions to 
expose their limitations and the insights they permit. The authors 
of these publications as well as of Ref.~\cite{Lawrie2020PRC}
introduced new terminology to replace the original termini transverse 
and longitudinal wobbling suggested by Frauendorf and 
D\"{o}nau~\cite{Frauendorf2014PRC}, which, in our view, may lead to 
an unfortunate confusion concerning the interpretation of the quantal 
PTR results. A clarification and justification of in our view 
appropriate terminology will be given. 

\section{The \texorpdfstring{$j$}{j}-space}
\label{s:j-space}

The Hilbert space of good absolute \am $j$ is spanned by the states with
projection $-j\leq k\leq j$ on the quantization axis 3. The quantum
system is one-dimensional: the motion is restricted to the sphere of
constant \amd~\footnote{We use $\hbar=1$.},
\beq\label{eq:jsphere}
 j^2=j_1^2+j_2^2+j_3^2=j(j+1).
\eeq
The pertaining operators of momentum $p$ and position $q$ (canonical
variables in the classical mechanics) obey the standard commutation
relation $[p,q]=-i$. One may take the \am projection $\hat{j}_3$
as the momentum operator $p$. Then the angle operator $\hat{\phi}$,
which fixes the orientation of the \am vector $\hat{\bm{j}}$ projection 
in the 1-2-plane, becomes the conjugate position operator $q$ and
\beq\label{eq:can}
 [p,q]=[\hat j_3,\hat\phi]=-i.
\eeq
As the momentum takes only the $2j+1$ discrete values $k=-j$, ..., $j$,
the angle can also take only $2j+1$ discrete values on the unit circle.
These are the eigenvalues of $\hat{\phi}$, which can be chosen as
\beq
 \phi_n=\frac{2\pi}{2j+1}n,~~~n=-j,~-j+1,~...,~j-1,~j.
\eeq
It is common to use the momentum eigenstates $\vert k\rangle$ as a
basis, which are related to the angle eigenstates $\vert n\rangle$
by the transformation
\begin{align} \label{eq:phi-k}
 \vert k\rangle &=\frac{1}{\sqrt{2j+1}}\sum\limits_{n=-j}^{j}
 e^{ik\frac{2\pi}{2j+1}n }\vert n\rangle,\\
 \vert n\rangle &=\frac{1}{\sqrt{2j+1}}\sum\limits_{k=-j}^{j}
 e^{-i\frac{2\pi}{2j+1}n k}\vert k\rangle.
\end{align}
The amplitude
\beq\label{eq:phin}
 \langle n\vert k\rangle
  =\frac{1}{\sqrt{2j+1}}e^{ik\frac{2\pi}{2j+1}n}
  =\frac{1}{\sqrt{2j+1}}e^{ik\phi_n}
\eeq
is the discrete version of the well known expression $1/\sqrt{2\pi}\exp{[i k \phi]}$ 
for the $\hat{j}_3$ eigenfunctions in the full orientation space. The 
discreteness of $j_3$ and $\phi$ should be kept in mind in the semiclassical 
visualization.

Working in the discrete $k$ basis, the operator $\exp[i\hat{\phi}]$
is more convenient than  $\hat{\phi}$, because its matrix elements
are simply
\beq
 \langle k\vert e^{i\hat{\phi}}\vert k'\rangle=\delta_{k',k+1},
\eeq
which is seen using Eq.~(\ref{eq:phi-k}) and noticing
their orthonormality. Then the matrix of the operator
\begin{align}
 &\quad \langle k\vert\sqrt{j-\hat{j}_3} e^{i\hat{\phi}}\sqrt{j+\hat{j}_3}
 \vert k'\rangle \notag\\
 &=\sqrt{(j-k)(j+k+1)}\delta_{k',k+1}
\end{align}
is recognized as the matrix of the $\hat{j}_+$ operator, and
the ladder operators $\hat{j}_\pm$ are
\begin{align}
 \hat{j}_+ 
  &= \sqrt{j-\hat{j}_3} e^{i\hat{\phi}}\sqrt{j+\hat{j}_3}, \label{eq:j+}\\
 \hat{j}_- 
  &=(\hat j_+)^\dagger=\sqrt{j+\hat{j}_3} e^{-i\hat{\phi}}
 \sqrt{j-\hat{j}_3}\label{eq:j-}.
\end{align}
The standard commutation relations $[\hat{j}_3,\hat{j}_\pm]=\pm \hat{j}_\pm$
and $[\hat{j}_+,\hat{j}_-]=2\hat{j}_3$ are fulfilled because they hold for
the pertaining matrices within the basis $\vert j,k\rangle$. They can
also be directly verified using $\big[\hat{j}_3,\exp[\pm i\hat{\phi}]\big]
=\pm \exp[\pm i\hat{\phi}]$.

In the following we use the SCS for visualization. They are
the most classical representation consistent with the uncertainty
relation between $\hat{j}_3$ and $\hat{\phi}$. They were introduced by
Atkins, Donovan, and Radcliffe~\cite{Atkins1971PRSLA, Radcliffe1971JPA}
for the SU(2) group. At variance to the original papers we introduce the
SCS $\vert \theta\phi\rangle$ in a way that is more instructive for the
purpose of visualization~\cite{Loh2015AJP},
\begin{align}\label{eq:SCS}
 \vert j \theta\phi\rangle 
  &={\cal R}^\dagger(\theta\phi 0) \vert j 00\rangle
   =\sum_kD^j_{kj}(\phi\theta0)\vert j k\rangle, \\
 \vert j 00\rangle 
  &=\vert j, k=j\rangle.
\end{align}
The basic SCS is the state with the maximal projection on the quantization 
axis 3, $\vert j 00\rangle=\vert j, k=j\rangle$. The complete set of SCS 
is generated by rotating the basic SCS by the polar angle $\theta $
and the azimuthal angle $\phi$. The basic SCS is not strictly
aligned with the 3-axis. There is an uncertainty in orientation
given by the root mean square of $\De j_1=\De j_2=\sqrt{\langle
jj\vert j_2^2\vert jj\rangle}=\sqrt{j/2}$. Hence, instead of an
arrow one should think  of a cone with the opening angle of
$ \sim 1/\sqrt{j/2}$. To replace it by an arrow is good enough for
many illustrations provided $j$ is large enough. For the lowest angular
momenta when $j$ is comparable with $\sqrt{j/2}$ such replacement
misses essential features. The consequences of this final resolution
will be discussed.

The set of the SCS is generated by rotating the basic SCS $\vert j 00\rangle$
over the whole \am sphere. It is normalized but
non-orthogonal,
\begin{align}
  \langle j \theta\phi\vert j\theta'\phi'\rangle
  &=\langle j 00\vert \mathcal{R}(\theta\phi 0) \mathcal{R}^\dagger(\theta'\phi' 0) 
    \vert j 00\rangle \notag\\
  &=\langle j 00\vert \mathcal{R}^\dagger(\theta''\phi'' 0) \vert j 00\rangle
   =D^j_{jj}(\theta''\phi''0).
\end{align}
Here, $\theta''$, $\phi''$ are the angles of the rotation resulting from
first rotating by $\theta'$, $\phi'$ and then backward by $\theta$, $\phi$.
The SCS set is massive over complete, because the dimension of the $j$-space
is $2j+1$ whereas the dimension of the SCS space is infinite. There are infinite
many possible transformations from the SCS basis back to the $k$-basis.
One obvious transformation is
\beq\label{eq:SCStoIK}
 \vert j k\rangle=\frac{(2j+1)}{4\pi}\iint d\theta d\phi~\sin\theta
 D^{j*}_{kj}(\phi\theta0)\vert j \theta\phi\rangle.
\eeq
Accordingly the resolution of the identity operator is not unique either.
Particular simple is
\begin{align}\label{eq:SCSto1}
 \hat 1
 &=\sum_k\vert jk\rangle\langle jk\vert \notag\\
 &=\frac{(2j+1)}{4\pi}\iint d\theta d\phi~\sin\theta
 \vert j \theta\phi\rangle \langle j \theta\phi \vert.
\end{align}

The expectation value of the \am operator with the SCS agrees with
the classical value
\beq
 \langle j \theta \phi\vert \hat j_i\vert j
 \theta \phi\rangle =(j \sin \theta \cos \phi, 
 j \sin \theta \sin \phi, j \cos \theta).
\eeq
Thus when drawing arrows to illustrate the \am composition, they represent 
the SCS $ \vert j\theta\phi\rangle$, which have an orientation uncertainty 
that agrees with the one of the state $\vert jj\rangle$. It is determined 
by the uncertainties of $\De j_1=\De j_2=\sqrt{\langle jj\vert j_2^2\vert
jj\rangle}=\sqrt{j/2}$, which correspond to an angle uncertainty of 
$\De \theta=\arctan(\De j_2/j)\approx 1/\sqrt{2j}$. That is, the orientation 
angles of the \am vectors cannot be determined better than $\De \theta$. 
A classical arrow corresponds to a point in the $\theta$-$\phi$-plane. The 
SCS represents a fuzzy blob with the mean radius $\sim 1/\sqrt{2j}$. One may 
also think about it as a precession cone with the opening angle $\sqrt{2/j}$. 
A pedagogic introduction of the SCS can be found in Ref.~\cite{Loh2015AJP},
which contains many complementary details.

It is important to keep in mind the mutual non-orthogonality
of the SCS, which will cause deviations from classical vector
geometry.
 The overlap
\beq\label{eq:overlap}
 \vert\langle j 00\vert j\theta\phi\rangle\vert
 =d^j_{jj}(\theta)
 =\cos^{2j}\left(\frac{\theta}{2}\right)
 \approx \exp\left(-\frac{j}{4}\theta^2\right)
\eeq
has a width of $\De \theta = \sqrt{2/j}$, which is the angle where 
the overlap is reduced to $\exp(-0.5)=0.607$. Instead of the continuous 
over-complete set, one may consider the discrete set of $2j+1$ SCS states, 
which are uniformly distributed over the unit sphere. Each of the SCS 
covers an area of $4\pi/(2j+1)$. To cover the sphere one needs hexagons 
and pentagons. Their area is close to a circle with the radius 
$r \approx \sqrt{2/j}$. The distance between the centers of two 
adjacent SCS is $2r$, and thus the angle between the two adjacent SCS 
is $\Delta\theta_r=2r/1=2\sqrt{2/j}$. Their overlap 
$\exp[-j(\Delta \theta_r)^2/4]=\exp[-2]=0.135$ is much smaller 
than that of continuous SCS. This discrete set of SCS is nearly 
orthonormal and complete. When inspecting the SCS maps, one 
should keep in mind this underpinning coarse grained basis.

Janssen~\cite{Janssen1977SJNP} generalized the SCS to the product group
SU(2)$\times$SU(2), the irreducible representations of which are the
eigenstates of the axial rotor $\vert IMK\rangle$, where $M$ is the
\am projection on the laboratory frame $z$-axis and $K$ the projection 
on the symmetry 3-axis. Discussing the physics it is sufficient to
restrict to the subspace of the states $\vert IIK \rangle$ that are
maximally aligned with the laboratory frame $z$-axis. The subspace is
spanned by the states with $-I\leq K \leq I$. The structure is quite
analogue to the above discussed $j$-space. The only difference is 
the commutation relations between the intrinsic components of the 
\am $[\hat{J}_1,\hat{J}_2]=-i\hat{J}_3$ (cyclic) which differ from 
the standard relations $[\hat{j}_1,\hat{j}_2]=i\hat{j}_3$ (cyclic) by 
the sign of $i$. This changes the commutators to $[\hat{J}_3,\hat{J}_\pm]
=\mp \hat{J}_\pm$ and $[\hat{J}_-,\hat{J}_+]=2\hat{J}_3$, that is, 
$\hat{J}_-$ becomes the raising operator and $\hat{J}_+$ the lowering 
operator. The exchange of the role of the ladder operators is taken 
into account by changing $\hat{\phi} \rightarrow -\hat{\phi}$ in 
Eqs.~(\ref{eq:j+}) and (\ref{eq:j-})
\begin{align}
 \hat{J}_+ &=\sqrt{I-\hat{J}_3} 
  e^{-i\hat{\phi}}\sqrt{I+\hat{J}_3}, \label{eq:J+}\\
 \hat{J}_- &=\sqrt{I+\hat{J}_3} 
  e^{ i\hat{\phi}}\sqrt{I-\hat{J}_3}. \label{eq:J-}
\end{align}
Otherwise, all said above holds for the SU(2) factor group of the 
intrinsic components of the total \am $\bm{\hat{J}}$.

For the purpose of visualization we construct set of intrinsic SCS
by rotating the basic state $\vert I,M=I,K=I\rangle$ over the \am
sphere
\begin{align}\label{eq:SCSI}
 \vert I \theta\phi\rangle &=\mathcal{R}^\dagger(\theta\phi 0)
 \vert I 00\rangle=\sum_KD^I_{KI}(\phi\theta0)\vert IIK  \rangle, \\
 |I00\rangle &=\vert I,M=I, K=I\rangle.
\end{align}
The full set of SCS for the $\textrm{SU(2)}\times \textrm{SU(2)}$ 
group~\cite{Janssen1977SJNP} encompasses the additional rotation 
of the basic state  $\vert I 00\rangle$ over the \am sphere in 
the laboratory frame. The intrinsic SCS (\ref{eq:SCSI})
comprises the SU(2) subset of states the \am of which is aligned with
the laboratory frame $z$-axis.

The basis spanned by the discrete angles $\phi_n$ has an counter-intuitive 
property that will be discussed below with the examples. It is avoided by 
using the over-complete, non-orthogonal Spin Squeezed States (SSS)
\beq\label{eq:squeezed1}
 \vert \phi \rangle=\frac{1}{\sqrt{2j+1}}
 \sum_{k=-j}^je^{i\phi k}\vert k\rangle.
\eeq
More intuitively, the SSS states are generated by rotating
the $n=-j$ state of the discrete $\phi_n$ basis about the 3-axis
\beq\label{eq:squeezed2}
 \vert \phi \rangle=\exp{\left[i\hat j_3 \Big(\phi-\frac{2\pi j}{2j+1}\Big)\right]}
 \Big\vert n=-j \Big\rangle.
\eeq
The SSS states are normalized $\langle \phi\vert \phi\rangle=1$.
The overlap $\langle \phi\vert \phi'\rangle$ has a width of
$\Delta \phi=2\pi/(2j+1)\approx 3/j$. The notation squeezed states comes 
from quantum optics. They are a generalization of coherent states. For 
coherent states the uncertainty product $\Delta x \Delta p$ is minimized. 
For squeezed states either $\Delta x$ or $\Delta p$ is smaller than for 
the optimal coherent states (and the complementary width larger). The SSS 
width is smaller than the SCS width $\sqrt{2/j}$ because $k$ has a large 
width of $2j+1$. 

\section{Triaxial rotor model} \label{s:TR}

\begin{figure}[th]
  \includegraphics[width=0.8\columnwidth]{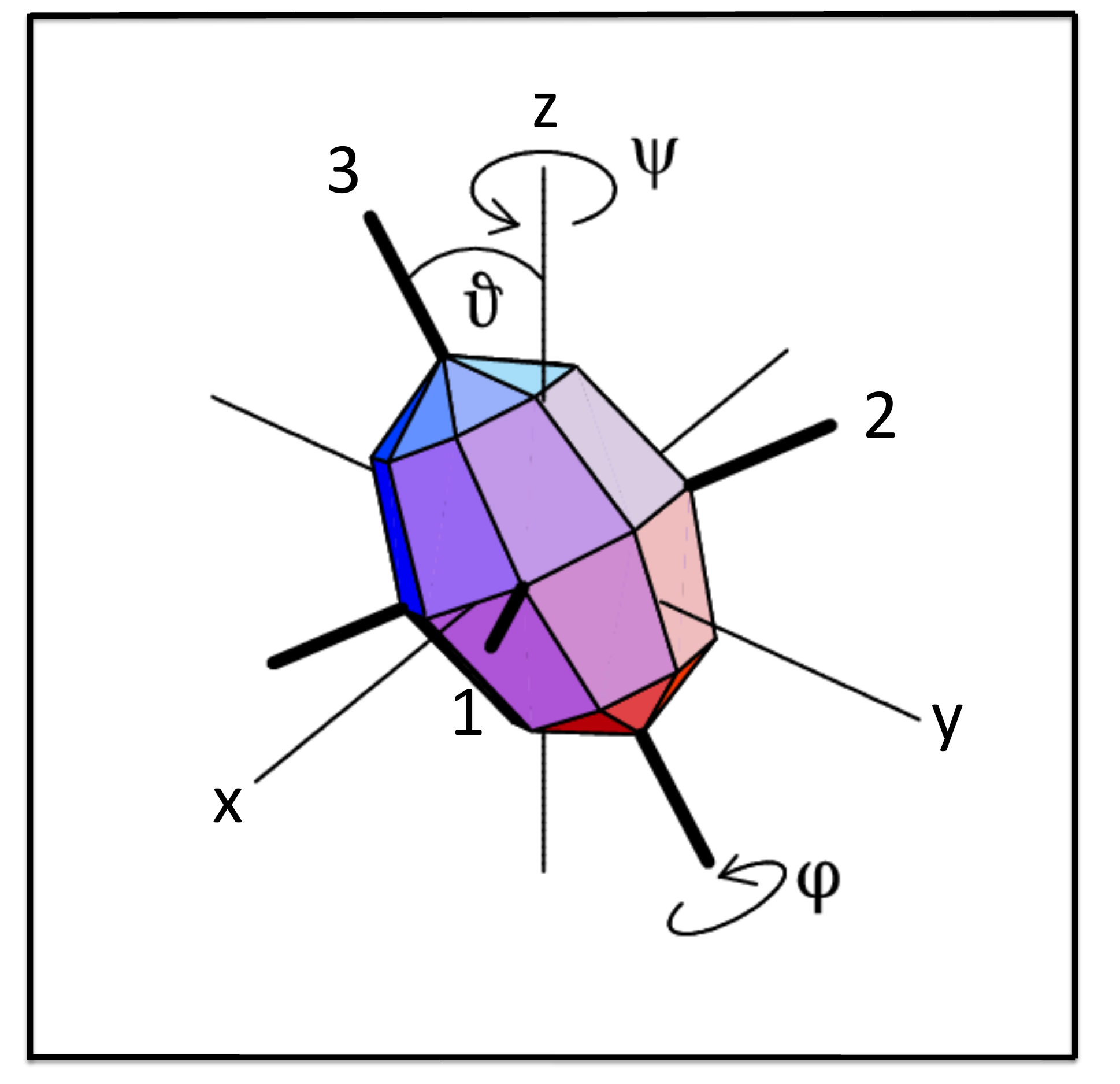}
  \caption{\label{f:angles} Euler angles specifying the orientation 
  of the triaxial reflection symmetric density distribution. Reproduced 
  from Ref.~\cite{Frauendorf2018PS}.}
\end{figure}

First we discuss the collective triaxial rotor model (TRM)~\cite{Bohr1975}, 
because the orientation angles are the only degrees of freedom. This makes 
it a transparent case to illustrate the restrictions imposed by 
non-commutativity of the \am components and to compare the various ways 
to illustrate the \am structure. The orientation of the triaxial body 
is specified by the Euler angles illustrated in Fig.~\ref{f:angles}. 
The dynamics is determined by the Hamiltonian
\bea\label{eq:HTR}
 H_{\textrm{TR}}=\sum_{i=1,2,3}\frac{\hat{J}_i^2}{2\mathcal{J}_i(\beta,\ga)},
\eea
where $\hat{J}_i$ is the collective \am of the rotor. The orientation
of the rotor is specified by the set of basis states
\begin{equation}
 \langle \f\vth\psi\vert IMK\rangle 
 =\sqrt\frac{2I+1}{8\pi^2} D^I_{K,M}(\f\vth\psi),   
\end{equation}
in which $D^I_{K,M}(\varphi \vth \psi)$ are the Wigner D-functions.
The rotor eigenstates
\beq\label{eq:rotorstate}
\vert IM\nu\rangle=\sum_{K=-I}^{I} C_{IK}^{(\nu)}\vert IMK\rangle
\eeq
are given by the amplitudes $C_{IK}^{(\nu)}$, which depend only on
the \am projection on one of the body-fixed principal axes. 
In the case of even-even nuclei, the states are completely symmetric 
representations of the $\textrm{D}_2$ point group. The symmetry restricts 
$K$ to be even and requires $C_{I-K}^{(\nu)}=(-1)^IC_{IK}^{(\nu)}$. This 
means the $K$ in the sum runs from $-I$ to $I$ for even $I$ and  
$-I+1$ to $I-1$ for odd $I$, and $C_{IK=0}^{(\nu)}=0$ for odd $I$.
In the following the restrictions on $K$ will not  be explicitly 
indicated. 

The \momis are parameters that must be determined by additional
considerations. The ratios between the \momis of the three
principal axes are assumed to be the ones of irrotational flow
\beq\label{eq:momiIF}
 \mathcal{J}_i(\beta,\ga)=\mathcal{J}\sin^2\left(\gamma-\frac{2\pi i}{3}\right),
\eeq
where the scale $\mathcal{J}$ is adjusted to the experimental energies.
The analysis of rotational spectra~\cite{Allmond2017PLB} and microscopic
calculations by means of the cranking model~\cite{Frauendorf2018PRC}
demonstrated that the dependence of moments of inertia on the
triaxiality parameter $\ga$ is well accounted for by Eq.~(\ref{eq:momiIF}).
In the following we use $s$, $m$, and $l$ to denote the short, medium, 
and long axes of the density distribution.

\begin{figure}[ht]
 \includegraphics[width=0.95\columnwidth]{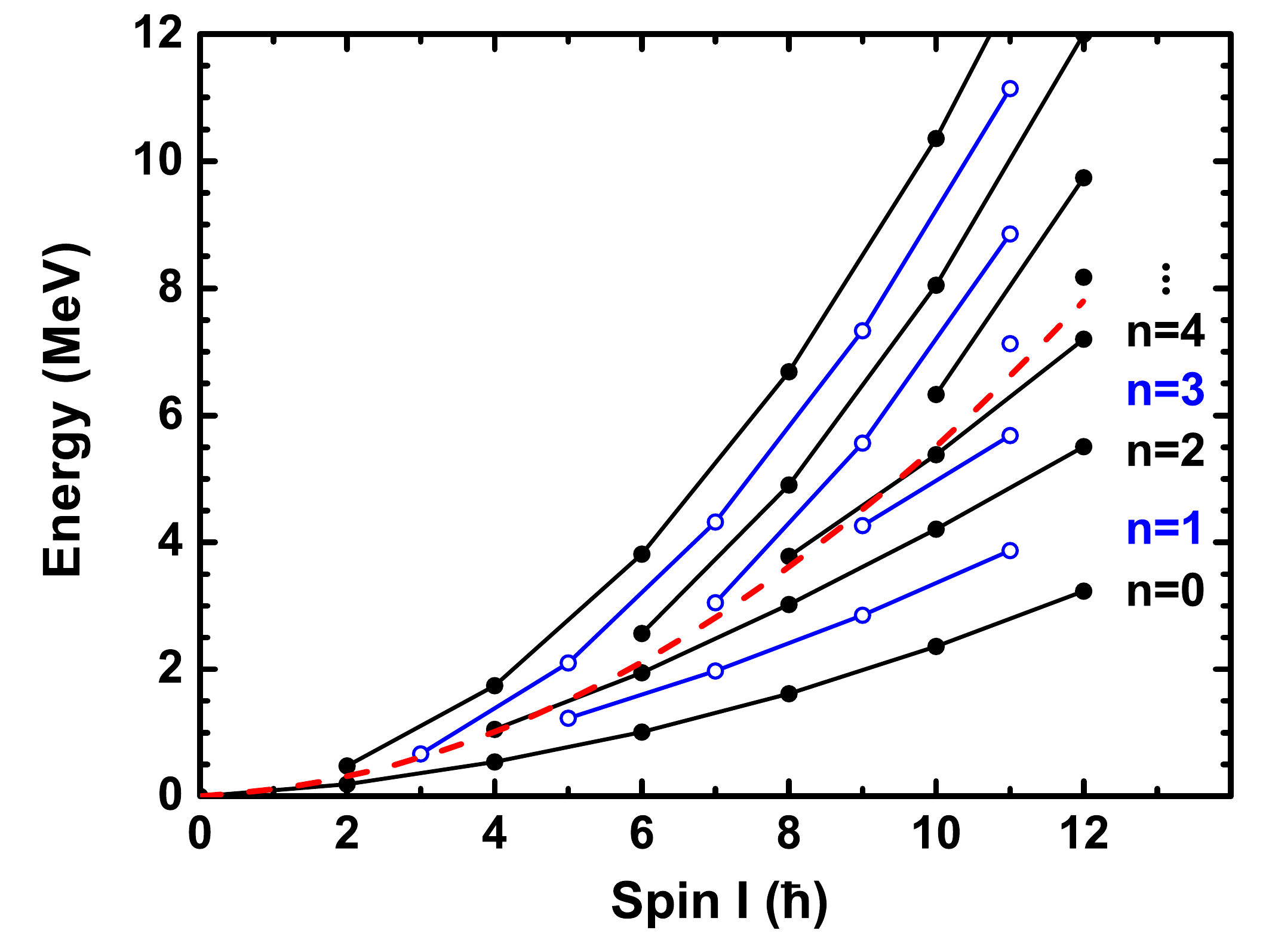}
 \caption{\label{f:TREnergy} Energies of a triaxial rotor as 
 function of its \am $R$. The states are labelled by the wobbling 
 quantum number $n$ and referred to as zero-, one-, two-, ... phonon 
 states. It is understood that this labelling does not imply that 
 the wobbling motion is harmonic.}
\end{figure}

As an example we show in Fig.~\ref{f:TREnergy} the energies of the 
triaxial rotor with the \momis $\mathcal{J}_m=30~\hbar^2$/MeV, 
$\mathcal{J}_s=10~\hbar^2$/MeV, $\mathcal{J}_l=5~\hbar^2$/MeV, 
which was studied in Refs.~\cite{Frauendorf2014PRC, W.X.Shi2015CPC}.

\begin{figure}[ht]
 \includegraphics[width=0.95\columnwidth]{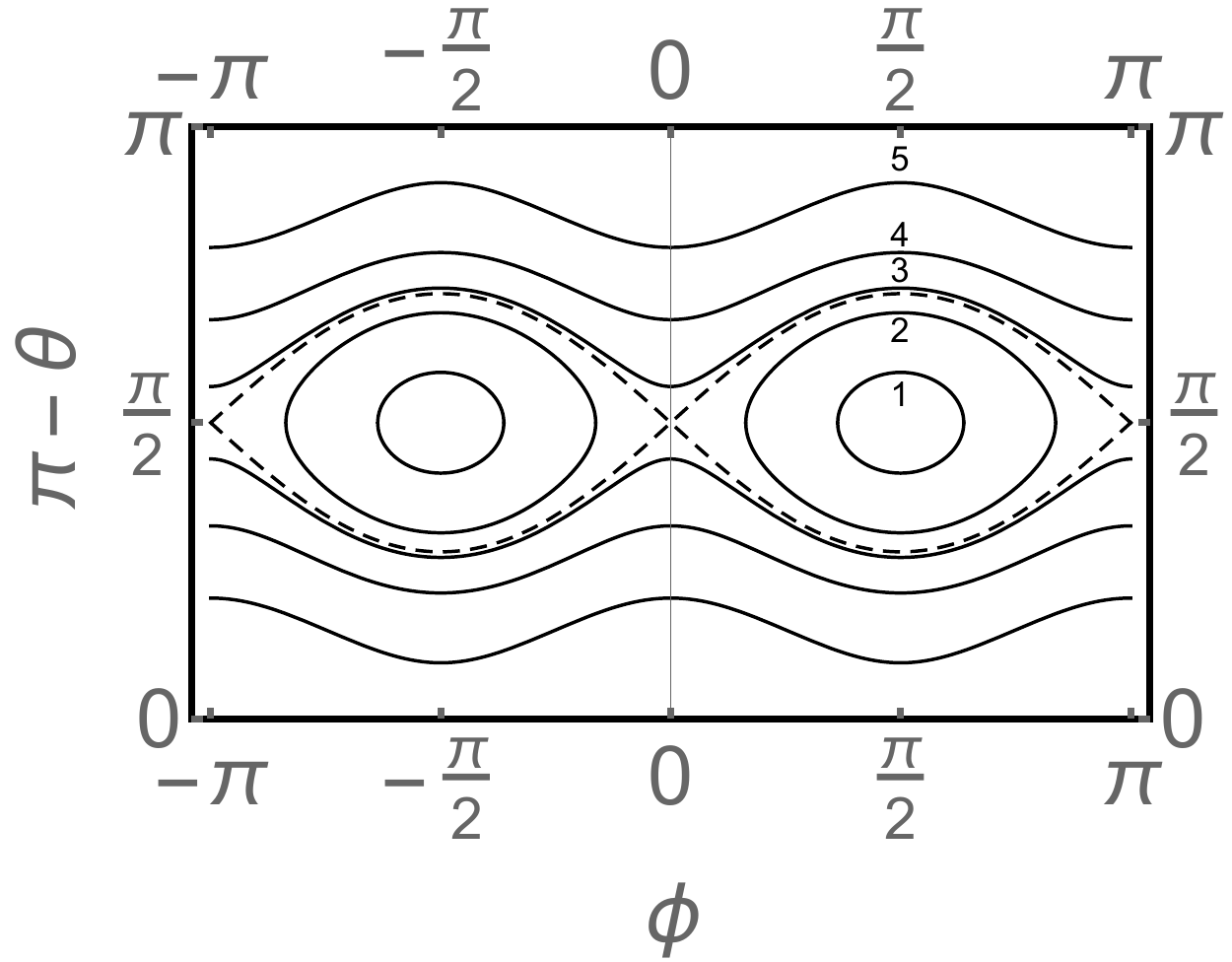}
 \caption{\label{f:TRorbits} Classical orbits of a triaxial rotor. The
  states 1-5 for $I=8$ are shown. The energies are equal to their quantal
  energies in Fig.~\ref{f:TREnergy}, labelled with in ascending order.}
\end{figure}

Classical mechanics of gyroscopes provides a classification of the 
quantal states. The classical orbits of the \am vector with respect 
to the body-fixed principal axes are the intersection lines between 
the sphere of constant \am (\ref{eq:jsphere}) and the ellipsoid of
constant energy (\ref{eq:HTR}). The construction of the orbits was 
discussed in detail in Refs.~\cite{Frauendorf2018PS, Frauendorf2014PRC}. 
For the triaxial rotor the orbits are given by the two angles 
$\theta$ and $\phi$ as 
\begin{align}
& J_1={\sqrt{I(I+1)}}\sin \theta\cos \phi,\nonumber\\
& J_2={\sqrt{I(I+1)}}\sin \theta\sin \phi,\\
& J_3={\sqrt{I(I+1)}}\cos\theta.\nonumber
\end{align}
The orbits on the unit sphere are determined by the implicit equation
\begin{align}
\cos^2 \theta(\phi)
  &=\frac{\frac{2E_i}{I(I+1)}-\left(\frac{\cos^2\phi} {\mathcal{J}_1}
   +\frac{\sin^2\phi} {\mathcal{J}_2}\right)}
   {\frac{1}{\mathcal{J}_3}-\frac{\cos^2\phi}{\mathcal{J}_1}
    -\frac{\sin^2\phi}{\mathcal{J}_2}}.
   \label{eq:J3class}
\end{align}
They are shown in Fig.~\ref{f:TRorbits} for $I=8$ as an example~\footnote{Note 
that the figure shows the polar angles $\theta$ and $\phi$ relative to 
the $s$-axis, whereas the plot in Ref.~\cite{Frauendorf2014PRC} shows 
the phase space $J_3$ and $\phi$ relative to the $m$-axis.}. To connect 
with spectra in Fig.~\ref{f:TREnergy} the quantal energies are taken 
for $E_i$ and the square of classical \am is replaced by its semiclassical 
corrected value of $I(I+1)$. The axes 1, 2, and 3 are associated 
with $s$, $m$, $l$, respectively. Three-dimensional illustrations of 
the intersection between the \am sphere and the energy ellipsoid can 
be found in Classical Mechanics textbooks or more specific for nuclei, 
e.g., in Ref.~\cite{Frauendorf2014PRC}.

There are two topologically different types of orbits. For low energy
they revolve the $m$-axis with the maximal \momid, and for
high energy they revolve the $l$-axis with the minimal \momid. The
separatrix shown by the dashed line in Fig.~\ref{f:TREnergy} divides
the two regions. It is the unstable orbit of uniform rotation about the
$s$-axis with the intermediate \momid. Its energy $I^2/(2\mathcal{J}_s)$ 
is shown by the red dashed line in Fig.~\ref{f:TREnergy}, where the
connected states correspond to classical orbits centered about the
$m$-axis (below) and $l$-axis (above). The distance between the orbits 
is determined by an increase of the action $\iint d J_3 d\phi$ by one 
unit, which is approximately $\sin\theta/{\sqrt{I(I+1)}}$ of the area 
between two orbits on the cylinder shown in Fig.~\ref{f:TRorbits}. 
Because the area shrinks with increasing \amd, new orbits are born 
above and below the separatrix, as seen in Fig.~\ref{f:TREnergy}.

The lowest states correspond to the 
orbits that revolve the $m$-axis with the largest \momid. These orbits 
represent the ``wobbling" mode. The name is quite appropriate, because 
the orbit of the \am with respect to the body-fixed axes coincides with 
the orbit of the $l$-axis of the density distribution in the laboratory 
system, where the \am vector stands still. ``Wobbling" is used to describe 
the staggering motion of a thrown baseball or of the swaying motion of 
the earth axis. 
For small amplitude the wobbling mode becomes harmonic, that is a precession 
cone that is generated by two harmonic oscillations in $J_3$ and $\phi$ 
directions with a phase difference of $\pi/2$. Bohr and Mottelson discuss 
the harmonic limit in Nuclear Structure II \cite{Bohr1975} p.190 ff.

\subsection{Root mean square values}\label{s:TRRMS}

As a consequence of the D$_2$ symmetry, the  expectation values of 
the \am components on the intrinsic principal axes are zero. Usually
their root mean square expectation values are used to visualize the
\am geometry,
\bea\label{eq:ANG}
 J_i=\sqrt{\langle \hat J^2_i\rangle}
    =\sqrt{\sum_{KK'} \rho_{KK'} \langle IK'\vert\hat J^2_i\vert IK\rangle}.
\eea
where we introduced the density matrix $\rho^{(\nu;i)}_{KK'}$. 
For the considered case of the triaxial rotor without coupled particles 
it has the simple form
\beq\label{eq:RDM}
 \rho^{(\nu;i)}_{KK'}=C^{(\nu;i)}_{IK}C^{(\nu;i )*}_{IK'}.
\eeq
The amplitudes $C^{(\nu;i)}_{IK}$ in the three basis sets are
\begin{align}
C^{(\nu;s)}_{IK} &=\sum_{K'} 
 D^{I*}_{K'K}(0 \frac{\pi}{2} 0)C^{(\nu)}_{IK'},\\
C^{(\nu;m)}_{IK} &=\sum_{K'} 
 D^{I*}_{K'K}(\frac{\pi}{2} \frac{\pi}{2} 0)C^{(\nu)}_{IK'},\\
C^{(\nu;l)}_{IK} &=C^{(\nu)}_{IK},
\end{align}
with the amplitudes $C^{(\nu)}_{IK}$ obtained from Eq.~(\ref{eq:rotorstate}).

\begin{figure}[ht]
 \includegraphics[width=8.0 cm]{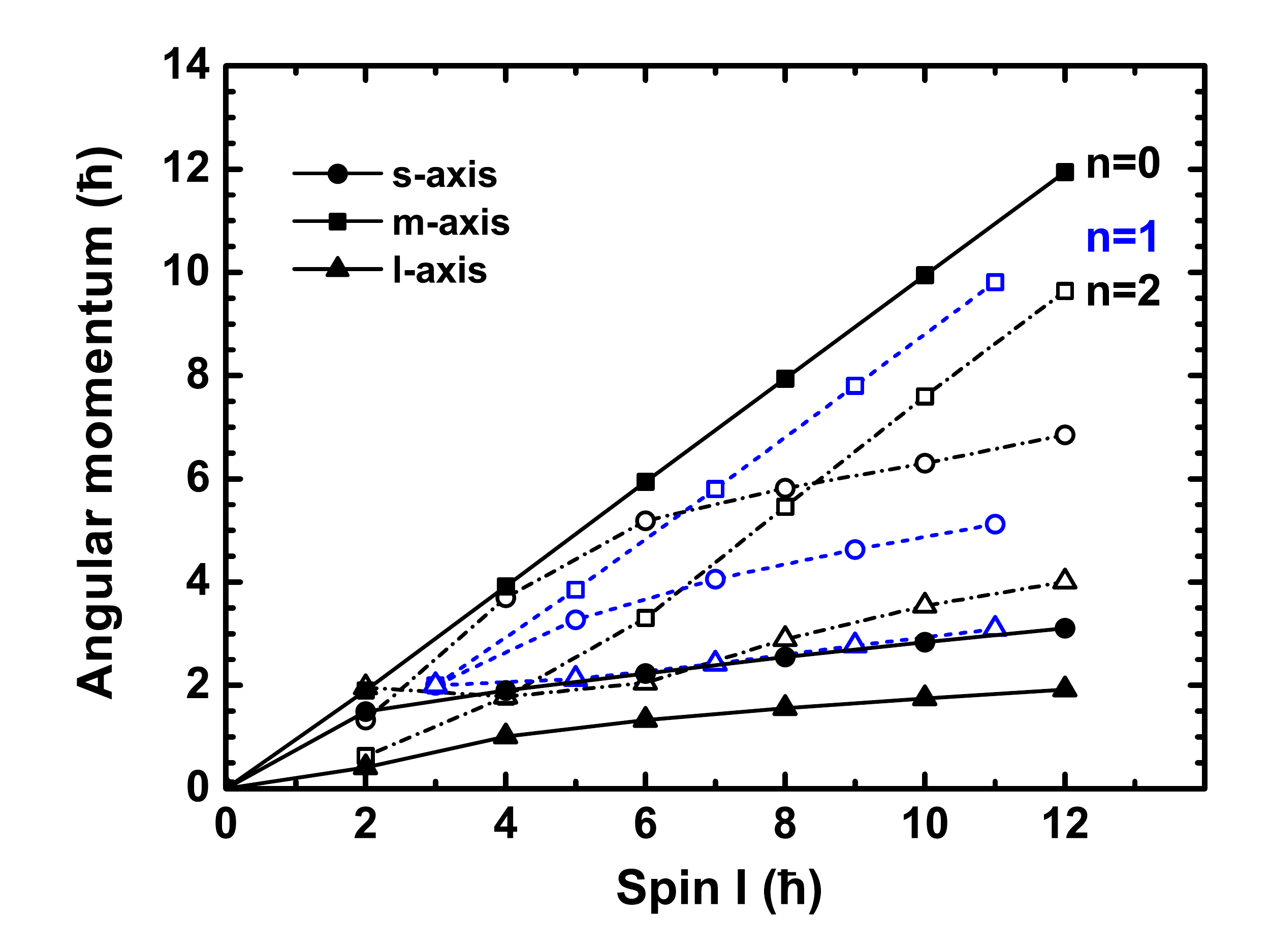}
 \caption{\label{f:ANG_TR} Root mean square expectation values (\ref{eq:ANG}) 
 of the \am components for the yrast and one- and two-phonon wobbling 
 excitations of the triaxial rotor.}
\end{figure}

Fig.~\ref{f:ANG_TR} shows the root mean square expectation values
(\ref{eq:ANG}) for the yrast and one- and two-phonon wobbling excitations
of the triaxial rotor. The \am of yrast states is best aligned with the 
$m$-axis, i.e., $J_m \approx I$. The finite $J_s$ and $J_l$ values manifest 
the quantum uncertainty of the \am orientation, where $J_s>J_l$ because
$\mathcal{J}_s>\mathcal{J}_l$.

The one-phonon state has one unit less aligned with the $m$-axis
$J_m\approx I-1$ and the two-phonon state two units less $J_m\approx
I-2$. The ratios $J_l/\sqrt{J_m^2+J_s^2}=\cot \theta$ and $J_s/J_m=\cot \phi$ 
are approximately equal to the lengths of the semi axes of the elliptical
orbits on the surface of the unit sphere that revolve the
$m$-axis (c.f. the states $9_1$ and $8_2$ in Figs.~\ref{f:TRmaps1}
and \ref{f:TRmapsCS}). This relation becomes only
obvious when the geometry of the orbits is taken into account. 

 \subsection{\texorpdfstring{$K$}{K}-plots}\label{s:TRKplot}
 
\begin{figure}[ht]
 \includegraphics[width=0.90\columnwidth]{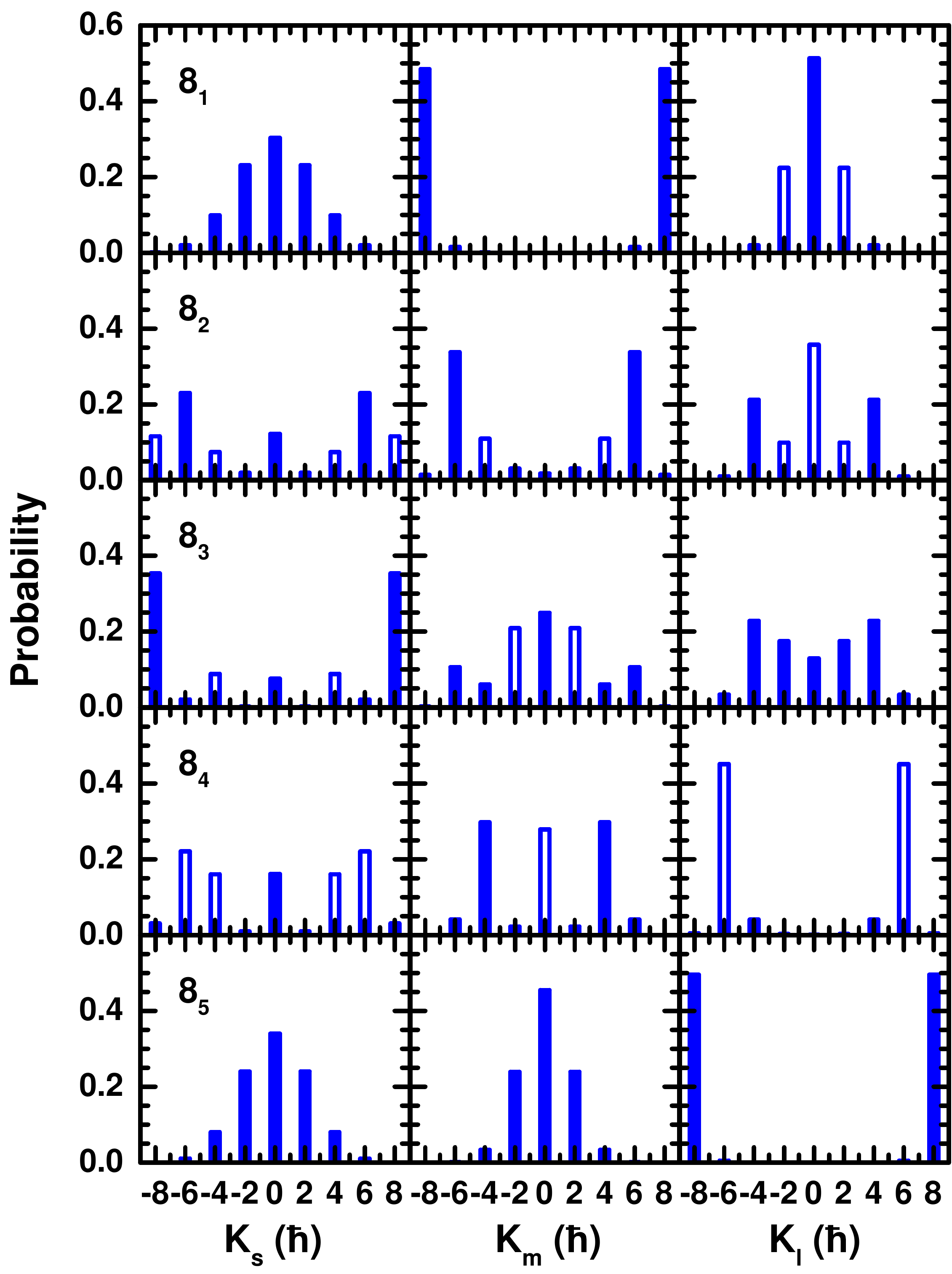}
 \caption{\label{f:TRWF} Probabilities of the basis states with good \am
 projection $J_i$ on the $s$-, $m$-, $l$-axes (columns 1, 2, 3, respectively)
 for the triaxial rotor states 1-5 with $I=8$. The sign of the amplitude
 is indicated: full-positive and open-negative. The energies are equal to 
 their quantal energies in Fig.~\ref{f:TREnergy}, labelled with in 
 ascending order.}
\end{figure}

For $I=8$, Fig.~\ref{f:TRWF} shows the probability distributions
\beq
P(K)_{I\nu;i}= \rho^{(\nu;i)}_{KK}
\eeq
of the \am  projection on the three principal axes, which were used in
Refs.~\cite{Frauendorf1996ZPA, J.Peng2003PRC, B.Qi2009PLB, Q.B.Chen2010PRC,
W.X.Shi2015CPC} and called $K$-plots or $K$-distributions by the authors. 

The \am of the lowest state $8_1$ is aligned with the $m$-axis, which
is displayed by the probabilities in the $m$-basis. Complementary,
the amplitudes in the $l$- and $s$-basis are broad. The \am of highest 
state $8_5$ is aligned with the $l$-axis, as seen by the probabilities 
in the $l$-basis and the very broad distribution and the alternating phase 
factor in the $\phi$-basis~\footnote{The state $\frac{1}{\sqrt{2}}(\vert
IIK=I\rangle +\vert IIK=-I\rangle)$ state gives $P(n)=\frac{1}{2I+1}
\cos(\frac{2\pi n}{2I+1})$.}. As seen in the left column, the \am of state 
$8_3$ is well aligned with the $s$-axis, which is reflected by the peaks 
centered at $\phi=0$, $\pm\pi$ (c.f. Fig.~\ref{f:TRphi-plot}). This 
distribution reflects the close neighborhood of state $8_3$ to the 
classical separatrix orbit in Fig.~\ref{f:TRorbits}, which corresponds 
to the uniform, yet unstable rotation about the $s$-axis with the 
intermediate moment of inertia (one could call it also Dzhanibekov 
orbit). The wobbling about $m$-axis nature of state $8_2$ (c.f. 
Fig.~\ref{f:TRorbits}) can be recognized in the $m$-basis as the 
shift of the strong component from $K=\pm 8$ to $\pm 6$. The analog 
holds for the state $8_4$ with respect to the $l$-axis.

The probability distributions $P_{KK}=\vert C^{(\nu)}_{IK}\vert^2$ 
have been used in illustrating the \am geometry of the triaxial rotor 
discussed here~\cite{W.X.Shi2015CPC} (see Figs.~6 and 7) and chiral 
bands~\cite{J.Peng2003PRC, B.Qi2009PLB, Q.B.Chen2010PRC}. They are 
quite instructive in case of good alignment with respect to 
one of the principal axes. They are less instructive when the wave 
function is delocalized (and possibly oscillates), one sees just 
a broad distribution.

\subsection{\texorpdfstring{$\phi$}{phi}-plots}\label{s:PRphi-plot}

\begin{figure}[ht]
 \includegraphics[width=0.95\columnwidth]{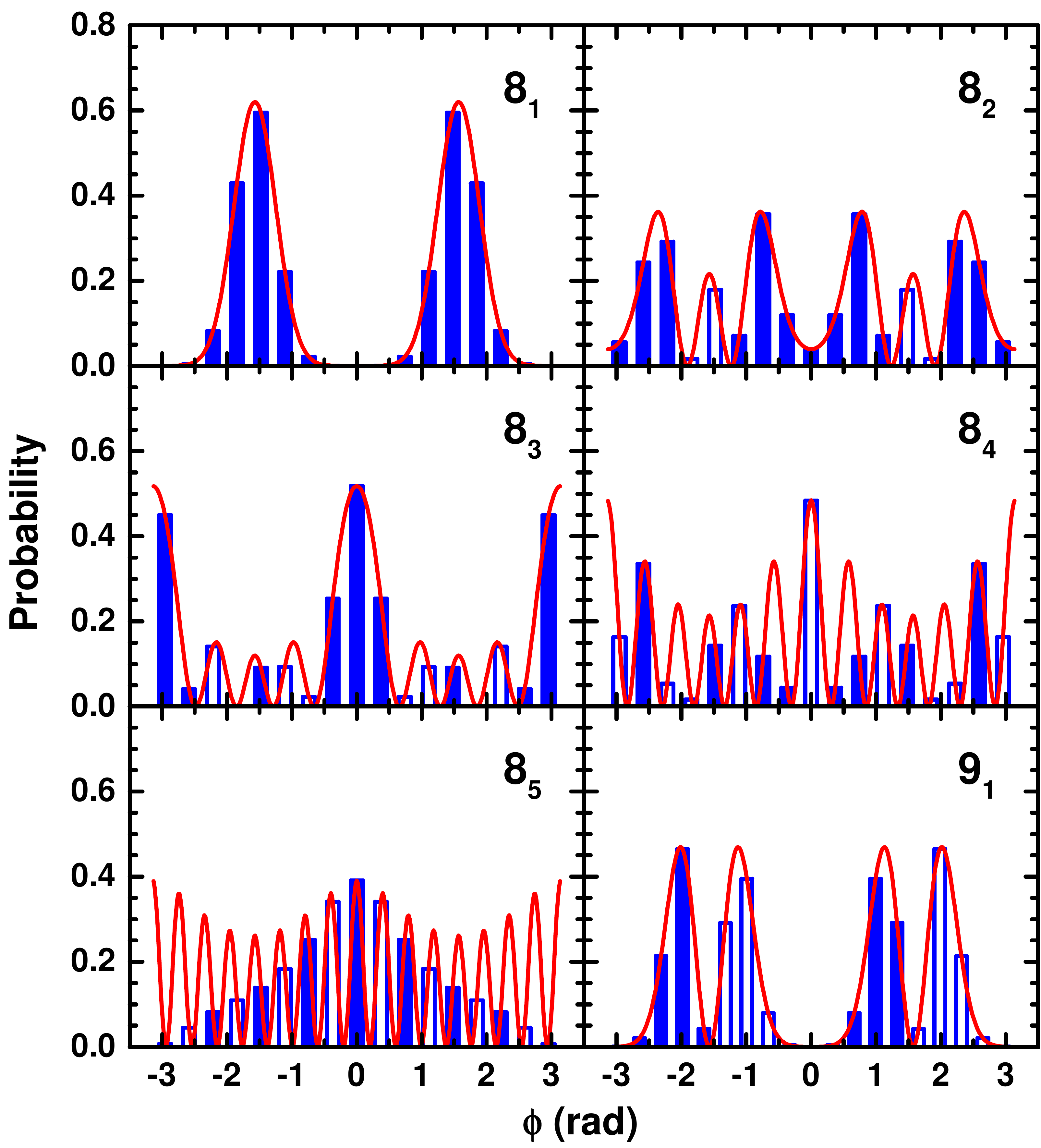}
 \caption{\label{f:TRphi-plot} Probabilities of the basis states with 
 good azimuthal angle $\phi_n$ (\ref{eq:Pphin}) for the triaxial 
 rotor states 1-5 with $I=8$ and state 1 with $I=9$. The sign of 
 the amplitudes is indicated: full-positive and open-negative. The 
 energies are equal to their quantal energies in Fig.~\ref{f:TREnergy}, 
 labelled in ascending order. The smooth curves show the 
 probability density of the spin squeezed states SSS (\ref{eq:PSSS}). 
 The figure shows $P(\phi_n)\frac{2I+1}{2 \pi}$ in order to match the 
 area under its envelop with the area under the SSS curve.}
\end{figure}

The probability distribution of the discrete $\phi_n$ states is given by 
\begin{align}\label{eq:Pphin}
P(\phi_n)_{I\nu} &=\rho^{I\nu}_{nn}, \\
\rho^{I\nu}_{nn'} &=\frac{1}{2J+1}\sum_{K,K'}e^{iK\frac{2\pi}{2J+1}n}
\rho^{(\nu;l)}_{KK'}e^{-iK'\frac{2\pi}{2J+1}n'},
\end{align}
where $\phi$ is angle with the $s$-axis in the $s$-$m$-plane and the 
amplitudes $\langle n|K\rangle$ are given by Eq.~(\ref{eq:phin}). 
Fig.~\ref{f:TRphi-plot} shows the same states displayed in the 
$K$-plots, where it retains the information about the phase 
factors by showing open and full symbols for $+$ or $-$.

The probability distributions of the SSS (\ref{eq:squeezed1}),
\beq\label{eq:PSSS}
 P(\phi)_{I\nu}= \frac{1}{2 \pi} 
  \sum_{K,K'}e^{i(K-K')\phi}\rho^{(\nu;l)}_{KK'},
\eeq
are shown in addition.
 
The localization of the zero- and one-phonon wobbling states $8_1$ 
and $9_1$ is well displayed in the $\phi$-basis as the peaks around 
$\phi=\pm \pi/2$. The vibrational character of the $9_1$ state is 
seen as the zeros at $\phi=\pm \pi/2$. The discrete $P(\phi_n)$ 
distribution of the $8_5$ state is counter intuitive. As 
$K=8$ is almost a good quantum number (c.f. Figs.~\ref{f:TRorbits} 
and \ref{f:TRWF}), the probability for all $\phi_n$ ought to be 
about the same. Instead a decrease toward $\phi=\pm \pi$ is seen. 
It is caused by the symmetrization of the state, 
$\left(\vert IIK\rangle+\vert II-K\rangle \right)/\sqrt{2}$, 
which causes an interference between the $\phi_{\pm n}$ states. 
The $P(\phi)$ density of the squeezed states looks like expected. 
It oscillates in terms of $\propto \cos^2(K \phi)$ while the envelop 
is roughly constant.  

\subsection{Spin Coherent State (SCS) maps}

A new elucidating visualization is to map probability distribution of the 
SCS. Such SCS maps have first been produced for the two-particle triaxial 
rotor model in order to illustrate the chiral geometry~\cite{Frauendorf2015Conf}. 
Later the authors of Ref.~\cite{F.Q.Chen2017PRC} used  SCS maps, 
which they called ``azimuthal plots", to visualize the appearance of chiral 
geometry in the results of quantal calculations in the framework of 
the angular momentum projection (AMP) method and in the two-particle 
triaxial rotor model~\cite{Q.B.Chen2018PRC_v1}. In Ref.~\cite{Streck2018PRC}, 
the azimuthal plots were first used to study the wobbling geometry. 
In this section we demonstrate the potential of the method to extract 
the classical mechanics underpinning of the quantal triaxial rotor 
model from the numerical results. 

\begin{figure*}[th]
 \includegraphics[width=\linewidth]{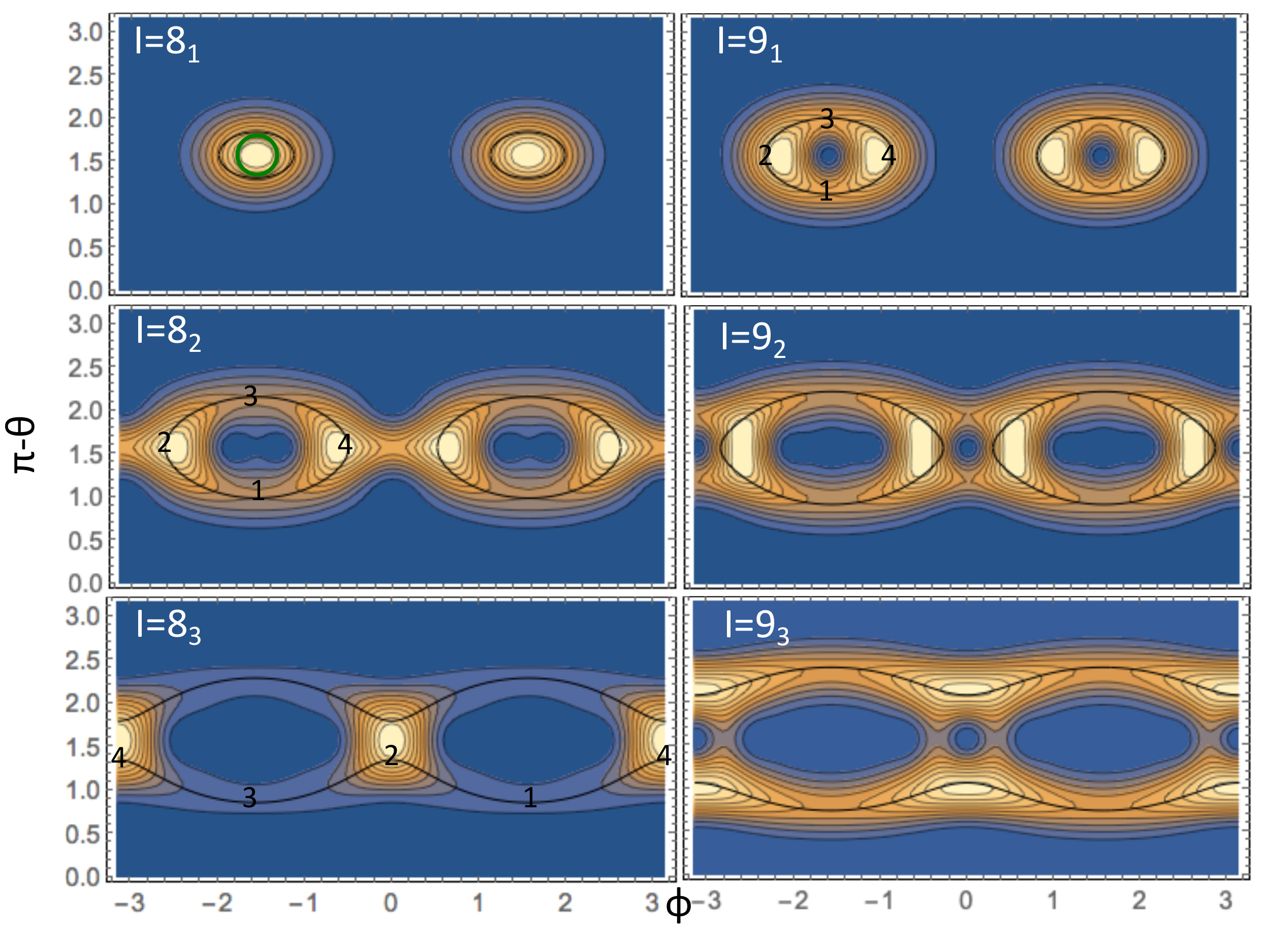}
 \caption{\label{f:TRmaps1} SCS probability densities $P(\theta \phi)_{I\nu}$
 of the triaxial rotor states for $I=8$ and 9 in
 cylinder projection. The states are labelled by $I_m$ in direction 
 of ascending energy. Color sequence with increasing probability: 
 dark blue $-$ zero level, light blue, dark browns, light brown, white. 
 The densities are  normalized. The classical orbits are shown as 
 black full curves. The energies are equal to their quantal energies in
 Fig.~\ref{f:TREnergy}. The numbers indicate the turning points.}
\end{figure*}

\begin{figure*}[th]
 \includegraphics[trim=0 5.8cm 0 0,clip,width=\linewidth]{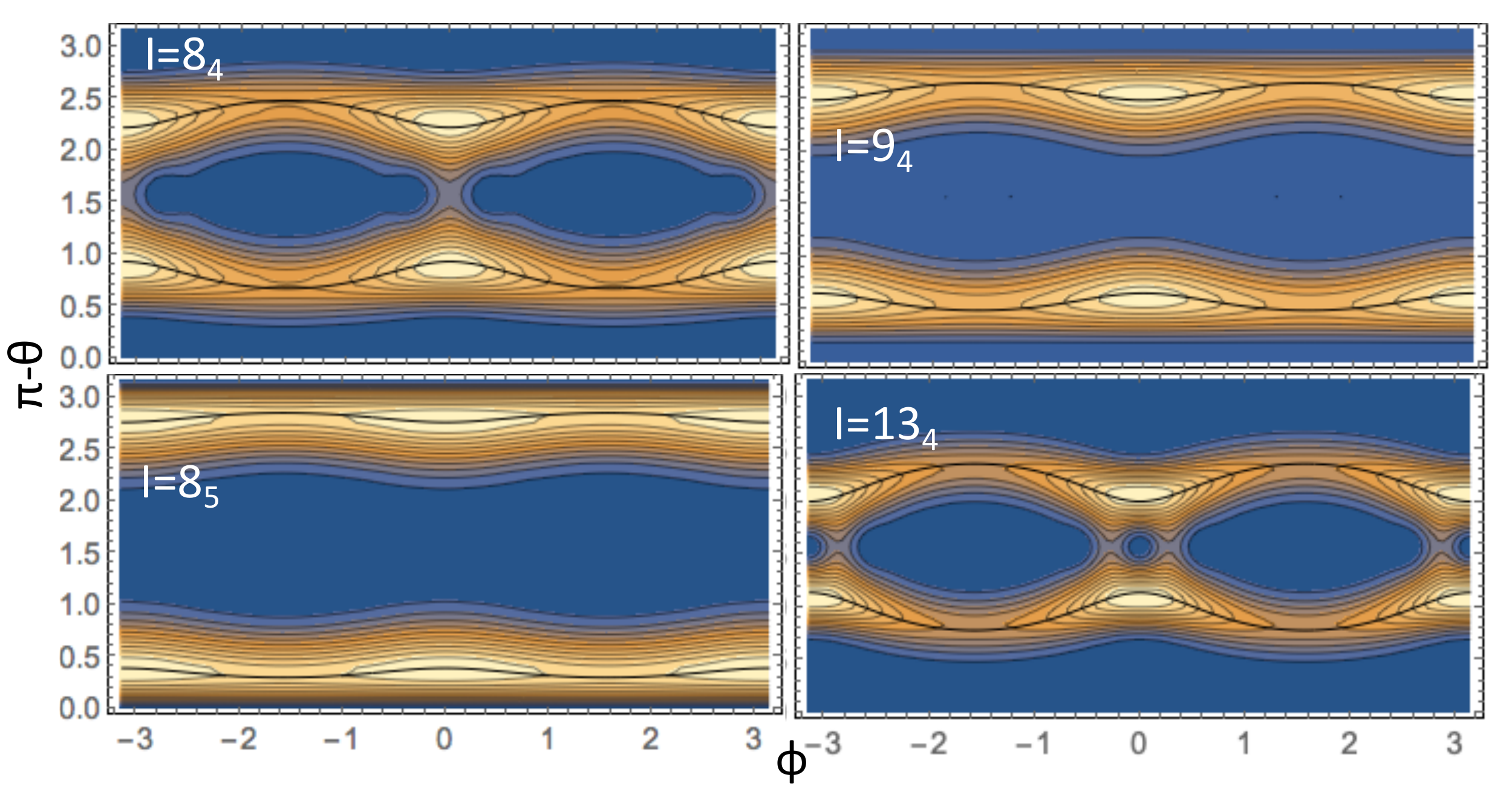}
 \caption{\label{f:TRmaps2} Continuation of  Fig.~\ref{f:TRmaps1} and 
  the results for $13_4$ state.}
\end{figure*}

\begin{figure*}[ht]
\includegraphics[angle=0, clip, width=\columnwidth]{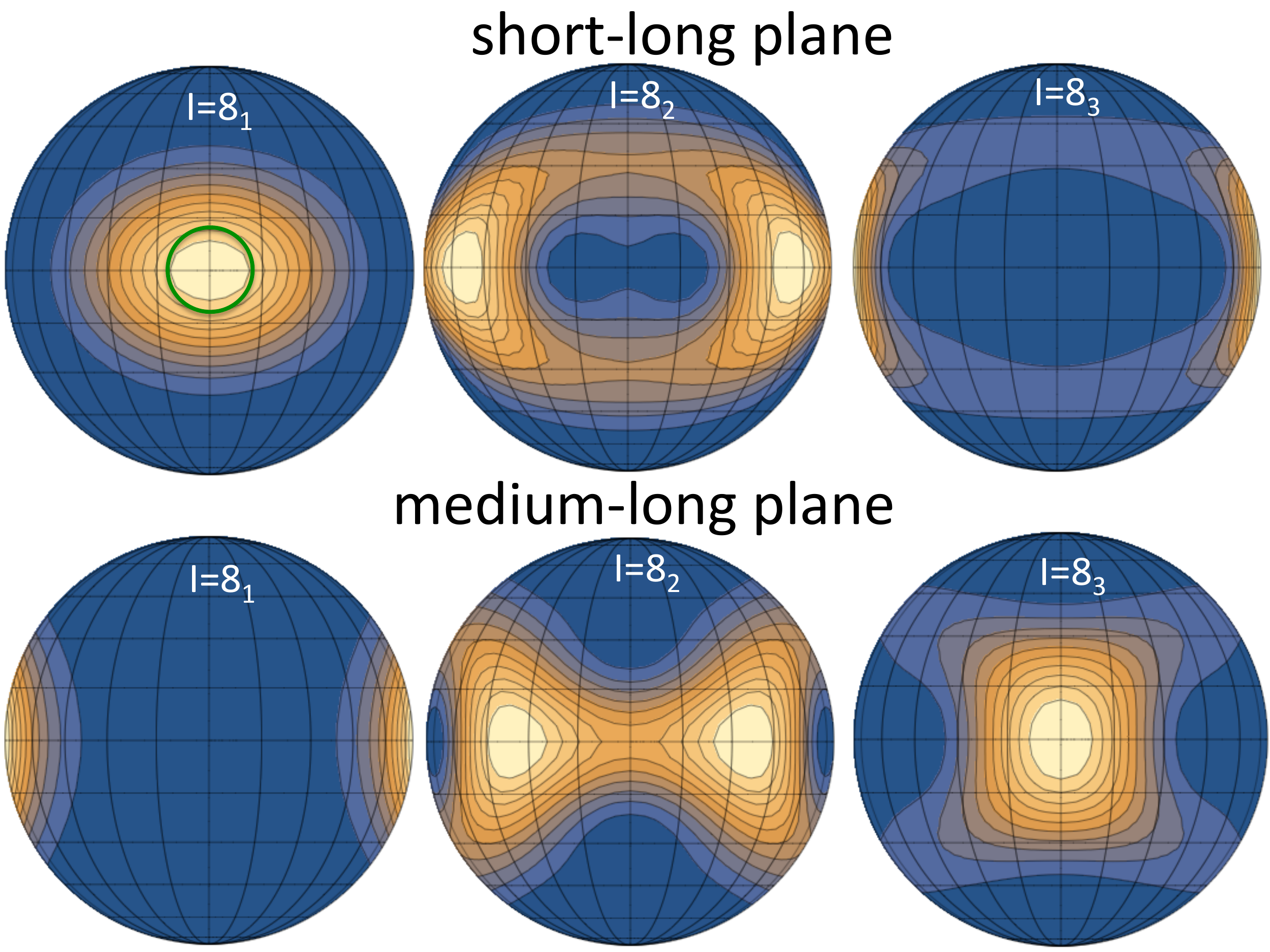}\quad
\includegraphics[angle=0, clip, width=\columnwidth]{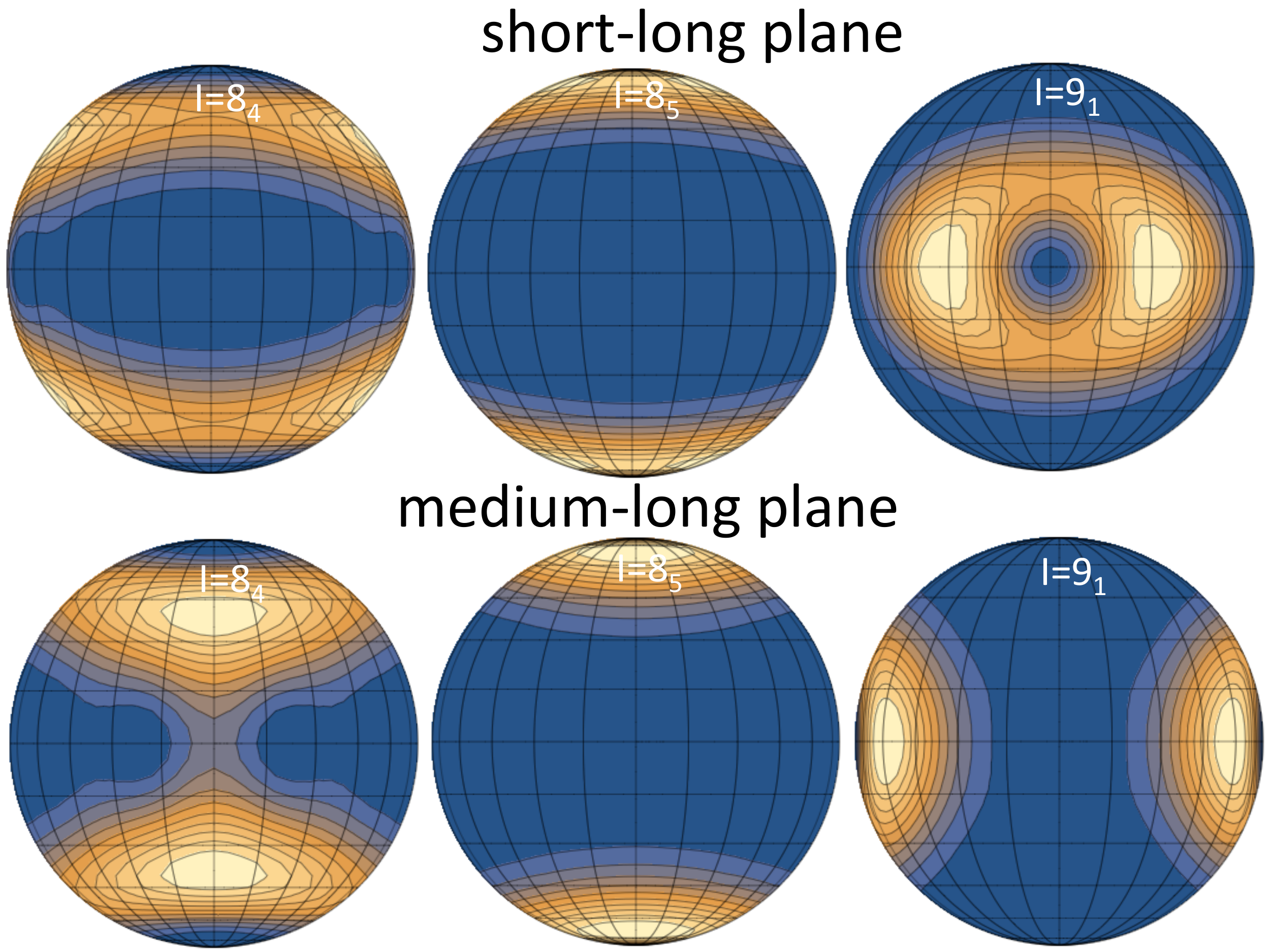}
 \caption{\label{f:TRmapsCS} SCS probability distributions 
 $P(\theta \phi)_{I\nu}$ of some  the states shown in Fig.~\ref{f:TRmaps1} 
 projected on the $s$-$l$-plane (viewpoint on $m$-axis) and the 
 $m$-$l$-plane (viewpoint on $s$-axis). Identical color code is used.}
\end{figure*}

The rotor states in the SCS basis are
\bea\label{eq:SCSTR}
 \langle I\theta\phi\vert II \nu\rangle
 =\sqrt\frac{2I+1}{8\pi^2} \sum_K D^{I*}_{KI}(\phi \theta 0)C^{(\nu)}_{IK}.
\eea
Their probability distributions are
\begin{align}
 P(\theta \phi)_{I\nu} &=\frac{2I+1}{4\pi}\sin\theta\notag\\
 &\quad \times \sum_{KK'} D^{I}_{KI}(\phi \theta 0)
 \rho^{(I\nu)}_{KK'}D^{I*}_{K'I}(\phi \theta 0),
 \label{eq:PSCS}\\
 \int_0^\pi d\theta & \int_0^{2\pi} d \phi~P(\theta \phi)_{I\nu}=1.
 \label{eq:PSCSn}
\end{align}

Fig.~\ref{f:TRmaps1} shows contour plots  of the probability distributions
$P(\theta \phi)_{I\nu} $ for the five $I=8$ rotor states and the first $I=9$
state, which we call SCS maps. Note, the maps shown in Refs.~\cite{Frauendorf2015Conf, 
F.Q.Chen2017PRC, Q.B.Chen2018PRC_v1, Streck2018PRC} do not contain 
the scale factor $\sin \theta$ of the surface element on the unit sphere. 
We include it to ensure that the integral (\ref{eq:PSCSn}) is equal to one, 
and $P(\theta \phi)_{I\nu}$ is the probability density on the cylindrical 
projection.

Preparing a SCS map, one has to decide how to project the surface of
the sphere of constant \am onto the map, which is analog to displaying 
the topography of the earth. Figs.~\ref{f:TRmaps1} and \ref{f:TRmaps2} use 
the equidistant cylinder projection (invented by Marinus of Tyre 100 AD). 
Fig.~\ref{f:TRmapsCS} re-displays three of the states using the orthographic
projections on the $m$-$l$-plane, where the  view point lies on the
$s$-axis on infinite distance, and the orthographic projections on
the $s$-$l$-plane, where the  view point lies on the $m$-axis
at infinite distance (invented by Hipparchos 200 AD). In the
orthographic projection $P(\theta \phi)_{I\nu} \Delta \theta \Delta \phi$
is the probability for the SCS to be located in the small trapezoidal
patch delineated by the  coordinate lines.

The SCS are eigenstates of the \am projection on the $\theta$-$\phi$-axis,
\begin{align}
(\cos \theta & \hat J_3  +\sin \theta\cos\phi \hat J_1+\sin \theta\sin\phi \hat J_2)
\vert I \theta\phi\rangle \notag\\
&=I\vert I \theta\phi\rangle,
\end{align}
where the axes are assigned as 3-$l$, 1-$s$, 2-$m$. That is, Figs.~\ref{f:TRmaps1}, 
\ref{f:TRmaps2}, and \ref{f:TRmapsCS} show the probability for the \am being 
oriented in a specific direction with respect to the principal axes of the 
body-fixed frame. However, ``oriented'' has to be understood in a restricted 
sense. The orientation is only specified within a distribution with the  
width discussed in the context of Eq.~(\ref{eq:overlap}). We illustrate 
the uncertainty of the orientation as a green circle with the radius 
$\arcsin [1/\sqrt{2I}]$. In other words, the SCS maps are blurred by the 
Uncertainty Principle like a poorly resolved picture of the Earth's 
surface taken from large distance (Zoom out Google Earth).

\subsubsection{Ridges and classical orbits}\label{s:ridges}

The full lines in Figs.~\ref{f:TRmaps1} and \ref{f:TRmaps2} depict the 
classical orbits from Fig.~\ref{f:TRorbits} for the angular momentum $I+1/2$ 
and the quantal energies as given in Fig.~\ref{f:TREnergy}. The ridges of 
the SCS probability density distributions trace the corresponding classical
orbits, that is, the SCS distribution is a fuzzy reproduction of the
classical orbit. Note, the close correspondence between the ridges
and the classical orbits appears only when the geometric scale factor
$\sin\theta$ is included in $P(\theta \phi)_{I\nu}$. The SCS maps 
filter out the classical mechanics underpinning of the quantal TRM 
with the best resolution permitted by the Uncertainty Principle. 

The exact location of a ridge is found by determining the minimum of
the square of the gradient $(\partial P/\partial \theta)^2+
(\partial P/\partial \phi)^2$ as function of $\theta$ for given
$\phi$ or as function of $\phi$ for given $\theta$. Alternatively,
one may search for the maximal curvature~\footnote{The curvature of
the contour line $P(x,y)=\textrm{const}$ is given by $(-P_{xx} P_y^2
+2P_{xy}P_xP_y-P_{yy}P_{x}^2)/P_y^3$, where $P_x$, $P_y$ denote
first order and $P_{xx}$, $P_{xy}$, $P_{yy}$ second order partial
derivatives. The expression holds when the contour line is a single-valued
function $y=f(x)$. In case it is a single-valued function $x=f(y)$
the curvature is $(-P_{xx} P_y^2+2P_{xy}P_xP_y-P_{yy}P_{x}^2)/P_x^3$.}
of the contour line. In principle, the two methods ought to give the
same result~\footnote{Approximate the region near the tong tip of a 
contour by an ellipse. The tip is located at the long semi axis where 
the curvature is maximal and distance to a slightly larger ellipse that
approximates a nearby contour is maximal.}. However, the equivalence
holds only for infinitesimal distances. Approximating derivatives by
finite differences on a grid of 1 degree, we found slight
differences that are not visible on the scale of the figures.

\begin{figure}[t]
 \includegraphics[width=\columnwidth]{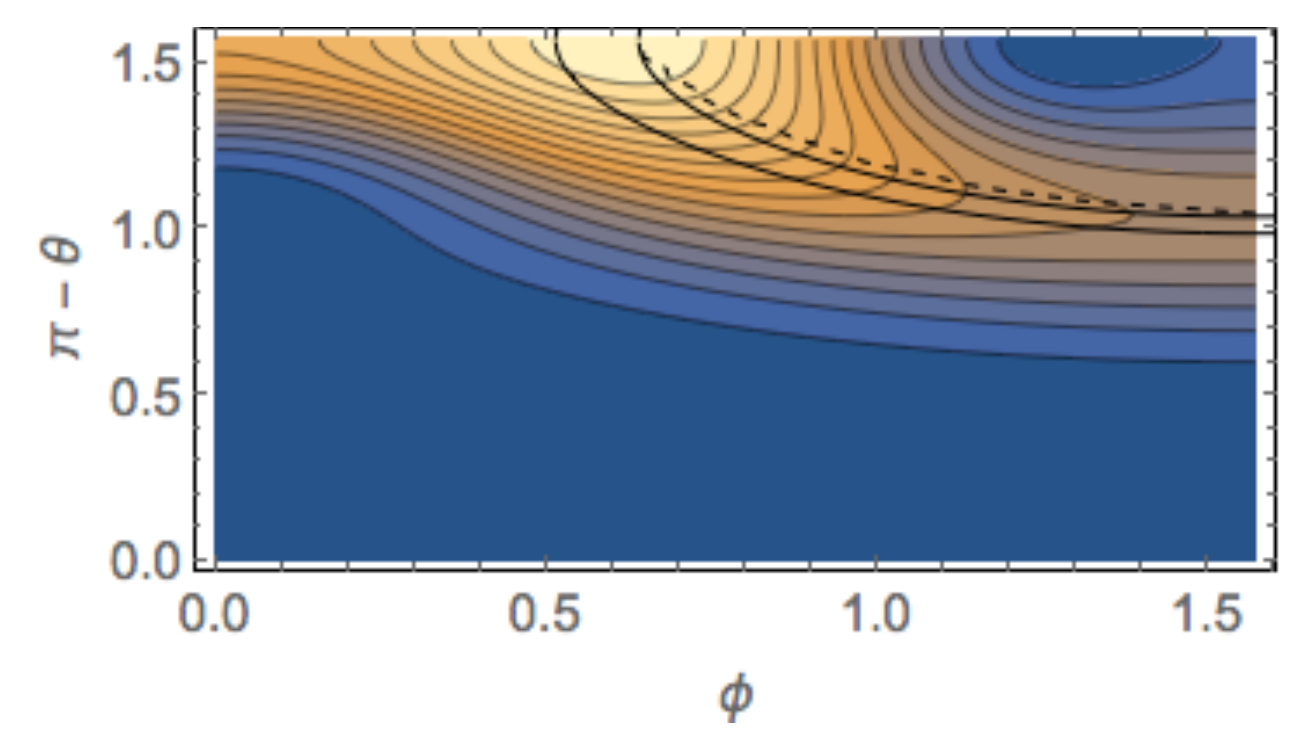}
 \caption{\label{f:TRmap82} Close-up of SCS probability distribution
 $P(\theta \phi) $ of the state 8$_2$ shown in Fig.~\ref{f:TRmaps1}.
 Only one quarter of the distribution is shown. The dashed curve is
 the numerically determined top of the ridge. The full curves show
 the classical orbit for the exact quantal energy $E(8_2)$ (turning
 point at smaller $\phi$) and $E(8_2)/1.1$ (turning point at larger
 $\phi$).}
\end{figure}

As an example, Fig.~\ref{f:TRmap82} shows the ridge for the state $8_2$ 
determined numerically in this way. The SCS map filters the corresponding 
classical orbits from the quantal state as the location of the ridge.
Although being close, the ridges deviate somewhat from the classical
orbits with energy equal to the quantal energy of the rotor. As seen in
Fig.~\ref{f:TRmap82} the classical orbit with the energy $E(8_2)/1.1$
comes very close to the numerical determined ridge. The deviations 
are caused by the symmetrization of the quantal wave function, 
which couples the two distinct classical orbits.

As seen in Fig.~\ref{f:TRmaps1}, the ridge of the $9_1$ state is somewhat 
smaller than the classical orbit, because the state contains only the 
components $\vert K\vert$=8, 6, 4, 2. The classical orbit also contains 
the projection $J_3=9$, which has a larger distance from the center. 
For the 8$_2$ state, the ridge  is closer to the classical orbit because 
the wave function contains the $\vert K\vert$=8 component with $J_3=8$.
The turning points 2 and 4 of the ridge are somewhat closer to the 
center than the ones of the classical orbit. This is caused by the 
interaction with the turning points of the backside orbit (centered 
at $\phi=-\pi/2$), which is seen as the bridges that connect them 
through $\phi=0$, $\pm \pi$.

\subsubsection{Velocity and phase}

Janssen~\cite{Janssen1977SJNP} has shown that the expectation values
$\langle I \theta(t)\phi(t)\vert \hat{J}_i \vert I\theta(t)\phi(t)\rangle$
obey the Euler equations which govern the classical motion of the triaxial
top~\footnote{ This is a special case of the time development of coherent
states being governed by the classical equations of motion.}. This suggests an
extension of the interpretation in classical terms: the scale of the SCS
probability density represents the fraction of the period time that
the rotor stays in a section of the orbit.

The probability to be in a square that encloses $(\theta, \phi)$ and has
the edges $dl_\parallel $ parallel to the ridge to and $dl_\perp$ 
perpendicular to it is given by
\begin{align}
 dW(\theta,\phi)=P(\theta \phi)_{I\nu} dl_\parallel dl_\perp.
\end{align}
Classically, the probability $dW(\theta,\phi)$ to be in the interval
$dl_\parallel$ of the orbit that encloses $(\theta,\phi)$ is given by
\beq
 dW(\theta,\phi)=\frac{dt}{T}=\frac{dl_\parallel}{\omega_\parallel T},
\eeq
where $\omega_\parallel$ is the angular velocity tangential to the
orbit and $T$ the period. That is, if the SCS states obey the classical
equations of motion, the probability of the SCS to be in the interval
$dl_\perp $ perpendicular to the ridge top is
\beq
 dW=dl_\perp \frac{dl_\parallel}{\omega_\parallel T},
\eeq
which means $1/P(\theta \phi)_{I\nu} \propto \omega_\parallel (\theta \phi)$.
The classical angular velocity tangential to the orbit is
\begin{align}
 \omega_\parallel(\phi)
 &=\frac{dl_\parallel}{dt}=\frac{dl_\parallel}{d\phi}\frac{d\phi}{dt},\\
 \frac{dl_\parallel}{d\phi}
 &= \sqrt{1+\left(\frac{d\theta(\phi)}{d\phi}\right)^2},\\
 \frac{d\phi}{dt}
 &=J_3(\phi)\left(\frac{1}{{\cal J}_3}-\frac{\cos^2\phi}{{\cal J}_1}
  -\frac{\sin^2\phi}{{\cal J}_2}\right),
\end{align}
where $J_3(\phi)$ and $\theta(\phi)$ are given by Eq.~(\ref{eq:J3class}).
The result of $d\phi/dt$ is obtained from the fact that the 
angles satisfy Euler's differential equations. 

\begin{figure}[t]
 \includegraphics[width=0.9\columnwidth]{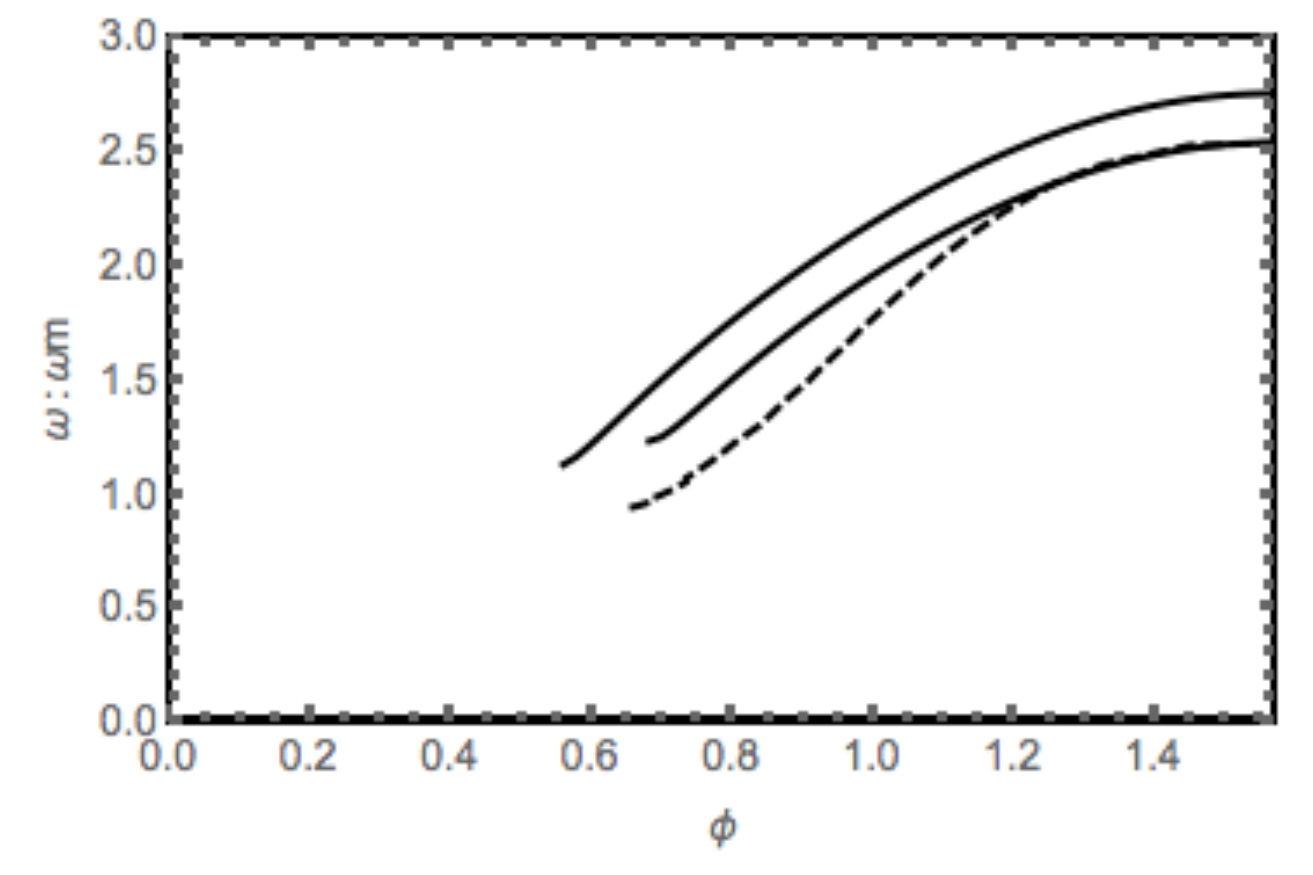}
 \caption{\label{f:TRomega82} Full curves: classical tangential
 angular velocity $\omega_\parallel(\theta\phi) $ for the exact
 quantal energy $E(8_2)$ (turning point at smaller $\phi$) and
 $E(8_2)/1.1$ (turning point at larger $\phi$) (See Fig.~\ref{f:TRmap82}).
 The dashed curve shows $a/P(\theta \phi)_{I\nu}$ along the
 numerically determined  top of the ridge shown in Fig.~\ref{f:TRmap82},
 where $a$ is chosen to match the classical velocity at $\phi =\pi/2$.
 The classical velocity is divided by $\omega_m=(I+1/2)/\mathcal{J}_2$,
 the angular velocity of uniform rotation about the $m$-axis.}
\end{figure}

Fig.~\ref{f:TRomega82} compares the classical angular velocity
$\omega_\parallel(\phi)$ for the two orbits shown in Fig.~\ref{f:TRmap82}
with the scaled inverse probability density $a/P(\theta \phi)_{I\nu}$
along the numerically determined top of the ridge shown in Fig.~\ref{f:TRmap82}.
The inverse of the probability density well approximates the angular velocity
of the classical orbit that comes closest to the ridge top. The deviations
near the turning point reflect the deviation of the classical orbit from the
ridge top seen in Fig.~\ref{f:TRmap82}.

The phase difference $\Phi(\theta\phi,\theta_0\phi_0)$ relative 
to some chosen point $\theta_0\phi_0$ (only relative phases have
a physical meaning), is given by
\begin{align}
\label{eq:phaseSCS}
 &\quad \Phi(\theta \phi,\theta_0\phi_0)_{I\nu} \notag \\
 &=-\Im\Big[ \log  \sum_{KK'} D^{I}_{KI}(\phi \theta 0)
 \rho^{(I\nu)}_{KK'}D^{I*}_{K'I}(\phi_0\theta_0 0) \Big].
\end{align}

Conservation of flux relates the phase $\Phi$ of the wave function
to its probability density $P$. For the SCS maps the conservation
implies that  $\int da P \vec n \cdot \nabla \Phi$ is constant,
where the integral is taken perpendicularly across the fuzzy orbit.
That is, when $P$ goes down $d\vec n \cdot \nabla \Phi$ must go up.
Therefore the phase change can be estimated from the probability
density $P$ in a qualitative way. As $d\vec n \cdot \nabla \Phi$ is
the momentum density in direction of the orbit, its connection with
$P$ just tells us that the rotor passes the regions of
low density faster than regions of high density, which repeats
the above discussed interpretation of the inverse of the probability
density as the tangential angular velocity.

\begin{figure}[t]
\includegraphics[width=\columnwidth]{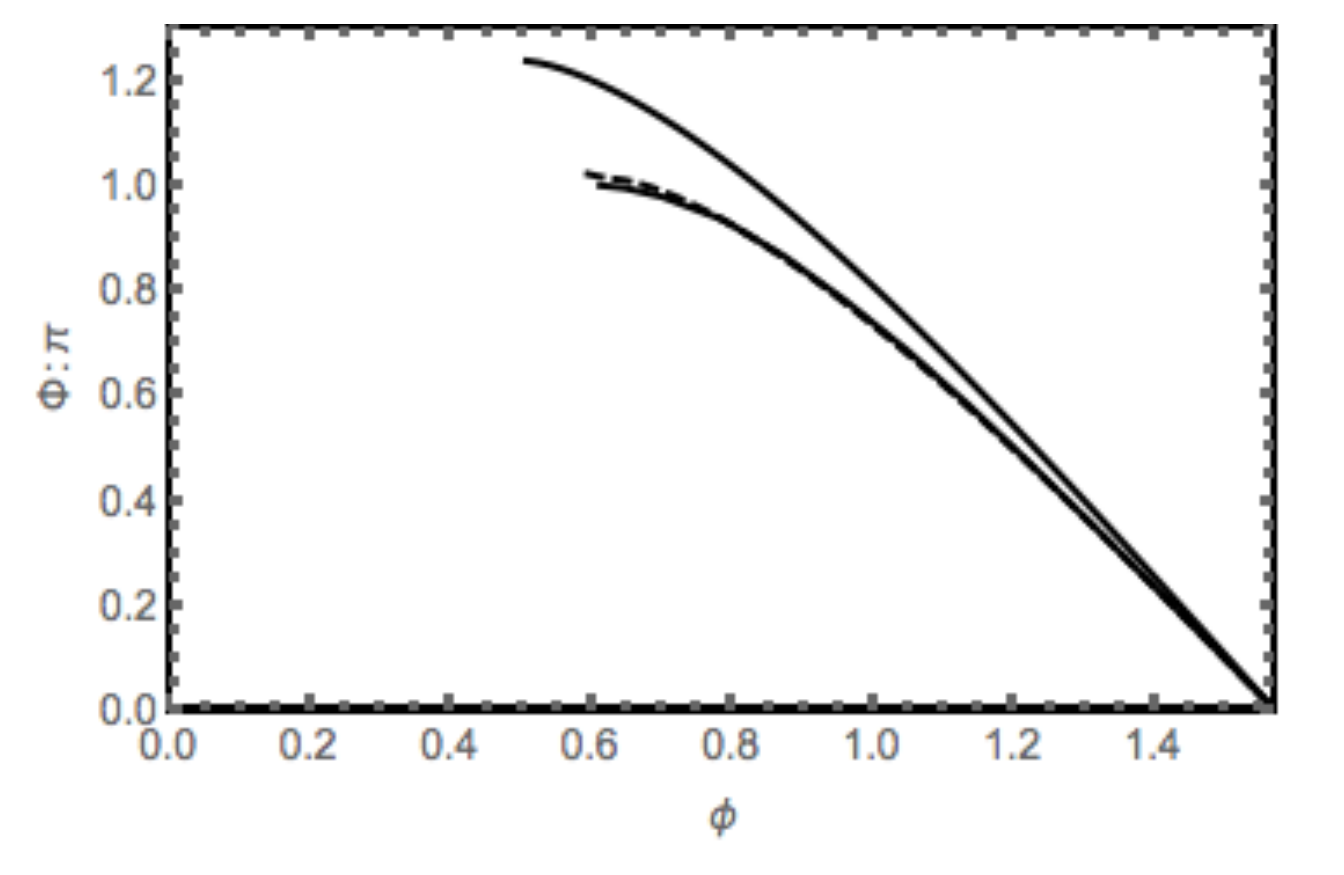}
\caption{\label{f:TRphase82} Phase difference $\Phi(\theta\phi,\theta_1\phi_1)$
relative to the turning point 1 of the state 8$_2$ shown in Fig.~\ref{f:TRmaps1}.
The dashed curve is calculated by means of Eq.~(\ref{eq:phaseSCS}) along the
numerically determined  top of the ridge shown in Fig.~\ref{f:TRmap82}.
The full curves show the classical phase difference for the
exact quantal energy $E(8_2)$ (turning point at smaller $\phi$)
and $E(8_2)/1.1$ (turning point at larger $\phi$) (See Fig.~\ref{f:TRmap82}).}
\end{figure}

Semiclassically, the phase is the mechanical action in units of $\hbar$,
that is
\beq\label{eq:phaseCL}
 \Phi(\phi)=\int_{\phi_0}^\phi J_3(\phi')d\phi',
\eeq
if $J_3$ is in units of $\hbar$ as commonly assumed. For the state
$8_2$, Fig.~\ref{f:TRphase82} compares the classical phase difference
(\ref{eq:phaseCL}) with the SCS phase difference (\ref{eq:phaseSCS})
along the path on top of the ridge. The latter nearly agrees with the classical phase of
the orbit that come closest 
to the ridge in Fig.~\ref{f:TRmap82}. That is, one can estimate the phase 
from the SCS map in orthographic projection (like Fig.~\ref{f:TRmapsCS}) 
as the area between the meridians $\phi$ and $\phi_0$ multiplied by
$I+1/2$.

\begin{figure}[th]
 \includegraphics[width=0.9\columnwidth]{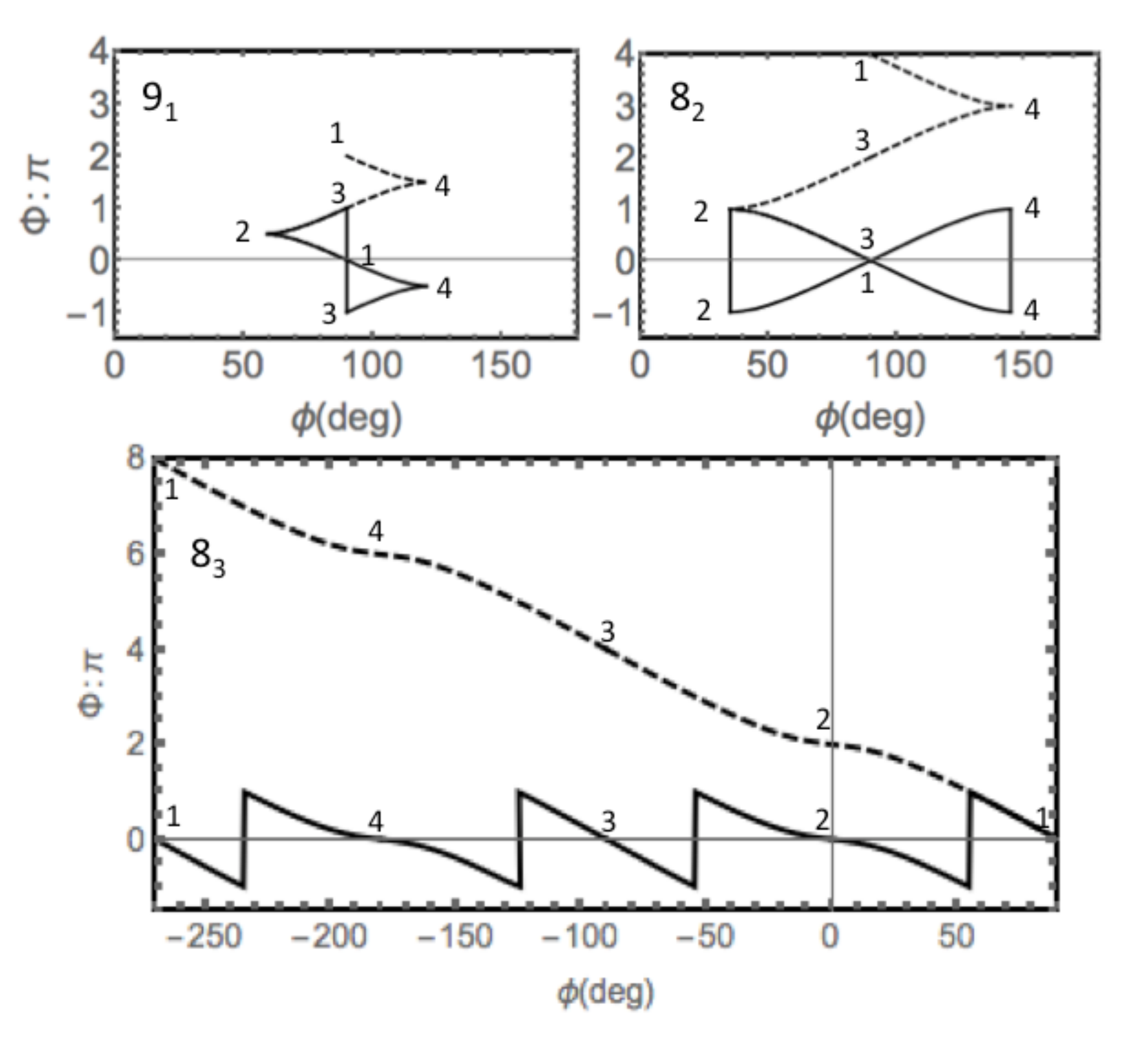}
 \caption{\label{f:TRphase} Phase difference $\Phi(\theta\phi,\theta_1\phi_1)$
 relative to the turning point 1 of the states $9_1$, $8_2$, and $8_3$ shown
 in Fig.~\ref{f:TRmaps1}. The numbers indicate the turning points.
 The full curve is calculated by means of Eq.~(\ref{eq:phaseSCS}) along the
 numerically determined top of the ridge shown in Fig.~\ref{f:TRmap82}
 (Around the turning points 2 and 4 of state $8_2$ the classical path is used).
 The dashed curves are constructed by adding $2\pi$, $4\pi$, ... at each jump of
 the full curve such that the continuous classical phase difference is obtained
 (See Figs.~\ref{f:TRmap82} and \ref{f:TRphase82}).}
\end{figure}

Fig.~\ref{f:TRphase} displays the phase differences $\Phi(\theta \phi,
\theta_1\phi_1)_{I\nu}$ for the complete orbits of the states 9$_1$,
8$_2$, and 8$_3$. The $\log$ function in Eq.~(\ref{eq:phaseSCS}) jumps
from $\pi$ to $-\pi$ at its branch cut. This generates the jumps of the
full curves in the figure. The phase is only determined up to a multiple
of $2\pi$. The dashed curves are generated by adding $2\pi n$ to the
full curve such that a continuous curve results, which represents the
increment of the  action along the path. The jumps make two-dimensional 
maps of the phase generated by means of Eq.~(\ref{eq:phaseSCS}) very complex. 
We did not find them useful for interpretation. A calculation of the phase
along the orbit (ridge of the probability density) provides useful
insight in the quantal nature of the state. Semiclassical quantization
requires that the classical action for a full turn must be $2 \pi n$.
The number of phase jumps in a figure like Fig.~\ref{f:TRphase}
provides such kind of quantum number $n$ for the angular momentum motion.

The phase gain after passing one turn of the periodic orbit is of topological 
nature. We checked numerically that any closed path in the SCS plane of a 
state which encloses $\theta=\pi/2$, $\phi=\pi/2$ gives a phase gain of 
respectively $2n$. This holds not only for the wobbling states but also 
for the states above the separatrix, which revolve the poles 
$\theta=\pm\pi/2$.

\subsubsection{Detailed discussions of the SCS maps}\label{s:SCSdetail}

Now we discuss the SCS maps in Fig.~\ref{f:TRmaps1} in more detail. 
The $I=8$ yrast state 8$_1$ represents uniform rotation about the 
$m$-axis. Accordingly the probability distribution is a blob centered 
at $\theta=\pi/2,\ \phi=\pi/2$. Because of the D$_2$ symmetry being 
even with respect to a rotation by $\pi$ about the three principal 
axes, there is another blob at $\theta=\pi/2,\ \phi=-\pi/2$. The 
doubling caused by the D$_2$ symmetry is present in all other SCS
maps.

The states $I=9_1$ and 8$_2$ are, respectively, the one- and two-phonon
wobbling excitations about the $m$-axis. Accordingly, their probability
distributions are fuzzy ellipses centered at the $m$-axis, where the size
of the two-phonon 8$_2$ ellipse is larger. The probability is highest
at the two turning points 2 and 4 of the coordinate $\phi$, where
classically the angular velocity $\dot\phi$ is zero, and it is the lowest
at the two turning points 1 and 3 of $\theta$ where the angular velocity has
a maximum. For the one-phonon state $9_1$ the phase calculated by means
of Eq.~(\ref{eq:phaseSCS}) is $0$, $\pi/2$, $\pi$, $3\pi/2$ at the turning
points 1-4, respectively. For the two-phonon state $8_2$ the phase is
$0$, $\pi$, $2\pi$, $3\pi$ at the turning points 1-4, respectively. The phase
increment corresponds to the action increment along the classical orbit,
which is used for semiclassical quantization (see Fig.~\ref{f:TRphase}
and Ref.~\cite{Frauendorf2014PRC}). As expected, the one-phonon wobbling
state is odd and the two-phonon state even under $\phi\rightarrow\pi-\phi$.

The state $8_3$ is close to the classical separatrix orbit in Fig.~\ref{f:TRorbits}.
The separatrix contains the stationary points $\theta=\pi/2$, $\phi=0$ and
$\theta=\pi/2$, $\phi=\pi$, for which the rotor rotates uniformly about the 
$s$-axis. These stationary points are unstable with respect to the dashed orbits,
where it takes infinite time to approach or leave the stationary point.
Accordingly, the SCS probability is centered around the $s$-axis, and
it has four extrusions in direction of the separatrix branches. As seen
in Figs.~\ref{f:TRorbits} and \ref{f:TRmaps1}, the classical orbit with
its quantal energy lies slightly outside the separatrix. That is, it
revolves the 3-axis ($l$-), and it is topologically different from the
so far discussed states, which revolve the 2-axis ($m$-). This is
reflected in Fig.~\ref{f:TRphase} which shows a steady increase of
the phase difference from $\phi=\pi/2$, where $\Phi=0$ to
$\phi=2\pi+\pi/2$, where $\Phi=8\pi$. The phase increment is small near
at the points 2 and 4 and large  only a little away from them. 
That is, the angular velocity $\dot \phi$ is, respectively,  
small near and large away from these points. In terms of classical motion 
this means the rotor stays for a certain time rotating about the $s$-axis, 
then it quickly flips to opposite direction of the $s$-axis, remains 
for the same time rotating about it, then it flips back, etc. We suggest 
the name \textit{axis-flip wobbling} for this flipping motion, which is 
repeated periodically. For macroscopic objects it is called the Dzhanibekov 
effect after the Russian astronaut, who unscrewed a wing nut under zero gravity, 
which slipped his hand and executed the axis-flip motion while flying 
through the space station. The reader can watch it on an entertaining movie 
on~\footnote{\url{https://www.youtube.com/watch?v=L2o9eBl_Gzw}.}. A recent 
mathematical analyse has been published in Ref.~\cite{Mardesic2020PRL}, 
where further references can be found.

For odd $I$ the D$_2$ symmetry requires that the wave function is odd 
with respect to a rotation about the $s$-axis by $\pi$. This is only possible 
if it is zero on the $s$-axis, which is seen as the hole in the distributions 
for the states $9_3$, $9_4$, and $13_4$. The acceleration away from the 
$s$-axis increases with the distance from it. Due to the presence of the hole,
the rotor remains a shorter fraction of the total period in this position than 
for even $I$, which means it spends a larger fraction on the flip orbit.

As seen in Fig.~\ref{f:TREnergy}, the energy difference between axis-flip 
states of even and odd $I$ is much smaller than the wobbling energy 
$\hbar\omega_w$ between two adjacent harmonic oscillation (HO) multi-phonon 
states. The transition from the HO limit to the axis-flip regime is gradual. 
The distance between adjacent states of opposite signature decreases. The 
fraction of the period the rotor stays near the $\phi$-turning points 
increases.

The \am of the state $8_5$ is as far as possible aligned with the
$l$-axis. Accordingly, the SCS probability density is maximal for
$\theta=20^\circ$ and depends weakly on $\phi$. The $8_4$ state
has two-phonon  structure with respect to the $l$-axis.
The polar angle $\theta\approx \pi/4$ and $3\pi/4$, and the azimuthal
angle $\phi$ revolves the $l$-axis.

The SCS plots disentangle the states that are superposed to generate
the D$_2$ symmetry. They appear at different places in the map. The
pertaining classical orbits revolve the respective axis in opposite
direction. This is clear from the orientation of the constituent SCS
with respect to the axis and is reflected by the opposite sign of 
the phase calculated by Eq.~(\ref{eq:phaseSCS}).

Using the symmetry properties of D-function, Eq.~(\ref{eq:SCSTR}) can
be rewritten as
\begin{align}\label{eq:TRWFlab}
 \langle I\vth\f\vert II \nu\rangle 
 =\sqrt{\frac{2I+1}{8\pi^2}}&\sum_K D^{I}_{KI}(\f \vth 0 )C^{(\nu)}_{IK},\\
 ~~\vth=-\theta, ~\f &=-\phi.
\end{align}
As $ D^{I}_{KI}(\f \vth 0)$ describes the orientation of the body-fixed axes 
in the laboratory frame, one recognizes that the SCS map also shows the probability 
for the rotor (its body-fixed axes) being oriented with respect to the laboratory
frame. The two perspectives represent the ``active'' rotation of the state
vectors generating the SCS basis versus the ``passive" rotation of the
coordinate system.

The panels $8_1$, $9_1$, and $8_2$ display, respectively, for the zero-, 
one-, and two-phonon states the wobbling motion of the $m$-axis about 
the $z$-axis of the laboratory system, along which the \am is aligned. 
The state $8_3$ lies energetically quite close to the separatrix in 
Fig.~\ref{f:TRorbits}. The panel shows that it corresponds to rotation 
about the unstable $s$-axis with low probability to move away. The states 
$8_4$ and $8_5$ correspond to precession of the $l$-axis about the 
laboratory $z$-axis. (Fig.~\ref{f:TRmapsCS} is quite helpful in realizing 
the different types of shape wobbling.)  

\begin{figure*}[th]
 \includegraphics[width=\linewidth]{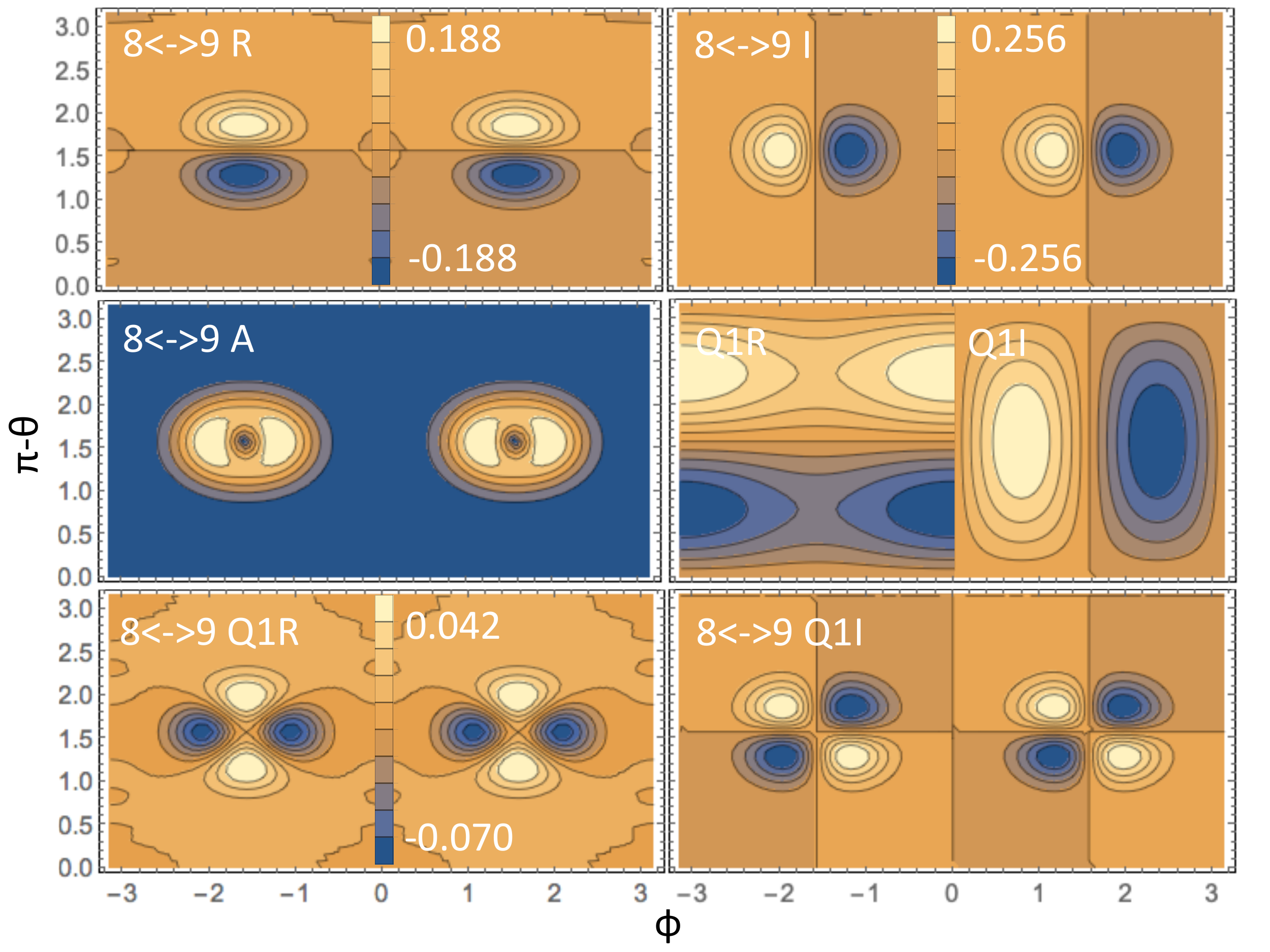}
 \caption{\label{f:TR9to8} SCS maps of the densities for the transition 
 $9_1\rightarrow8_1$ between the triaxial rotor states. 
 Upper panels: real and imaginary parts of the transition density; 
 middle panels: absolute value of the transition density (left) 
 and real and imaginary parts of the charge quadrupole moment (right); 
 lower panels: real and imaginary parts of the integral of 
 Eq.~(\ref{eq:Q1int}).}
 \end{figure*}
 
\subsection{Transition density maps}\label{sec:TDMTR}

The SCS probability density maps lose the information about the 
phase, which has to be exposed separately. Another way to retain 
this information is to map the transition density. It is well known 
from textbooks that the electromagnetic transition probabilities can 
be obtained from the classical radiation power by replacing the 
classical expression for the oscillating multipole by its the integral 
over the transition density obtained from the wave functions of the 
initial and final states (e.g., Ref.~\cite{Krane1988book}). Reversing 
the perspective, a SCS map of the transition density filters out 
the oscillating multipole that generates the transition. 

The $\Delta I=1$ $E2$ interband transition probability 
$B(E2)_{\textrm{out}}$ is calculated as
\begin{align}
 &\quad B(E2, I \to I-1)\notag\\
  &=\frac{2I-1}{(2I+1)\langle I I 2 -1|(I-1) (I-1)\rangle^2}\notag\\
  &\quad \times \frac{5}{16\pi} 
  \Big|\langle (I-1) (I-1) |Q_{2-1}|I I\rangle \Big|^2.
\end{align}
The matrix element is
\begin{align}
&\quad \langle (I-1) (I-1) |Q_{2-1}|I I\rangle \notag\\
 &=\sum_{KK^\prime} C_{(I-1)K^\prime}^*C_{IK}
  \langle (I-1)(I-1) K^\prime|Q_{2-1}|IIK\rangle\notag\\
 \label{eq3}
 &=2\pi\sum_{KK^\prime} C_{(I-1)K^\prime}^*C_{IK} 
  \iint d\theta \sin\theta  d\phi \notag\\
 &\quad \times \sqrt{\frac{2(I-1)+1}{8\pi^2}}
   D_{K^\prime(I-1)}^{(I-1)*}(\phi\theta 0)
   Q_{2,-1}(\phi\theta 0) \notag\\
 &\quad \times \sqrt{\frac{2I+1}{8\pi^2}}D_{KI}^{I}(\phi\theta 0),
\end{align}
with the quadrupole moment operator
\begin{align}
 &\quad Q_{2-1}(\phi\theta0)\notag\\
 & \propto
 D_{0-1}^2(\phi\theta0) \cos\gamma
 +[D_{2-1}^2(\phi\theta0)
 +D_{-2-1}^2(\phi\theta0)]\frac{\sin\gamma}{\sqrt{2}}\notag\\
 &= \sqrt{\frac{3}{8}}\sin2\theta \cos\gamma
 +\Big( -\cos2\phi \sin\theta\cos\theta \notag\\
 &\quad -i\sin2\phi\sin\theta \Big) \frac{\sin\gamma}{\sqrt{2}}.
\end{align}

We can define the transition density matrix as
\begin{align}\label{eq:TDM}
 \rho_{I \to I^\prime}(K, K^\prime)
  =C_{IK}C_{I^\prime K^\prime}^*, \quad I^\prime=I-1,
\end{align}
and define the transition SCS plot as
\begin{align}\label{eq:TDCS}
 & P(\theta\phi)_{I\rightarrow I^\prime}
  =\frac{\sqrt{(2I+1)(2I^\prime+1)}}{4\pi} \sin\theta \notag\\
  &\quad \times
  \sum_{KK^\prime} D_{KI}^I(\phi\theta0) \rho_{I \to I^\prime}(K, K^\prime)
  D_{K^\prime I^\prime}^{I^\prime *}(\phi\theta0).
\end{align}
The transition matrix element becomes an integral over 
the transition density and the quadrupole operator,
\begin{align}\label{eq:Q1int}
 &\quad \langle (I-1) (I-1) |Q_{2-1}|I I\rangle \notag\\
 &=\iint d\theta  d\phi~P(\theta\phi)_{I\rightarrow I^\prime} 
 Q_{2-1}(\phi\theta 0).
\end{align}
In classical radiation theory the matrix element (\ref{eq:Q1int}) 
is replaced by the corresponding integral, which contains the
time-dependent charge density $\rho(t)$ instead of the transition 
density. That is the SCS map of the transition density visualizes 
the corresponding classical motion of the charge density.

Fig.~\ref{f:TR9to8} illustrates the transition from the one-phonon 
wobbling state $9_1$ to the zero-phonon state $8_1$. The upper panels 
show the SCS maps of the transition density. The two blobs with 
opposite sign of the real part represent a linear oscillation of 
the body in $\theta$-direction and the two blobs of the imaginary 
part a linear oscillation in $\phi$-direction. The two linear 
oscillation with a relative phase shift of $\pi/2$ combine to 
an elliptical wobbling motion, which is shown by the absolute value 
in the middle panel. Multiplying the motion of the charged body
by the quadrupole operator in the right middle panel, gives the 
integral of the matrix element (\ref{eq:Q1int}). As the quadrupole 
matrix element is real, the integral over the imaginary part 
(lower left panel) is zero, as can easily be seen from its 
anti-symmetry. Only the integral over the real part is left.          

\section{Particle triaxial rotor model} \label{s:PTR}

Transverse and longitudinal wobbling as well as chiral modes are 
described by coupling particle(s) to the triaxial rotor, which 
leads to increasing complexity of the wave function. In order to 
visualize the \am constituents of interest the reduced density 
matrix is used, which is constructed by averaging over the degrees 
of freedom that are not of interest. In this section we discuss the 
particle-plus-triaxial rotor model (PTR), which couples one high-$j$ 
particle (hole) to the triaxial rotor core. In his seminal 
papers~\cite{MeyerTV1975NPA, MeyerTV1975NPA_v1}, Meyer-ter-Vehn 
generalized the approach to the quasiparticle triaxial rotor model 
and demonstrated its impressive capability to account for the 
experimental data in odd-$A$ nuclei available at the time. He 
interpreted the numerical results in the framework of a 
weak coupling approach. Here we focus on the analogies with 
classical mechanics of gyroscopes.  In particular, 
the concepts of transverse and longitudinal wobbling introduced 
by Frauendorf and D\"onau~\cite{Frauendorf2014PRC} will be 
substantiated by means of the various ways to analyse the 
quantal numerical results, which have been introduced in the 
preceding section.  

\subsection{Construction of the plots}

The PTR couples a high-$j$ particle to the triaxial rotor core. 
The corresponding Hamiltonian is
\begin{align}\label{eq:HPTR}
 H_{\textrm{PTR}}
 &=\sum\limits_{i=1,2,3}\frac{(\hat J_i-\hat j_i)^2}{2{\cal J}_i(\beta,\ga)}
  +h_{p}(\ga),\\
 \label{eq:hproton}
 h_p(\gamma)&=\kappa\left[\left(3j_3^2-\bm{j}^2\right)\cos\ga
     +\sqrt3\left(j_1^2-j_2^2\right)\sin\ga\right],
\end{align}
where $\hat{J}_i=\hat{R}_i+\hat{j}_i$ is the total angular momentum, 
$\hat{j}_i$ the \am of the particle, $\hat{R}_i$ the \am of the triaxial 
rotor, and $\kappa$ is the coupling strength to the deformed potential.

The PTR Hamiltonian is diagonalized in the product basis 
$\vert IIK\rangle\vert jk\rangle$, where $\vert IIK\rangle$ are the 
rotational states for half-integer $I$ and $\vert jk\rangle$ the high-$j$ 
particle states in good spin $j$ approximation. The eigenstates are
\begin{equation}
 |II\rangle=\sum_{K,k} C_{IKk} \vert IIK\rangle\vert jk\rangle.
\end{equation}
The coefficients $C_{IKk}$ of the states in the triaxially deformed 
odd-$A$ nuclei are not completely free. They are restricted by requirement 
 that collective rotor states must be symmetric representations 
of the D$_2$ point group: When the $K$ and $k$ in the sum run 
respectively from $-I$ to $I$ and from $-j$ to $j$, their 
difference $K-k$ must be even and one half of all coefficients 
is fixed by the relation $C_{I-K-k}=(-1)^{I-j}C_{IKk}$. 

From the amplitudes of the eigenstates $C_{IKk}$, the 
reduced density matrices
\beq\label{eq:TWrhoj}
 \rho_{kk'}=\sum_K C_{IKk}C_{IKk'}^*
\eeq
and
\beq\label{eq:TWrhoJ}
 \rho_{KK'}=\sum_k C_{IKk}C_{IK'k}^*
\eeq
are obtained, which
contain the information about the particle \am $\bm{j}$
and the total \am  $\bm{J}$, respectively.

Most commonly discussed quantities (e.g., in Refs.~\cite{B.Qi2009PLB, 
B.Qi2009PRC, Q.B.Chen2010PRC, Hamamoto2013PRC, H.Zhang2016CPC, 
Q.B.Chen2018PRC_v1, Streck2018PRC, Q.B.Chen2019PRC, Q.B.Chen2019PRC_v1, 
Q.B.Chen2020PLB, Q.B.Chen2020PLB_v1}) are the root mean square 
expectation values of the projections on the principal axes the 
rotor of the total angular momentum $\bm{J}$, the proton angular 
momentum $\bm{j}$ and the collective rotor angular momentum $\bm{R}$,
\begin{align}\label{eq:ANGPTR}
J_i&=\sqrt{\langle \hat J^2_i\rangle}=\sqrt{\sum_{KK'} 
    \rho_{KK'}\hat J^2_{i;K'K}}, \\
j_i&=\sqrt{\langle \hat j^2_i\rangle}=\sqrt{\sum_{kk'} 
    \rho_{kk'}\hat j^2_{i;k'k}}, \\
R_i&=\sqrt{\langle \hat R^2_i\rangle}\notag\\
   &=\sqrt{\sum_{KK'kk'} C_{IKk}^*\big(\hat J_i
    -\hat j_i\big)^2_{KK'kk'}C_{IK'k'}}.
\end{align}

One may define orientation angles of the classical vectors defined by means of 
the root mean square expectation values of the angular momentum components.
Another way to derive mutual orientation angles from PTR wave functions has
been introduced in Refs.~\cite{Starosta2002PRC, Tonev2007PRC}. It
consists of replacing the classical expression for the angle derived
from vector products by the corresponding quantal operator expression.
For example the angle between $\bm{J}$ and $\bm{j}$ the replacement is
\beq\label{eq:AngleJj}
 \phi_{Jj}=\arccos\Big[\frac{\bm{J}\cdot\bm{j}}{|\bm{J}||\bm{j}|}\Big]
 \rightarrow \arccos \Big[ \frac{\sum_i \langle \hat J_i \hat
 j_i\rangle}{\sqrt{I(I+1)j(j+1)}}\Big].
\eeq

The $K$-plots~\cite{B.Qi2009PLB} show the probability distribution 
with respect to three principal axes which are for the total 
angular momentum
\begin{align}
 l:&~P(K)=\rho_{KK}, \\
 m:&~P(K)= \sum_{K'K''} D^{I*}_{K'K}(0 \frac{\pi}{2} 0)
  \rho_{K'K''}D^{I}_{K''K}(0 \frac{\pi}{2} 0),\\
 s:&~P(K)= \sum_{K'K''} D^{I*}_{K'K}(\frac{\pi}{2} \frac{\pi}{2} 0)
  \rho_{K'K''}D^{I}_{K''K}(\frac{\pi}{2} \frac{\pi}{2} 0),
\end{align}
and the proton angular momentum
\begin{align}
 l:&~P(k)=\rho_{kk}, \\
 m:&~P(k)= \sum_{k'k''} D^{j*}_{k'k}(0 \frac{\pi}{2} 0)
  \rho_{k'k''}D^{I}_{k''k}(0 \frac{\pi}{2} 0),\\
 s:&~P(k)= \sum\limits_{k'k''} D^{I*}_{k'k}(\frac{\pi}{2}
  \frac{\pi}{2} 0)\rho_{k'k''}D^{I}_{k''k}(\frac{\pi}{2} \frac{\pi}{2} 0).
\end{align}
Alternatively, the probability distributions for the $m$- and $s$-axes
can be calculated by the simple expression for the $l$-axis and 
re-assigning the axes by changing $\gamma \rightarrow \gamma+2\pi/3$ 
and $\gamma\rightarrow \gamma+4\pi/3$. Fig.~\ref{f:TWK} shows the 
$K$-plots for selected yrast and wobbling states.

The SCS map for the total angular momentum $\bm{J}$ is given by 
Eq.~(\ref{eq:PSCS}) using the density matrix (\ref{eq:TWrhoJ}). 
For the particle angular momentum the map is given by
\beq\label{eq:TWmapsj}
 P(\theta \phi)=\frac{2j+1}{4\pi}\sin\theta
 \sum_{kk'}D^{j}_{kj}(\phi \theta 0) 
 \rho_{kk'}D^{j*}_{k'j}(\phi \theta 0),
\eeq
with the density matrix (\ref{eq:TWrhoj}). Fig.~\ref{f:TWmaps1} shows
the SCS maps in cylindric projection for the total \amd, which are
calculated by means of Eq.~(\ref{eq:PSCS}) using the density matrix
(\ref{eq:TWrhoJ}). Fig.~\ref{f:TWmapsCS} adds a selection of the
SCS maps in orthographic projection. Figs.~\ref{f:TWmapsjlow} 
and \ref{f:TWmapsjhigh} show the SCS maps for the odd proton in 
cylindrical projection. 

The phase changes between different angles of $\bm{J}$ is given by 
Eq.~(\ref{eq:phaseSCS}) with reduced density matrix using (\ref{eq:TWrhoJ}). 
It is important to realize that the reduced density matrix implies 
a certain degree of decoherence. The full density matrix (\ref{eq:RDM}) 
of the TRM represents one quantal state, which has fully coherent 
phase relations between different angles. In this case the phase 
differences are additive: $\Phi(\theta_2\phi_2,\theta_0\phi_0)
=\Phi(\theta_2\phi_2,\theta_1\phi_1)+\Phi(\theta_1\phi_1,\theta_0\phi_0)$. 
For the reduced density matrix this holds only approximately 
in case of weak decoherence or gets completely lost.

SCS plots of the transition density are obtained by means of 
Eq.~(\ref{eq:TDCS}) using the reduced transition density matrix (\ref{eq:TWrhoJ}).
The more detailed behavior of the core angular momentum $\bm{R}$ is 
visualized by the $K_R$ ($K_R=K-k$) and $R$ plots introduced in 
Refs.~\cite{Streck2018PRC, Q.B.Chen2019PRC}. The basis of the PTR 
eigenstates is re-coupled,
\begin{align}
 &\quad \vert IIjRK_R\rangle\notag\\
 &=\sum_{Kk}(-1)^{j+k}\langle j-kIK\vert RK_R\rangle 
 \vert IIK\rangle\vert j-k\rangle \sqrt{1+\delta_{K_R0}}\notag\\
 &\equiv \sum_{Kk}A^{IK}_{jk,RK_R} \vert IIK\rangle\vert j-k\rangle,
\end{align}
with
\begin{align}
 A^{IK}_{jk,RK_R}=(-1)^{j+k}\langle j-kIK\vert RK_R\rangle \sqrt{1+\delta_{K_R0}},
\end{align}
and a density matrix matrix is obtained from the components in 
the new basis
\bea
 \rho_{R,K_RK'_R}=\sum_{Kk,K'k'}A^{IK}_{jk,RK_R}C_{IKk}
  C^*_{IK'k'}A^{IK'}_{jk',RK'_R}.
\eea

The probability distribution for the projection $K_R$ on the 
3-axis is given by
\beq
 P_{K_R}=\sum_R \rho_{R,K_R K_R}.
\eeq
The distributions with respect to the 1- and 2-axes are obtained by
means of changing $\gamma\rightarrow-\gamma$ and $\gamma\rightarrow
2\pi/3-\gamma$. In contrast to $I$ and $j$ the absolute value of core
angular momentum is not constant. Its probability distribution
is given by
\beq
 P_{R}=\sum_{K_R} \rho_{R,K_R K_R}.
\eeq

The SCS map showing the orientation of the core angular momentum 
$\bm{R}$ with respect to the body-fixed coordinate system is 
constructed as
\begin{align}
 P_R(\theta \phi)
 &=\sin\theta\sum_{RK_RK'_R} \frac{2R+1}{4\pi}\notag\\
 &\quad \times D^{R}_{K_RR}(\phi \theta 0)
 \rho_{R,K_RK'_R}D^{R*}_{K'_RR}(\phi \theta 0),
\end{align}
and the distribution
\beq \label{eq:PRSCS}
P(\theta \phi)=\sum_RP_R(\theta \phi)
\eeq
shows the probability for the orientation of rotor angular 
momentum $\bm{R}$, where its length depends on the angles.

\begin{figure}[ht]
 \includegraphics[width=\columnwidth]{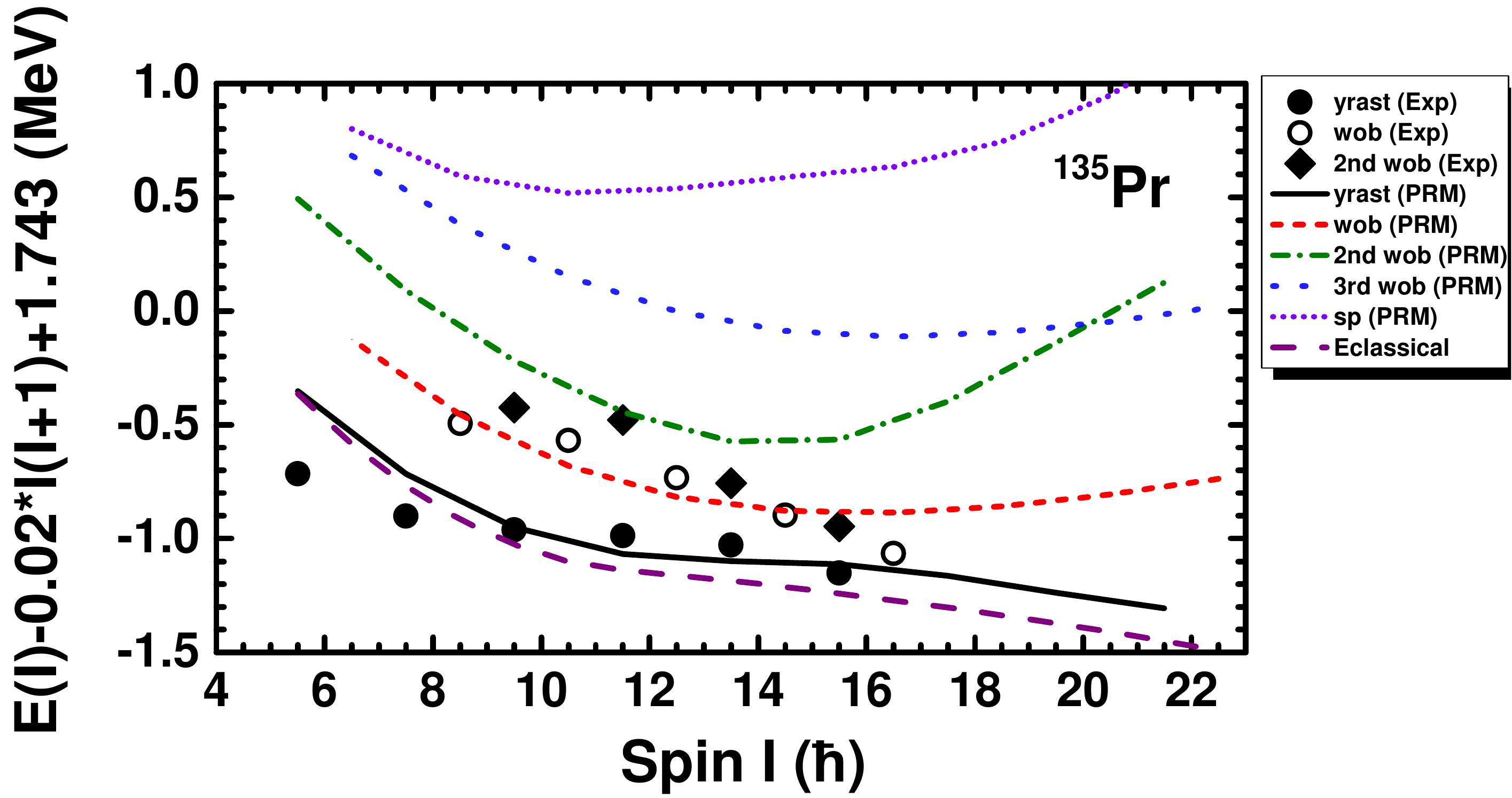}
 \caption{\label{f:E135Pr} Energy of the lowest states of the PTR
 Hamiltonian (\ref{eq:HPTR}) with the parameters for
 $^{135}$Pr~\cite{Frauendorf2014PRC}. The energies are shifted 
 by 1.743 MeV, which is the lowest energy from the diagonalization 
 of $h_p(\gamma)$ in Eq.~(\ref{eq:hproton}). In the following the 
 yrast states are denoted by $11/2_1$, $15/2_1$, $19/2_1$, ..., 
 the single wobbling excitations by $13/2_1$, $17/2_1$, $21/2_1$, ..., 
 and the double wobbling excitations by $15/2_2$, $19/2_2$, $23/2_2$, .... 
 The experimental energies~\cite{Matta2015PRL, Sensharma2019PLB} 
 of yrast states and single and double wobbling excitations are included.}

\end{figure}
\begin{figure}[ht]
 \includegraphics[width=0.75\columnwidth]{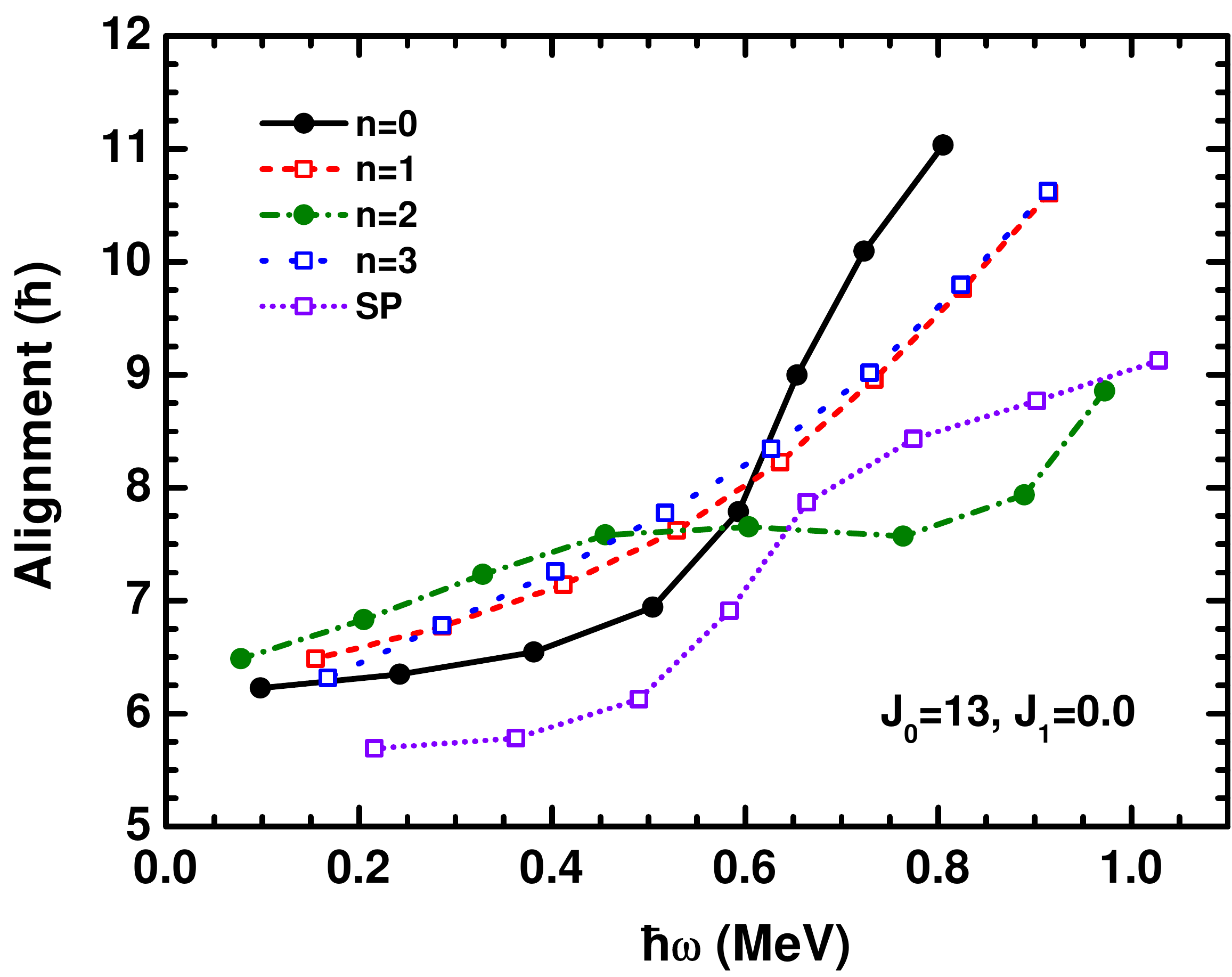}
 \caption{\label{f:i135Pr} Alignments of the
 $n=0$, 1, 2, 3 bands and SP band shown in Fig.~\ref{f:E135Pr}. 
 The alignment is calculated as 
 $i=I-(\mathcal{J}_0+\mathcal{J}_1\omega)\omega$, where 
 $\omega(I)=[E(I)-E(I-2)]/2$. The Harris parameters are taken as
 $\mathcal{J}_0=\mathcal{J}_s=13~\hbar^2/\textrm{MeV}$ and 
 $\mathcal{J}_1=0.0$.}
\end{figure}

We discuss the interpretation of the PTR model using $^{135}$Pr studied 
in Refs.~\cite{Frauendorf2014PRC, Streck2018PRC} as an example. The parameters 
of the PTR are $\beta=0.17$ (corresponds to $\kappa=0.038$), 
$\gamma=-26^\circ$, and $\mathcal{J}_{m,s,l}=21$, 13, $4~\hbar^2/\textrm{MeV}$. 
Fig.~\ref{f:E135Pr} shows the energies of the  lowest bands, which 
represent the zero-, one-, two-, and three-phonon wobbling states and the 
lowest excitation of the particle mode (signature partner state). 
Fig.~\ref{f:i135Pr} displays the standard angular momentum 
alignment~\cite{Bengtsson1979NPA} relative to a Harris reference, 
which corresponds to uniform rotation about the $s$-axis. 


\subsection{Geometry of the PTR states --- the 
 classical limit}\label{s:topology}

The classical analog to the quantal PTR corresponds to considering 
$J_3,~\phi_J$ and $j_3,~\phi_j$ as pairs of classical canonical variables, 
which means, in Eq.~(\ref{eq:can}) the commutator is replaced by the 
Poisson bracket. The classical PTR Hamiltonian is
\begin{align}\label{eq:HPTRcl}
 H_{\textrm{class}}
 &=\frac{(J_\perp\cos\phi_J- j_\perp\cos\phi_j)^2}{2{\cal J}_1}\notag\\
 &\quad+\frac{(J_\perp\sin\phi_J- j_\perp\sin\phi_j)^2}{2{\cal J}_2}\notag\\
 &\quad+\frac{(J_3- j_3)^2}{2{\cal J}_3}+h_{p}(\ga),\\
 h_p(\gamma)
 &=\kappa\Big[(3j_3^2-j^2)\cos\ga \notag\\
 &\quad+\sqrt{3}j_\perp^2(\cos\phi_j^2-\sin\phi_j^2)\sin\ga\Big],
\end{align}
with angular momentum projections on $s$-$m$-plane $J_\perp=\sqrt{J^2-J_3^2}$ 
and $j_\perp=\sqrt{j^2-j_3^2}$. Using the classical Hamiltonian, we replace 
$J^2\rightarrow I(I+1)$ and $j^2\rightarrow j(j+1)$, which is a semiclassical 
correction that brings the classical results closer to the quantal ones. 

\begin{figure}[ht]
 \includegraphics[width=0.95\columnwidth]{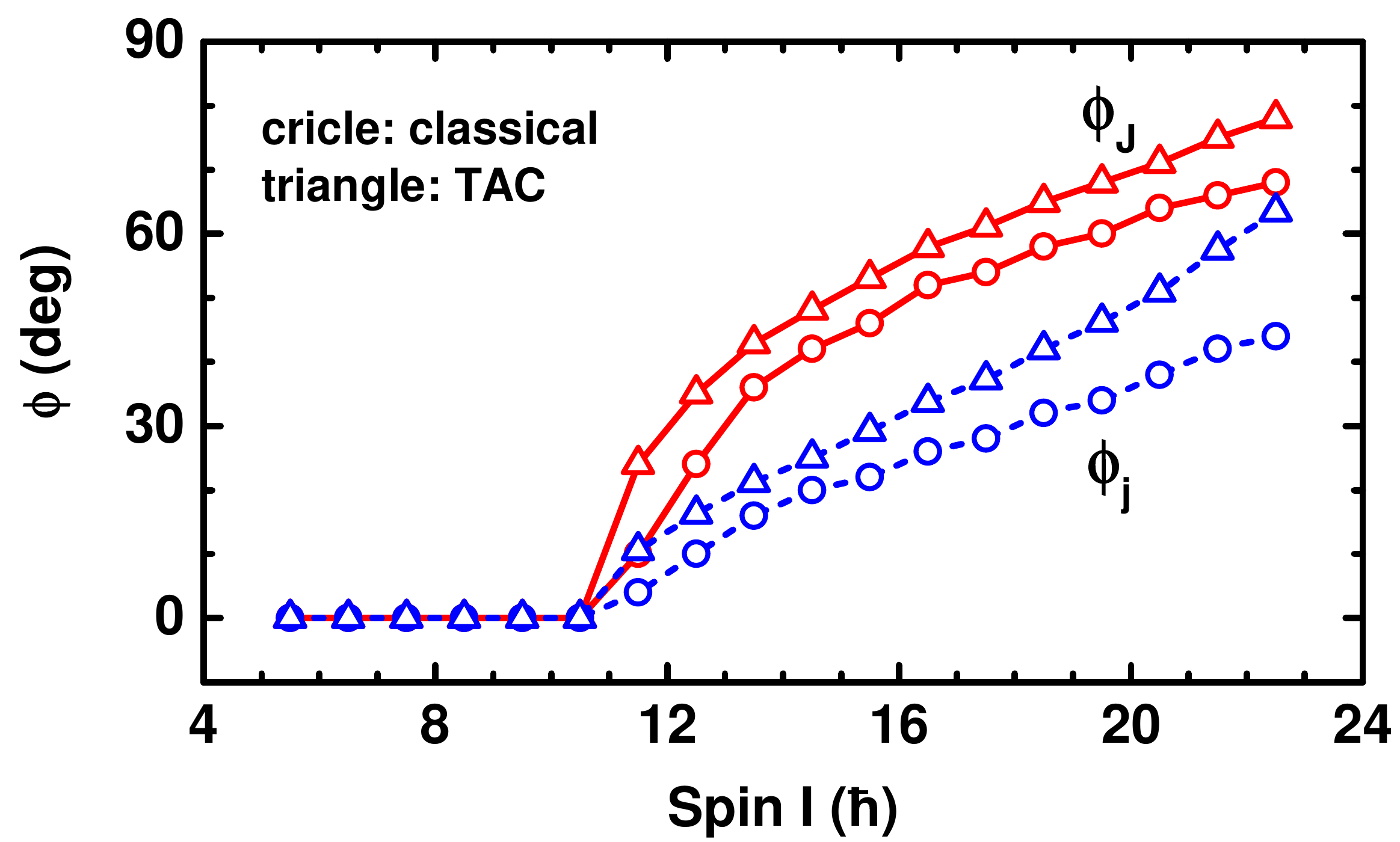}
 \caption{\label{f:ClassAngle} Orientation angles of $\bm{J}$ and
 $\bm{j}$ in the $s$-$i$-plane. The circles are obtained from minimizing the 
 classical energy. The triangles are calculated by means of the TAC method, 
 where $\phi_J$ is the angle of the minimum $E_{\rm{crank}}$ and $\phi_j$ 
 the angle obtained from the rms values of the $\bm{j}$ components shown 
 in Fig.~\ref{f:ANG135PrTAC}. The polar angles are $\theta_J=\theta_j=\pi/2$ 
 for all spin values.}
\end{figure} 

The system is two-dimensional and non-separable, that is, one 
cannot construct the orbits by conservation laws as for the TRM. 
Solving the classical equations of motion
\begin{equation}
 \dot{J_3}=\frac{\partial H}{\partial \phi_J},
 ~~\dot{\phi_J}=-\frac{\partial H}{\partial J_3},~~  
 \dot{j_3}=\frac{\partial H}{\partial \phi_j},
 ~~\dot{\phi_j}=-\frac{\partial H}{\partial \phi_j},
\end{equation}
is by far more complicated than finding the quantal states. Most
relevant for the interpretation of the latter is the topological 
classification of the classical motion.

\begin{figure}[ht]
 \includegraphics[width=0.95\columnwidth]{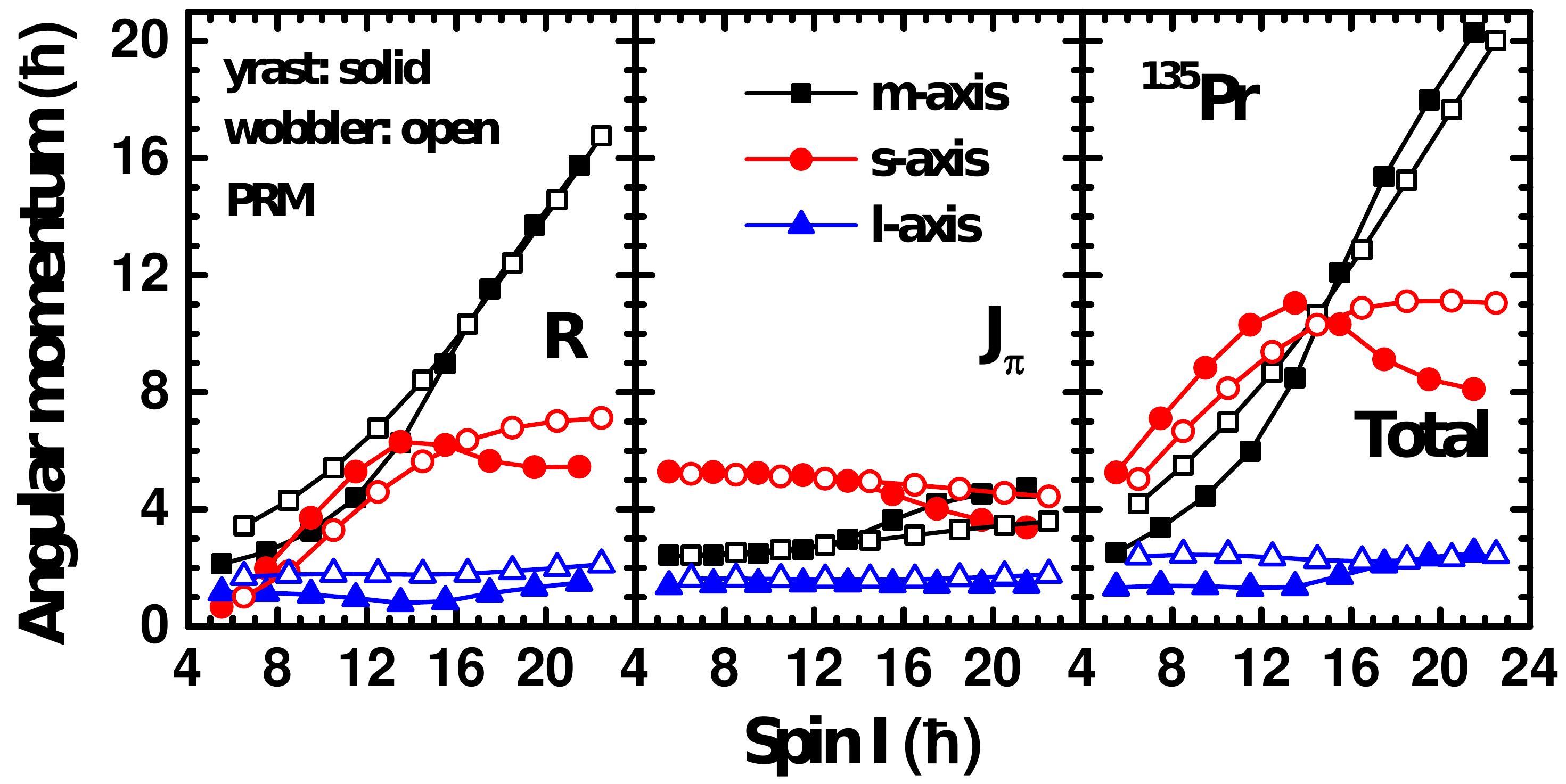}
 \caption{\label{f:ANG135Pr} Root mean square expectation values of the
 \am components by the PTR model. The \am of the $h_{11/2}$ proton is
 denoted by $\bm{j}_\pi$.}
\end{figure}

The static equilibrium configuration is found by minimizing  
$E_{\textrm{class}}(J, \theta_J, \phi_J; j, \theta_j, \phi_j)$ 
with respect to all four angles, which gives the classical 
correspondence of the yrast energy. The minimum lies at 
$\theta_J=\theta_j=\pi/2$ and at the $\phi_J$ and $\phi_j$ values 
shown in Fig.~\ref{f:ClassAngle}.
 
\begin{figure}[ht]
 \includegraphics[width=0.95\columnwidth]{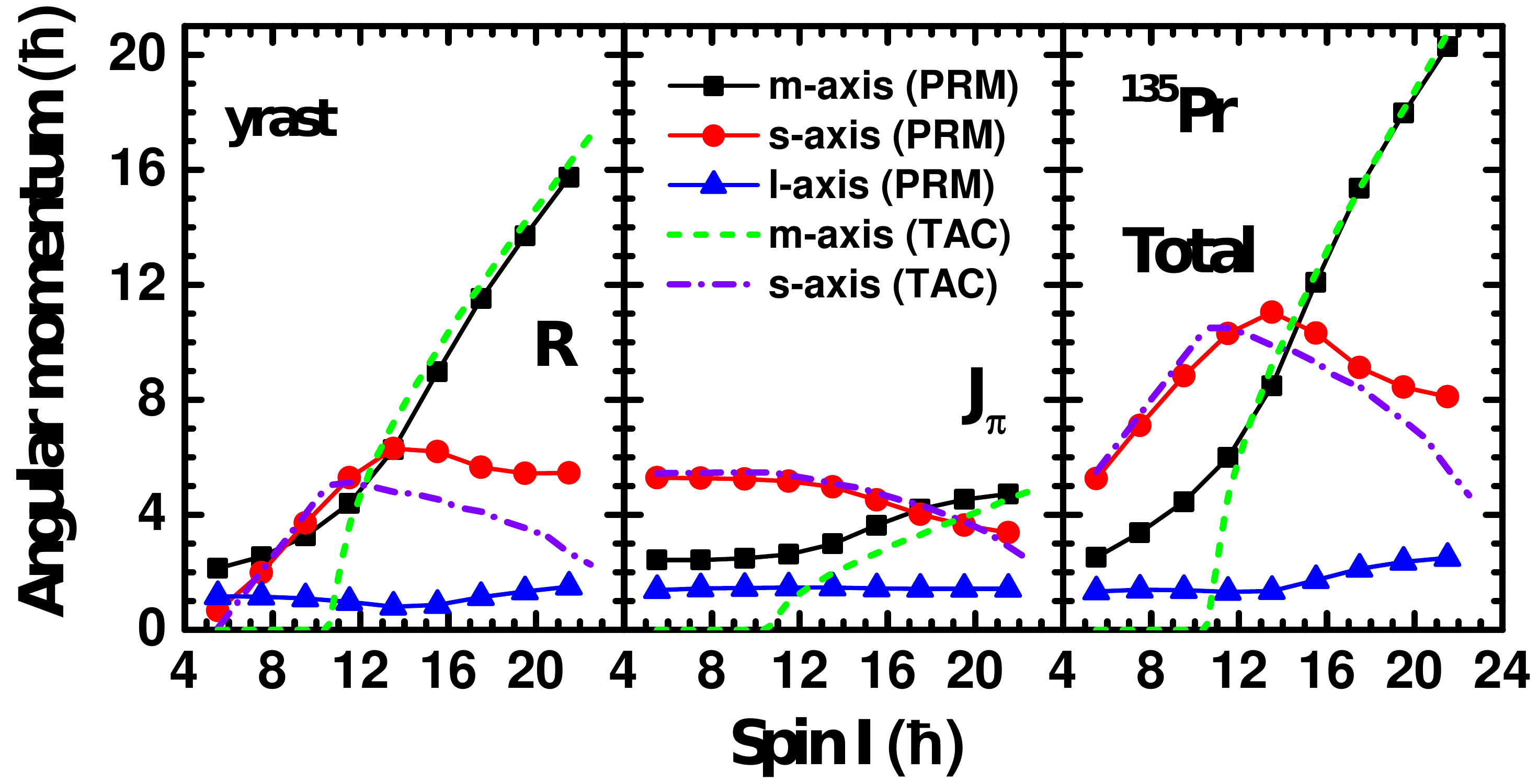}\\
 \includegraphics[width=0.95\columnwidth]{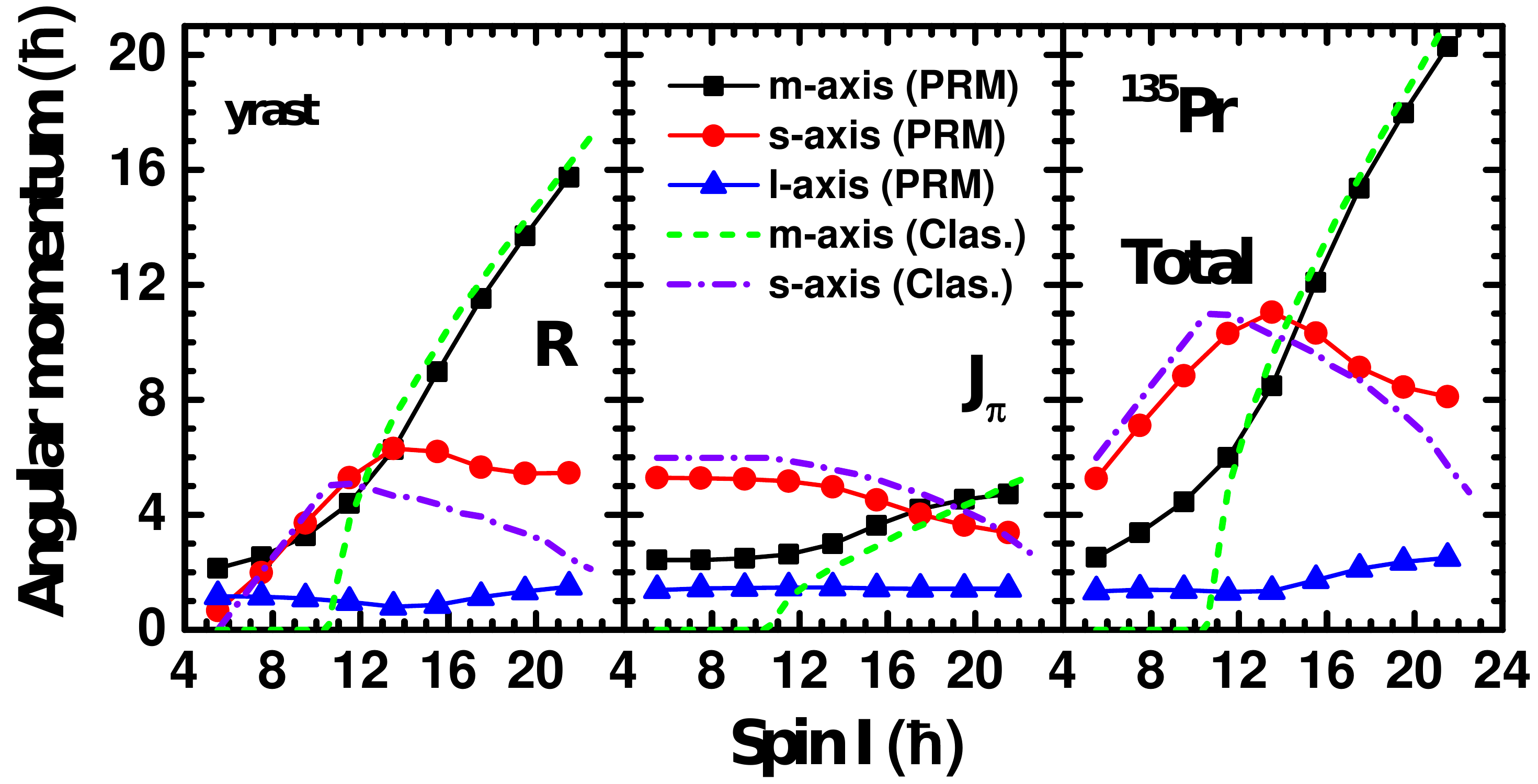}
 \caption{\label{f:ANG135PrTAC} Root mean square expectation values of
 the \am components of yrast states calculated by the PTR compared with 
 the values obtained by means of the TAC approximation and classical 
 Hamiltonian. The \am of the $h_{11/2}$ proton is denoted by 
 $\bm{j}_\pi$.}
\end{figure}

\begin{figure}[t]
 \includegraphics[width=0.90\columnwidth]{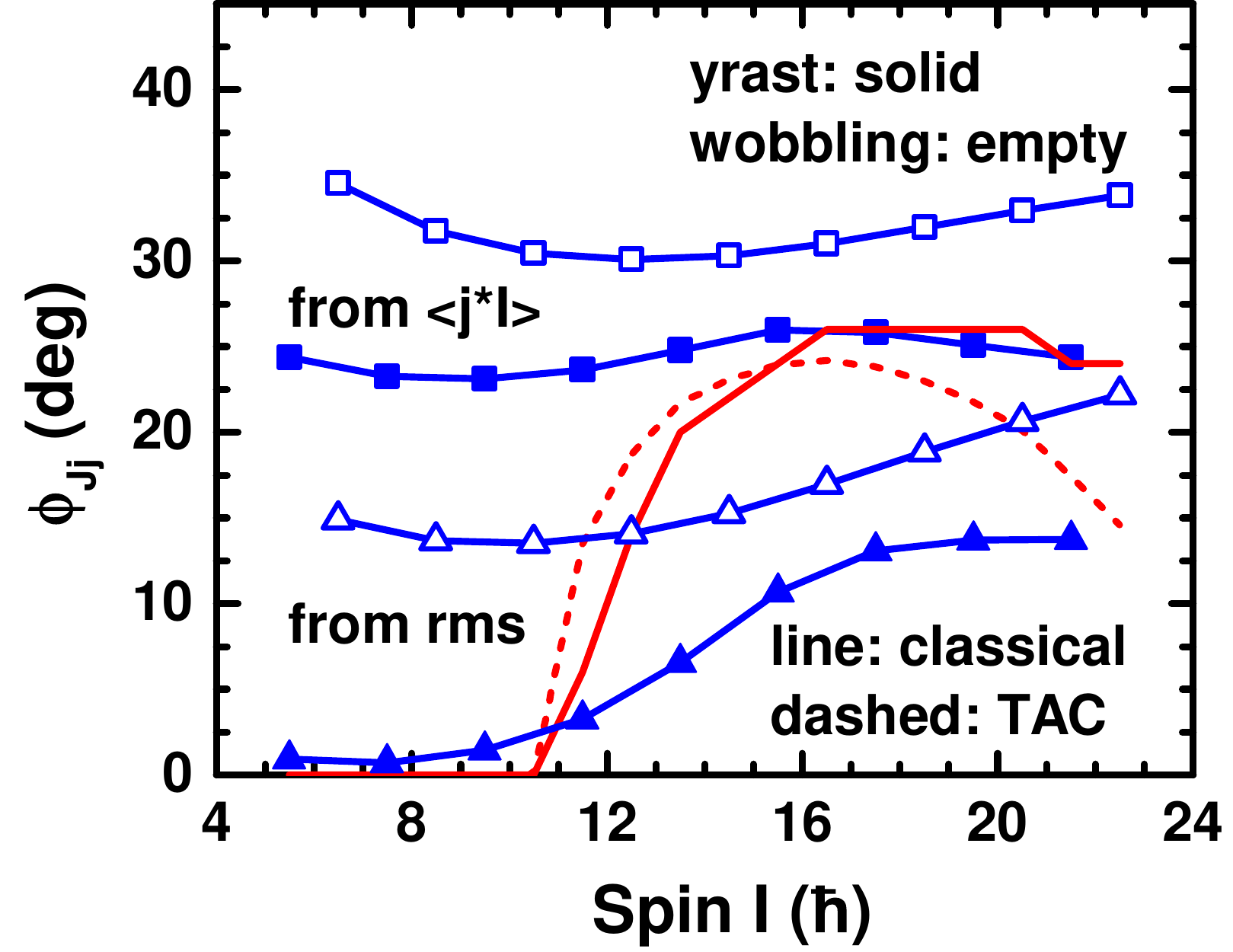}
 \caption{\label{f:Effective_angle} Angle between angular momenta 
 $\bm{J}$ and $\bm{j}$ of the yrast states and the first wobbling states. 
 The angles obtained by the quantal expectation value of the operator 
 scalar product (second term of Eq.~(\ref{eq:AngleJj})) are labeled 
 ``from $\langle j * J\rangle$". The angles of the classical 
 Hamiltonian and TAC are related to Fig.~\ref{f:ClassAngle} by calculating 
 $|\phi_J-\phi_j|$. The results of ``from rms'' are calculated as follows: 
 from the PTR root mean square expectation values of angular momentum 
 components in Fig.~\ref{f:ANG135Pr} we calculate $\phi_J=\arctan{(J_m/J_s)}$ 
 and $\phi_j=\arctan{(j_m/j_s)}$, and their difference $|\phi_J-\phi_j|$ 
 are $\phi_{Jj}$.}
\end{figure}

\begin{figure}[ht]
 \includegraphics[width=0.90\columnwidth]{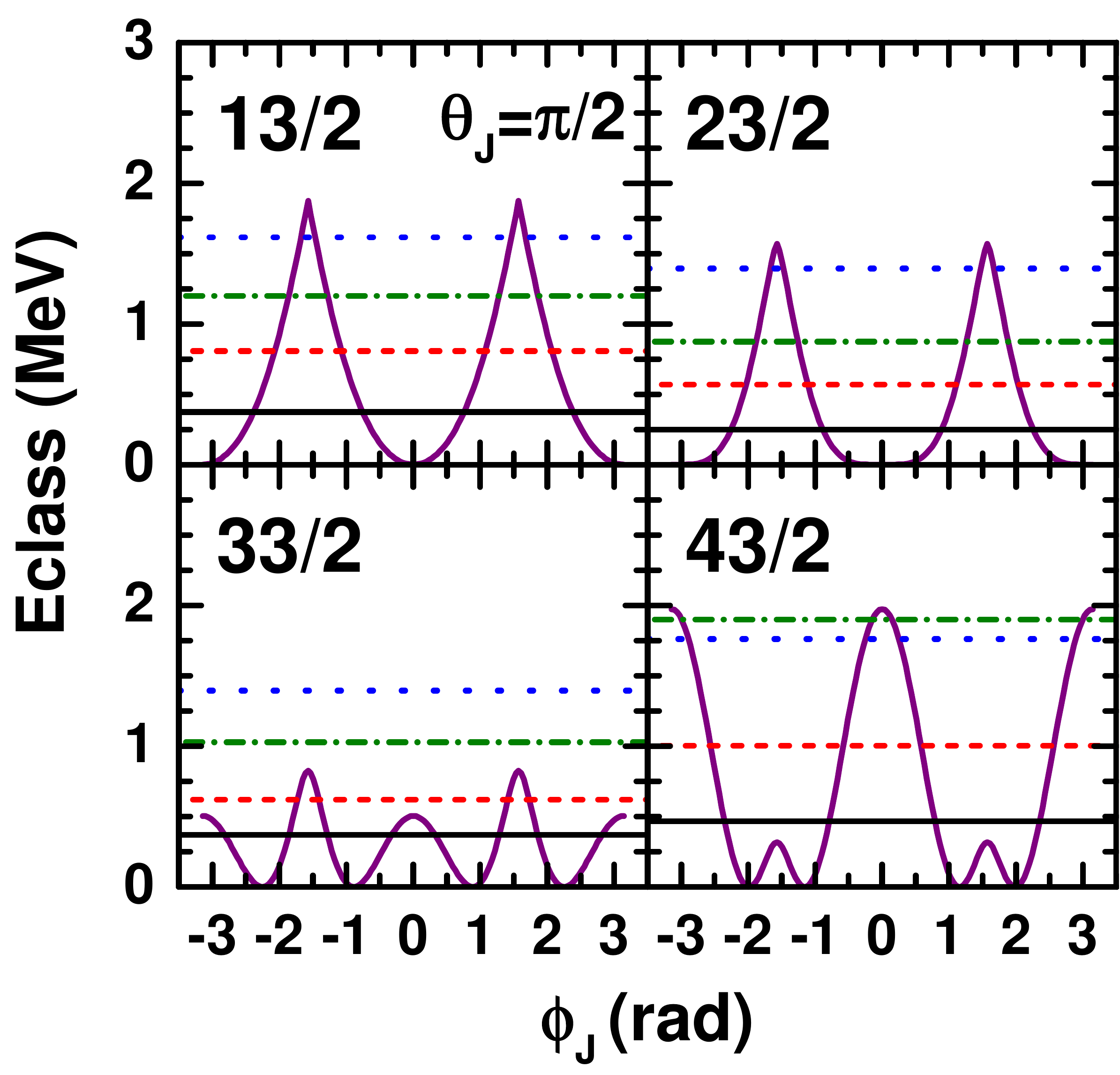}
 \caption{\label{f:Eclass} Classical energy as function of $\phi_J$ obtained
 by minimizing $E_{\textrm{class}}(J, \theta_J, \phi_J;j, \theta_j, \phi_j)$ 
 with respect to the angles $\theta_j$ and $\phi_j$ and taking 
 the $\theta_J=\pi/2$ for spins $I=13/2$, $23/2$, $33/2$, and $43/2$. 
 The horizontal bars (solid, short dash, dash-dot, dot) show the PTR 
 excitation energies form Fig.~\ref{f:E135Pr} with respect to the 
 minimum of the TAC energy. The energies of the states with opposite signature 
 are taken as $[E(I-1)+E(I+1)]/2$.} 
\end{figure} 

\begin{figure}[t]
 \includegraphics[width=0.9\columnwidth]{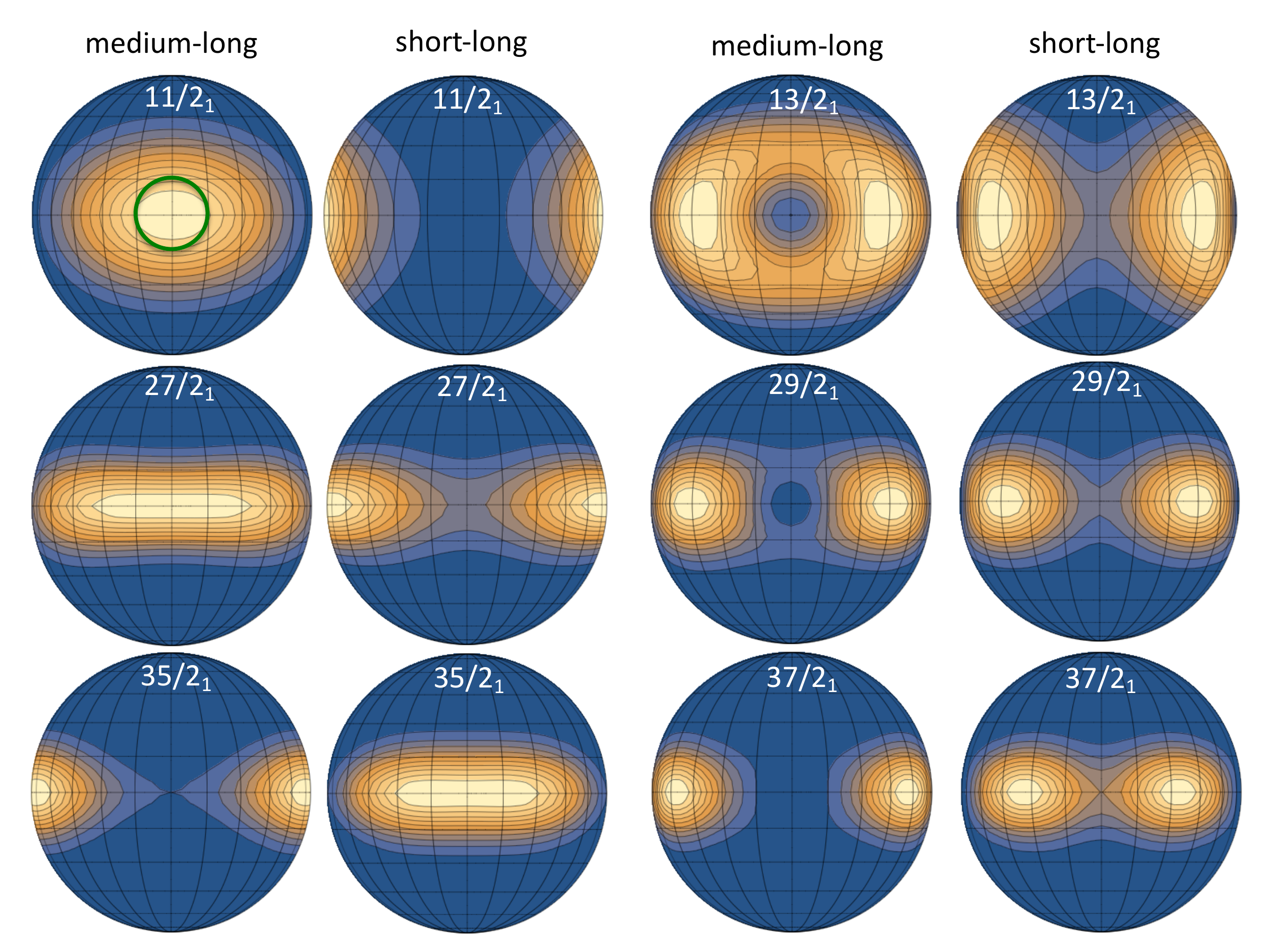}
 \caption{\label{f:TWmapsCS} SCS probability distributions
 $P(\theta \phi)_{I\nu} $ of the total angular momentum $\bm{J}$ of 
 six states shown in Fig.~\ref{f:TWmaps1} projected on the $m$-$l$-plane
 (viewpoint on $s$-axis) and $s$-$l$-plane (viewpoint on $m$-axis).
 Identical color code is used.}
\end{figure}

\begin{figure*}[t]
\includegraphics[width=\linewidth]{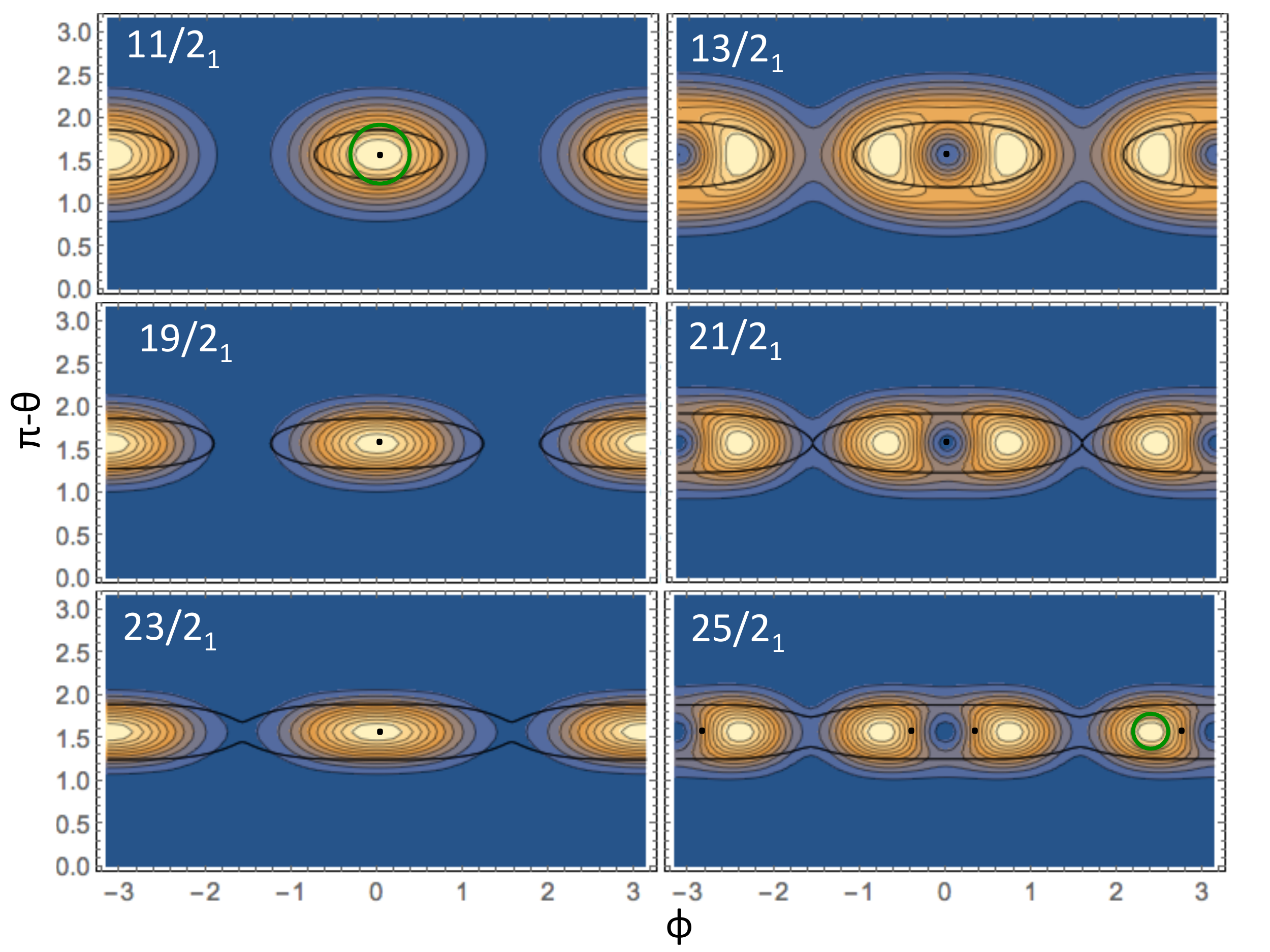}
\caption{\label{f:TWmaps1} SCS probability distributions $P(\theta \phi)_{I\nu}$
 of the total angular momentum $\bm{J}$ for some of the TW states $I\nu$ of the
 PTR states shown in Fig.~\ref{f:E135Pr}. See caption of Fig.~\ref{f:TRmaps1}
 for details. The dots are localized at the minimum of the classical energy
 $\phi_J$ shown in Fig.~\ref{f:ClassAngle}. The curves show the classical
 orbits calculated in FA where the orientation of $\bm{j}$ in the $s$-$m$-plane
 is given by the angle $\phi_j$ in Fig.~\ref{f:ClassAngle} and the energy
 is the PTR value shown in Fig.~\ref{f:E135Pr}.}
\end{figure*}

 \begin{figure*}[t]
 \includegraphics[width=\linewidth,trim= 0 0cm 0 0]{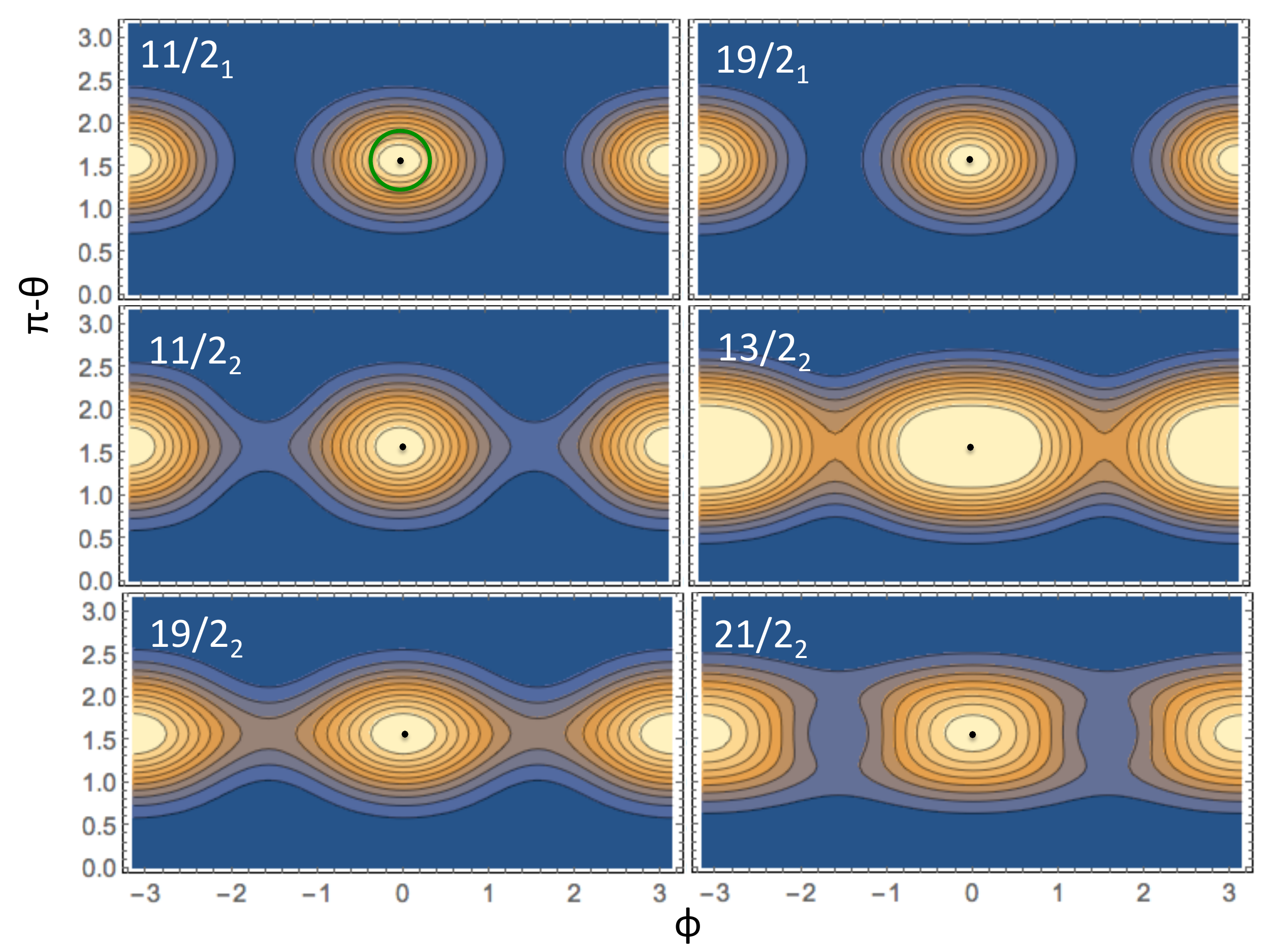}
 \caption{\label{f:TWmapsjlow} SCS probability distributions $P(\theta \phi)_{I\nu}$
 of the particle angular momentum $\bm{j}$ of some states $I\nu$ of the PTR states
 shown in Fig.~\ref{f:E135Pr}.}
\end{figure*}

\begin{figure}[!ht]
\includegraphics[width=1.0\columnwidth]{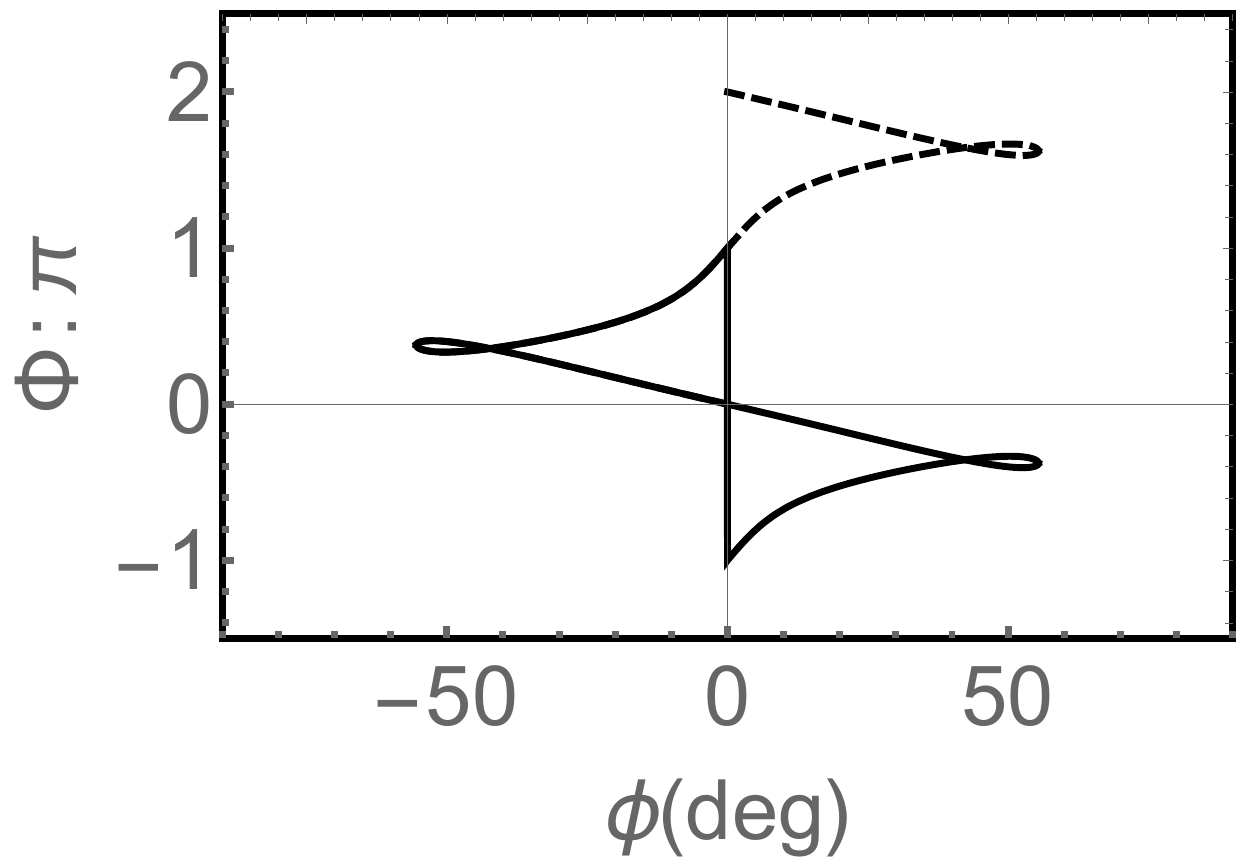}
\caption{\label{f:TWphase132} Phase difference $\Phi(\theta\phi,
\theta_0=\pi/2\phi_0=0)$ for the state $13/2_1$ along the classical 
orbit shown in shown in Fig.~\ref{f:TWmaps1} relative to the 
turning point $\phi_0=0$ at the upper branch.}
\end{figure}

The location of the minimum as function of \am is understood as follows.
The triaxial rotor is coupled with an $h_{11/2}$ proton. The $s$-axis
is its preferred orientation, because it maximizes the overlap of
the particle orbit with the triaxial core~\cite{Frauendorf1996ZPA}. 
The rotational energy of the rotor core prefers the $m$-axis with 
the largest moment of inertia. Fig.~\ref{f:ClassAngle} shows the 
result of the two competing torques. At low $J$ the torque of the 
quasiparticle wins. The orientation of $\bm{J}$ and $\bm{j}$ along the 
$s$-axis represents the stable configuration. The growth of total \am $J$ 
is generated by an increase the core \am $R_s$ along the $s$-axis. Above the 
critical angular momentum $J_c$ the torque of the rotor core takes over.
The energy minimum moves to the angle $\phi_J$ in the $s$-$m$-plane. 
The growth of total \am $J$ is essentially generated by
an increase the core \am $R_m$ along the $m$-axis while the $R_s$ 
components stays constant. The particle \am $\bm{j}$ is pulled toward 
to $m$-axis because the Coriolis force tries to minimize the angle 
between $\bm{j}$ and $\bm{J}$. The root mean square \am components 
shown in Fig.~\ref{f:ANG135Pr} reflect the position of the classical 
energy minimum.

The development of the \am geometry with increasing spin has already 
been discussed in Ref.~\cite{Frauendorf2014PRC}. In order to simplify, 
the authors assumed that the \am of the particle is rigidly aligned 
with the $s$-axis (frozen alignment approximation-FA). At low $J$ the 
rotor \am $\bm{R}$ aligns with the $s$-axis, in this way minimizing
the Coriolis force. For $J>J_c$ rotor \am $\bm{R}$ re-aligns toward
the $m$-axis, which has the largest \momid ~\footnote{For a detailed
discussion of the coupling high-$j$ orbitals with the rotating triaxial
potential see Ref.~\cite{Frauendorf2018PS}.}. The critical \am of $J_c=11$ 
in Fig.~\ref{f:ClassAngle} is lower than the FA estimate of $J_c=14$ in
Ref.~\cite{Frauendorf2014PRC} because taking into account the
finite de-alignment of the proton lowers the stability of the
minimum at $\phi_J=\phi_j=0$. 

The orientation angles can also be determined by applying the 
tilted axis cranking (TAC) approximation to the PTR Hamiltonian. 
In this case only the total \am operator is replaced by its 
classical vector $\bm{J}$ and the reduced PTR Hamiltonian is 
diagonalized in the subspace of the odd particle. The resulting 
energy $E_{\textrm{crank}}(J, \theta_J, \phi_J)$ is minimized 
with respect to the orientation angles $\theta_J$, $\phi_J$ of the 
total \amd. The resulting angles are shown in Fig.~\ref{f:ClassAngle}
and corresponding \am components in Fig.~\ref{f:ANG135PrTAC} together 
with their values at the minimum of the classical energy. 
For comparison, the root mean square components $J_i$, $j_i$, 
$R_i=\langle \hat J_i^2\rangle^{1/2}$, $\langle \hat j_i^2\rangle^{1/2}$, 
$\langle \hat R_i^2\rangle^{1/2}$ are included in Fig.~\ref{f:ANG135PrTAC}. 
The cranking \am components are not far from  the classical 
values as expected from the angles shown in Fig.~\ref{f:ClassAngle}. 
The root mean square \am components of the yrast
states behave roughly like their cranking and classical counterparts.
The contributions of the fluctuations
to the root mean squares wash out the characteristic kinks at the
critical \am and generate 1-2 units of \am for the vanishing components.

Fig.~\ref{f:Effective_angle} shows the angle $\phi_{Jj}$ between the proton
\am $\bm{j}$ and the total \am $\bm{J}$. In case of the angles calculated
by minimizing the classical yrast energy,  $\phi_{Jj}=|\phi_J-\phi_j|$
shown in Fig.~\ref{f:ClassAngle}, which is zero below $J_c$ and increases
afterwards. The angles obtained by the classic vector expression (\ref{eq:AngleJj})
from the root mean squares of the \am components of the yrast states
increase steadily. They approach the angles at the minimum of the classical
energy at large $I$, whereas the kink at $J_c$ is washed out. The quantal
expectation value of the operator scalar product (\ref{eq:AngleJj}) gives
a nearly constant angle, which is substantially larger than the angles
obtained from the root means square values of the \am components.
The quantal indeterminacy of the \am components leads to severe deviations 
from the classical vector scheme. The angles determined this way do not 
well reveal the underpinning \am geometry.

A topological classification can be found by considering the  adiabatic energy 
$E_{\textrm{class}}(J, \theta_J, \phi_J)$, which is obtained by minimizing 
classical energy $E_{\textrm{class}}(J, \theta_J, \phi_J; j, \theta_j, \phi_j)$ 
with respect to the angles $\theta_j$, $\phi_j$ for fixed angles $\theta_J$, 
$\phi_J$. The adiabatic energy along the path $\theta_J=\pi/2$ is displayed 
in Fig.~\ref{f:Eclass}, which represents the bottom of the valley in the surface 
$E_{\textrm{class}}(J, \theta_J, \phi_J)$. The figure includes the quantal PTR 
energies from Fig.~\ref{f:E135Pr} relative to the minimum of 
$E_{\textrm{class}}(J, \pi/2, \phi_J)$. The additional energy of the PTR 
states can be assigned to the collective wobbling motion. Assuming a constant 
mass parameter for the $\phi_J$ degree of freedom, the classical orbit is 
confined to the range $E_{\textrm{PTR}}> E_{\textrm{class}}$. As will be 
discussed in detail below, the character of the PTR wave function is closely 
related to the classical orbit. There are three topological regions.
\begin{enumerate}
 \item Transverse wobbling (TW, panel $I=13/2$): the total \am oscillates 
 around $\phi_J=0$, $\pi$, which is the $s$-axis that is transverse to the 
 $m$-axis with the largest \momid.
 \item Longitudinal wobbling (LW, panel $I=43/2$): the total \am oscillates 
 around $\phi_J=\pm \pi/2$, which is the $m$-axis with the largest \momid.
 \item Axis-flip wobbling (panel $I=33/2$): the total \am jumps over the whole
 range of $2\pi$.
\end{enumerate}
The TW and LW modes are restricted to the lowest states. For the higher 
states the total \am de-localizes.  

\subsection{Transverse wobbling}\label{s:TW}

The authors of Ref.~\cite{Frauendorf2014PRC} classified the excited wobbling 
states of the PTR system. In order to introduce the classification scheme,  
they started with assumption that odd quasiparticle is either
rigidly aligned with the $s$- or $l$-axis, which are transverse to the 
$m$-axis with the largest \momi (transverse wobbling TW), or that it 
is aligned with the $m$-axis (longitudinal wobbling LW). The assumption of 
``frozen alignment" FA makes the problem one-dimensional and the \am orbits 
are the intersection lines of the \am sphere and the energy ellipsoid, 
the center of which is shifted from origin~\cite{Frauendorf2014PRC}.  
For the TW regime the $\bm{J}$ revolves the $s$-axis if the quasiparticle 
is particle-like or the $l$-axis if it is hole-like. For the LW regime 
$\bm{J}$ revolves the $m$-axis. The classification of TW-LW was 
introduced in Ref.~\cite{Frauendorf2014PRC} based on the FA assumption 
because it leads to a transparent picture in terms of the classical 
orbits and to simple analytical expressions for the energy and $E2$ 
transition in the small amplitude limit, the harmonic frozen 
alignment (HFA) approximation. Although not explicitly stated, 
the TW classification was understood in the  topological sense 
illustrated by Fig.~\ref{f:Eclass}: the total \am oscillates around
the $s$-axis, which is transverse to the $m$-axis. Classically, 
the yrast states represent uniform rotation about the transverse $s$-axis. 
For the TW excitations $\bm{J}$ revolves the $s$-axis in the body-fixed 
frame and in the laboratory system the density distribution wobbles such 
the $s$-axis executes an ellipse with respect to the \am 
axis $\bm{J}$. Fig.~\ref{f:ClassAngle} shows that in TW regime, 
$J<J_c$ the classical $\bm{j}$ vector is aligned with the $s$-axis.  
As seen in Fig.~\ref{f:ANG135PrTAC} the alignment of $\bm{j}$ in the 
yrast states is not complete for the full quantal calculation. As expected,  
the $s$-component is smaller and the $m$-component larger for TW excitation
than for the adjacent yrast states. 

\begin{figure*}[!ht]
\includegraphics[width=0.85\columnwidth]{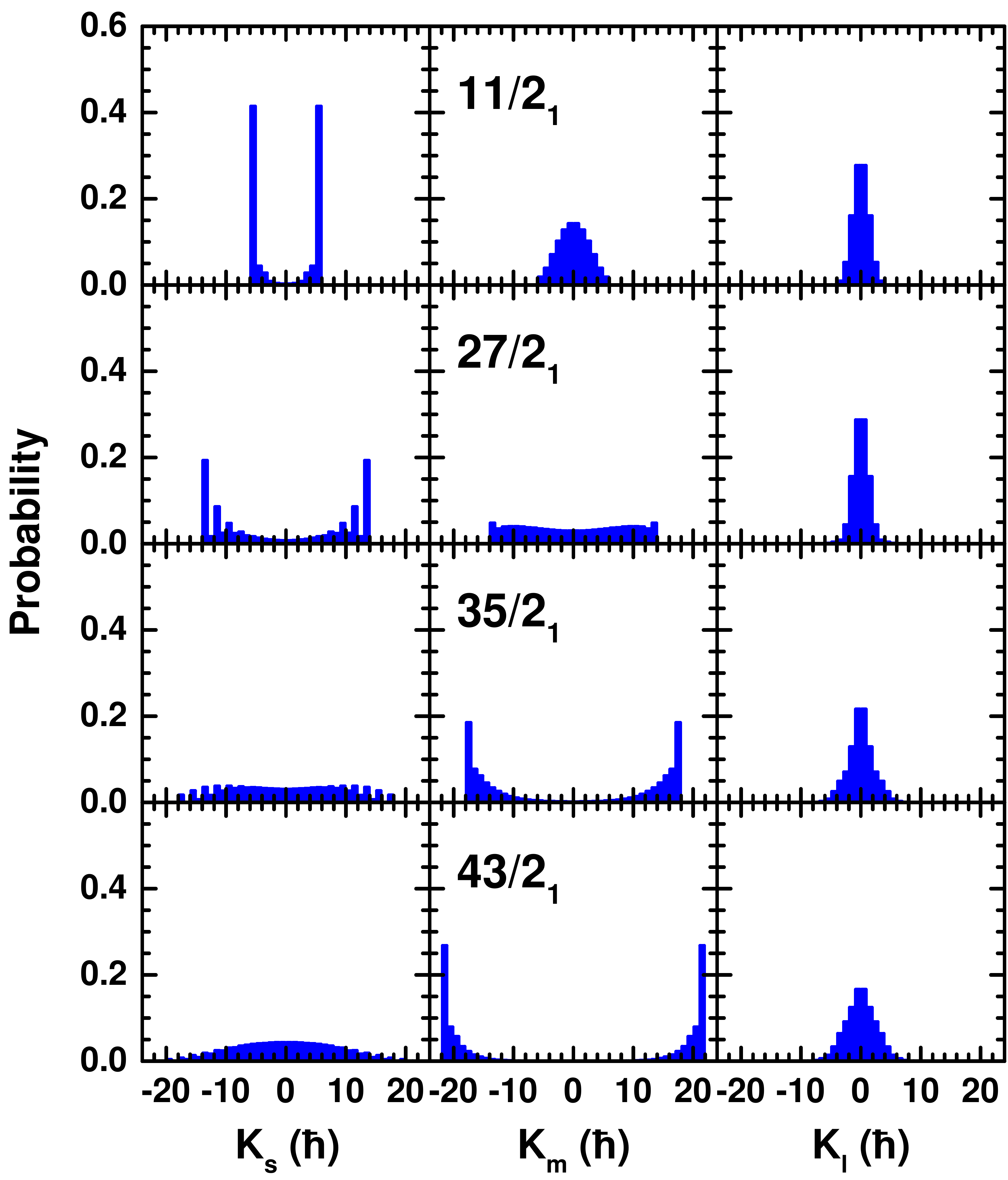}\quad\quad
\includegraphics[width=0.85\columnwidth]{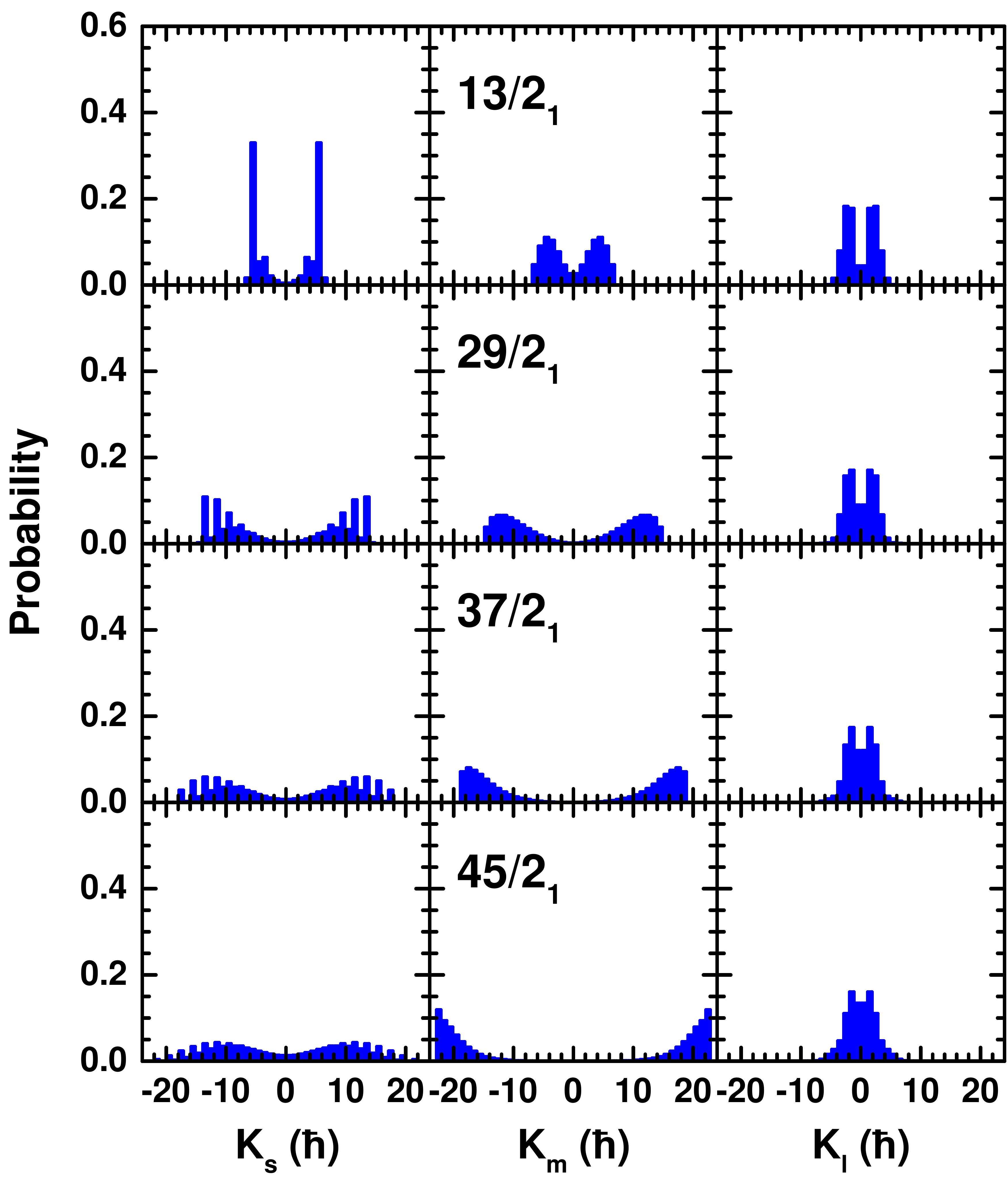}\\
~~\\
\includegraphics[width=0.85\columnwidth]{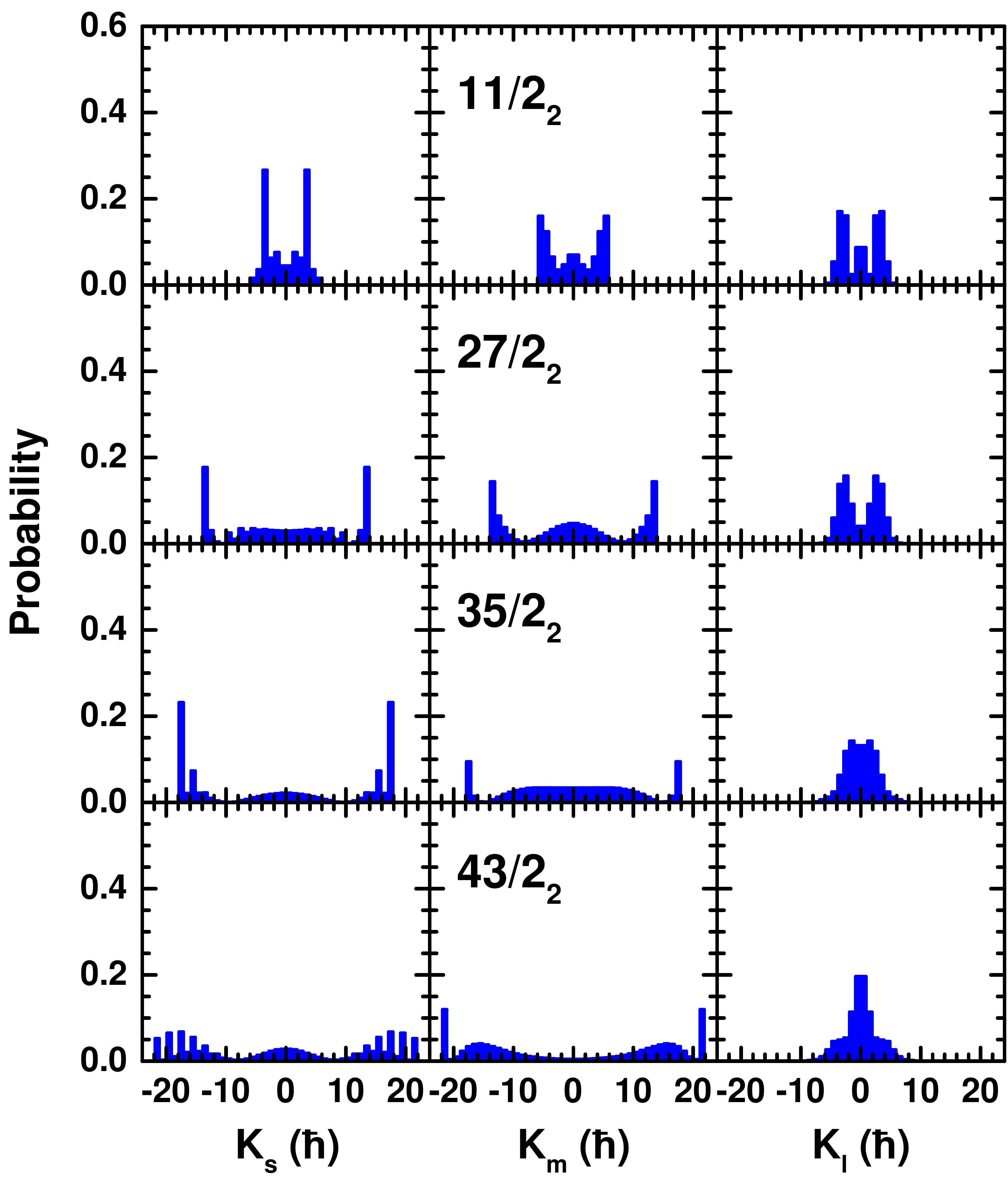}\quad\quad
\includegraphics[width=0.85\columnwidth]{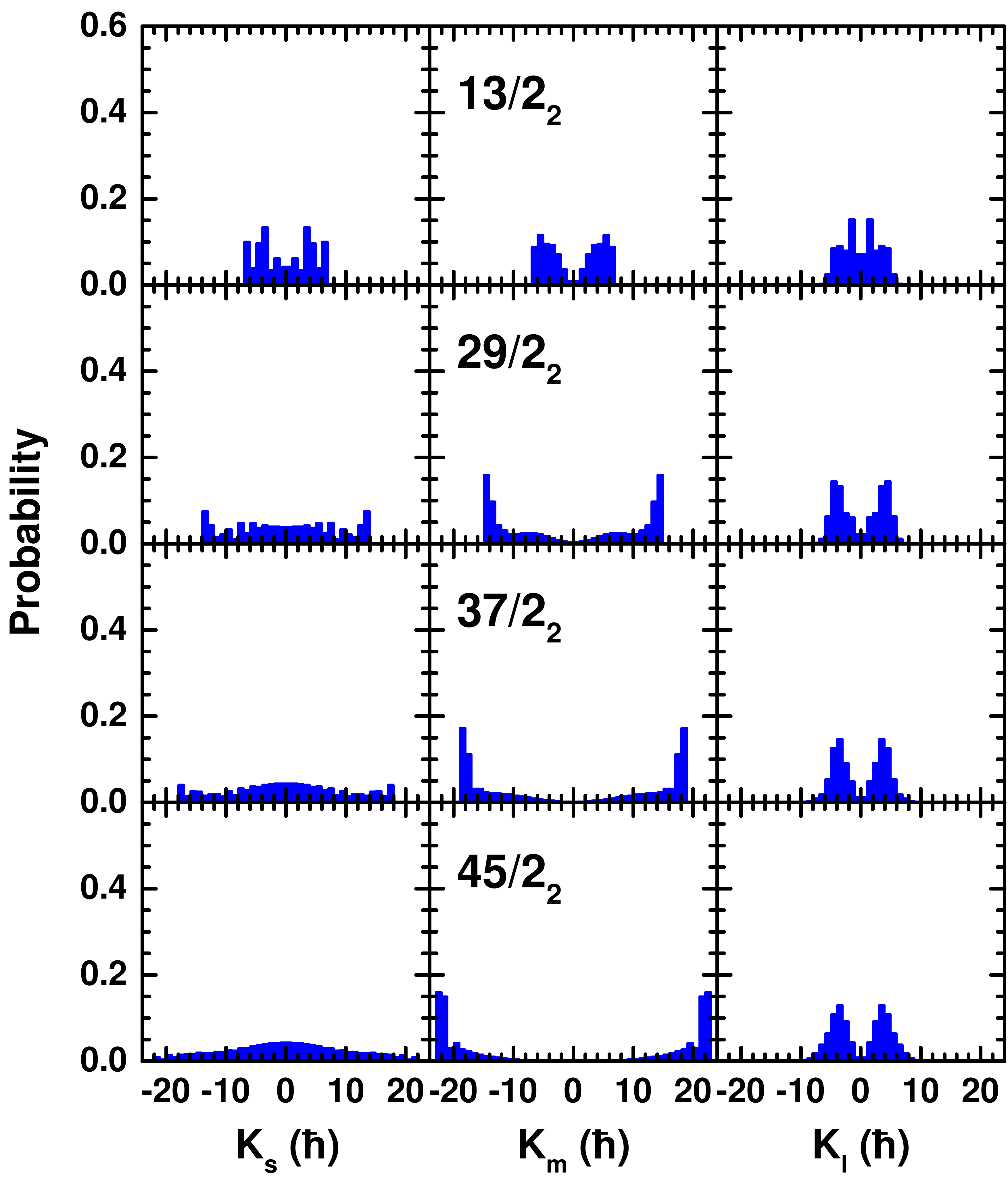}
\caption{\label{f:TWK} Probability distribution of the total angular
momentum projection $K_i$ on the three principal axes for some states
of the TW.}
\end{figure*}

\begin{figure*}[ht]
\includegraphics[trim= 0 6cm 0 0, width=\linewidth]{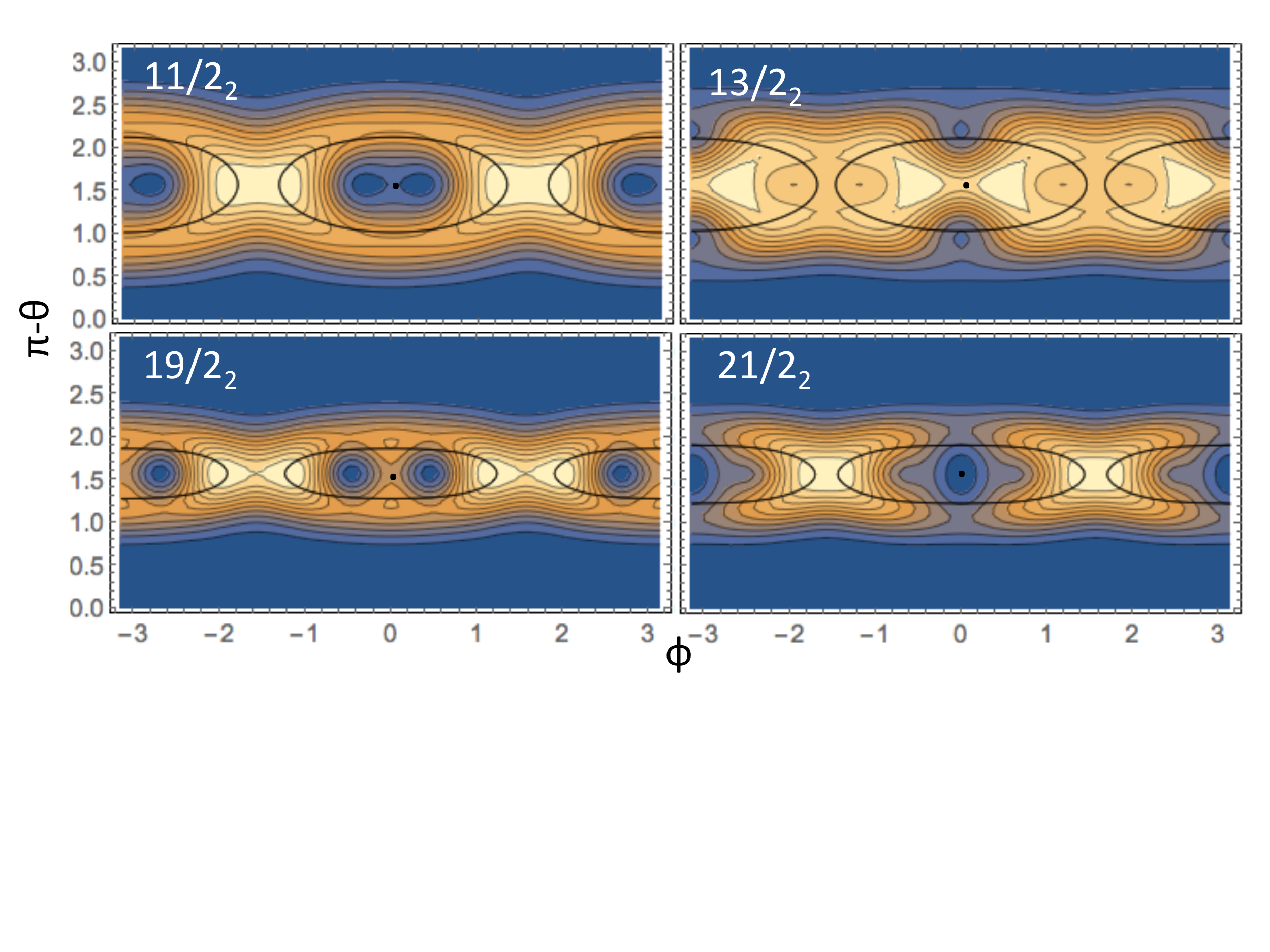}
\caption{\label{f:TWmaps1-2} SCS probability distributions 
$P(\theta \phi)_{I\nu}$ of the total angular momentum $\bm{J}$ 
for the second and third wobbling states.}
\end{figure*}

The SCS maps, which are directly generated from the quantal results, 
clearly illustrate the topology. Figs.~\ref{f:TWmapsCS} and \ref{f:TWmaps1}
show the maps for the total \am $\bm{J}$ obtained by Eq.~(\ref{eq:PSCS})
using the density matrix (\ref{eq:TWrhoJ}). The TW regime extends to the
critical \am $J_c\approx 11$. The probability  in the yrast maps $I=11/2_1$, 
$I=15/2_1$ (not shown), and $19/2_1$ is localized around the $s$-axis. The 
maps $I=13/2_1$,  $I=17/2_1$ (not shown), and $21/2_1$ for the one-phonon 
states show a rim that revolves the $s$-axis. The distribution is analog 
to one of the one-phonon states of the simple rotor in Fig.~\ref{f:TRmaps1}, 
which has a rim around the $m$-axis. 

The dots in Fig.~\ref{f:TWmaps1} display the minima of the classical
energy $E_{\textrm{class}}(J, \theta_J, \phi_J; j, \theta_j, \phi_j)$ 
given by the angles $\theta_J=\pi/2$ and $\phi_J=0$ according to 
Fig.~\ref{f:ClassAngle}. The SCS maps include the classical orbits, 
which are the contours of the adiabatic classical energy 
$E_{\textrm{class}}(\theta_J,\phi_j)=E_{\textrm{PTR}}
-E_{\textrm{TAC}}(\textrm{min})$. As discussed in context with 
Fig.~\ref{f:Eclass}, the TAC energy at the minimum contains the 
zero point energy of the proton. The additional PTR energy is 
assigned to the wobbling motion. 

The classical orbits trace the rim, although not quite as close as
in the case of the simple rotor. Fig.~\ref{f:TWmapsjlow} displays the 
probability distribution of $\bm{j}$ of the odd proton obtained by 
Eq.~(\ref{eq:TWmapsj}) for the state $11/2_1$. As expected for the 
TW regime, the density distribution is centered at the $s$-axis. 
The distribution is wider than the natural width of the SCS, 
which is indicated by the circle. This reflects the zero point 
fluctuations of $\bm{j}$. The $\bm{j}$-maps for the states 
$13/2_1$ and $15/2_1$ are nearly the same, as expected for 
the strong coupling of the proton.  

Fig.~\ref{f:TWphase132} shows the phase increment along the classical 
orbit of the one-phonon TW  state $13/2_1$. The phase development 
is similar to the triaxial rotor (TR) one-phonon wobbling 
state $9_1$ in Fig.~\ref{f:TRphase}. The phase gains $2\pi$ over 
one revolution. However, the phase gain over the four branches 
between the turning points in not symmetric as in Fig.~\ref{f:TRphase}. 
The origin has been mentioned before. The phase is calculated 
from the reduced density matrix, not from the complete density 
matrix as for the TR state $9_1$. The reduction generates some loss of 
phase information, which has the consequence that phase increment is no 
longer additive. The phase increment in Fig.~\ref{f:TWphase132} is 
calculated relative to the turning point at $\phi=0$ in the upper 
branch of the classical orbit. Relative to this starting point the 
lower branch differs from the upper one, and so does the phase 
increment. Nevertheless, the total gain after one revolution is 
$2\pi$, as expected.

The $K$-plots in Fig.~\ref{f:TWK} qualitatively agree with the distributions
along the three axes in the SCS maps, which is best visible by comparing with
the maps in the orthographic projection Fig.~\ref{f:TWmapsCS}. The existence
of the rim around the $s$-axis in Fig.~\ref{f:TWmaps1} can be guessed from
inspecting the $K$-plots for the three axes.

Fig.~\ref{f:TWmaps1-2} shows the SCS maps $P(\theta \phi)_{I\nu}$ of 
the total angular momentum $\bm{J}$ for the second and third wobbling 
states. In a harmonic vibration scenario, the two-phonon states $11/2_2$ 
and $19/2_2$ should appear as a wobbling mode with a wider rim than the 
one-phonon states (c.f. the $8_2$ state in Fig.~\ref{f:TRmaps1}), which is 
the case (c.f. $13/2_1$ in Fig.~\ref{f:TWmaps1}). However the wobbling cone 
that revolves the positive $s$-axis and the cone that revolves the negative 
$s$-axis are no longer as well separated as for the one- and zero-phonon 
states, which is expected from inspecting Fig.~\ref{f:Eclass}. The overlap 
causes deviations from the harmonic regime. Analogous to the $I=8_3$ state 
of the TR, the merging of the rims at $\phi=\pm \pi/2$ indicates the 
instability of the TW mode. The anharmonicities drive the TW wobbler 
toward the axis-flip regime like in the case of the simple TR discussed 
in Sec.~\ref{s:SCSdetail}. The role of the axes is reversed. In the 
TW regime the $s$-axis is the stable and $m$-axis the unstable axis. 
The concentration of the probability near the $m$-axis signals the 
approach to instability of the TW mode. The $\bm{j}$-distributions 
in the lower two rows of Fig.~\ref{f:TWmapsjlow} show the weak reaction 
of the proton \am caused by the Coriolis coupling with $\bm{J}$. 
The distribution becomes elongated toward the $m$-axis. 

The $\bm{J}$-distributions of the $11/2_2$ and $19/2_2$ states 
have bridges at $\phi=0$, $\pi$ between the upper and lower parts 
of the rim. They are a quantum feature. The $n=2$ state of the 
one-dimensional harmonic oscillator has probability distributions
$P_2(x)$ and $P_2(p)$, which are determined by the Hermite polynomial 
$H_2$. The polynomial has a maximum at $x=0$ and $p=0$, respectively. 
The corresponding bumps can by seen in the $K_m$ and $K_l$ in 
Fig.~\ref{f:TWK}. The bumps combine to the bridge in the 
two-dimensional SCS map. We will discuss the structure in 
more detail in Sec.~\ref{sec:TW-SP}. The bridge is less visible  
for the TR two-phonon state $8_2$ in Fig.~\ref{f:TRmaps1}.
The reason is the presence of the proton, which pulls $\bm{J}$ 
toward $\theta=\pi/2$. 
 
Fig.~\ref{f:Eclass} shows that the $n=3$ structures are not well 
confined by the classical potential. The map for the states $21/2_2$ in 
Fig.~\ref{f:TWmaps1-2} and $17/2_2$ (not shown, but looks alike) have 
axis-flip character as well. Their probability around $\phi=0$, $\pi$ 
is very small, because the state is odd under $\phi \rightarrow -\phi$. 
The $\bm{J}$-map for the $13/2_2$ state is complex. We attribute this 
to the admixture of the $13/2_3$ signature partner structure, which has 
maxima at $\phi=0$, $\pi$ (see discussions in the next section). The 
$\bm{j}$-distributions of the states in Fig.~\ref{f:TWmapsjlow} 
show some response of the proton \am to the inertial forces.

\subsection{Transverse wobbling and signature 
partner modes}\label{sec:TW-SP}

For \am well below the instability at $J_c$ the amplitudes of $\bm{J}$ 
and $\bm{j}$ are small, which allows one to approximate the PTR as a 
system of two coupled harmonic oscillators. As the coupling is not 
very strong in the TW regime, one may classify the excitations in terms 
of individual excitations of the uncoupled modes. The harmonic approximation 
has been worked out in Refs.~\cite{Tanabe2017PRC, Raduta2020PRC}. The authors 
applied boson expansions to the \am operators. The leading terms provide 
the coupled oscillator approximation. A review of the work 
in Ref.~\cite{Tanabe2017PRC} is given in the Appendix~\ref{sec:AppA}. 
 
Here we discuss the structure of the two-oscillator Hamiltonian. We 
derive it in a modified way, which was used in Ref.~\cite{Frauendorf2014PRC}. 
Assuming that amplitudes of $J_2$, $J_3$, $j_2$, $j_3$ are small one 
may approximate 
\begin{align}
 J_1 &=\sqrt{J^2-J_2^2-J_3^2}\approx J-\frac{J_2^2}{2J}-\frac{J_3^2}{2J}, \\
 j_1 &=\sqrt{j^2-j_2^2-j_3^2}\approx j-\frac{j_2^2}{2j}-\frac{j_3^2}{2j},
\end{align}
and the PTR Hamiltonian (\ref{eq:HPTR}) by the bi-linear expression
\beq\label{eq:H02}
 H_{02}=A_1(J-j)^2+h^\prime_{p}+H_{\textrm{HFA}}+H_c.
\eeq

With $A_ i=1/(2{\cal J}_i)$ the term 
\begin{align}
 H_{\textrm{HFA}} &=(A_2-\bar{A}_1)J_2^2+(A_3-\bar{A}_1)J_3^2,\notag\\
 \bar A_1 &=A_1\left(1-\frac{j}{J}\right),
\end{align}
is recognized as the HFA Hamiltonian introduced in Ref.~\cite{Frauendorf2014PRC},
which generates the pure harmonic TW excitation spectrum. Analytical 
expressions for the energies and $E2$ transition probabilities 
are given there.

The modified particle Hamiltonian 
\begin{align}\label{eq:Hcranking}
 h^\prime_{p} &=h_{p}+(A_2-A_1)j_2^2+(A_3-A_1)j_3^2-\omega (j_1-j),\notag\\
 \omega &=2A_1J,
\end{align}
has the form of a Hamiltonian in a frame rotating with the angular 
velocity $\omega$. The  cranking term $\omega j_1$ accounts for  
the inertial forces. The other two corrections are known as the 
``recoil terms" of the PTR. As the particle excitations are given 
by the states in the rotating potential, the authors of 
Ref.~\cite{Odegaard2001PRL} called them ``cranking mode". We will 
keep this name. Interpreting the spectra in the framework of the 
Cranked Shell Model, it has become customary to call ``signature 
partner" the excited quasiparticle Routhian of opposite signature 
which branches from the yrast Routhian with increasing $\omega$. 
The authors of Ref.~\cite{Matta2015PRL} kept the custom using 
the name ``signature partner band". In the following we will 
also use the name ``signature partner" to denote the first 
excited state of the odd proton. It should be noted that the 
authors of Ref.~\cite{Tanabe2017PRC} called the cranking mode 
``precessional mode" and the authors of Ref.~\cite{Raduta2020PRC} 
use the terminologies ``transversal", ``longitudinal", 
and ``signature partner" in  different ways, which may cause 
confusion.

The term 
\beq
H_c=-2A_2j_2J_2-2A_3j_3J_3
\eeq
couples the wobbling and cranking modes. 

\begin{figure}[t]
\includegraphics[width=0.8\columnwidth]{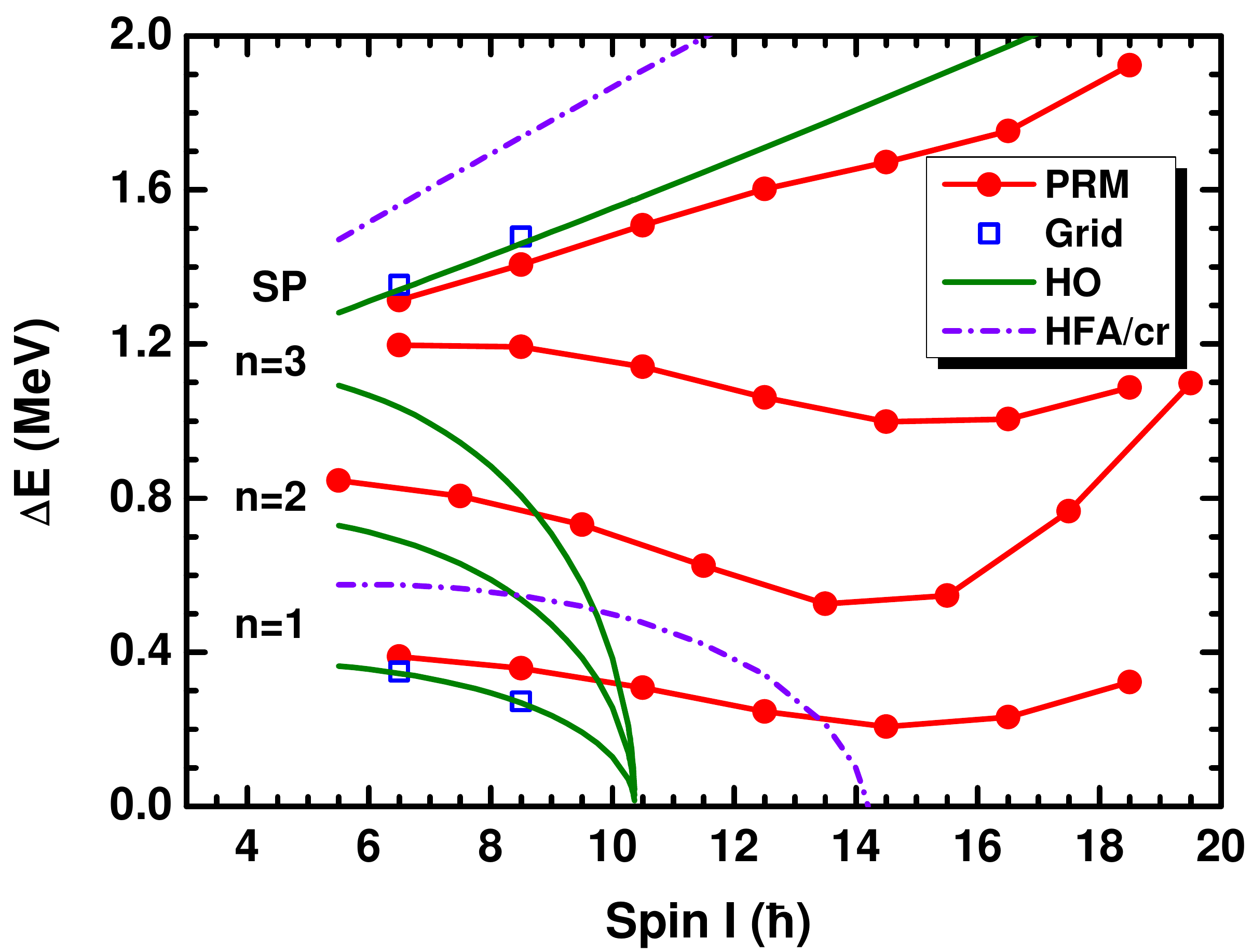}
\caption{\label{f:Ewob_grid} Excitation energies of the wobbling 
  band with phonon numbers $n=1$, $2$ and $3$ and signature 
  partner band with respect to the yrast band calculated by 
  PTR, grid method (presented in Appendix~\ref{sec:AppA}), 
  HP boson expansion (labeld as HO), HFA/cr (obtained by HP 
  boson expansion with $F=G=0$ in Eq.~(\ref{eq:Astab})) solutions.}
\end{figure}

 \begin{figure*}[ht]
\includegraphics[width=13.0 cm]{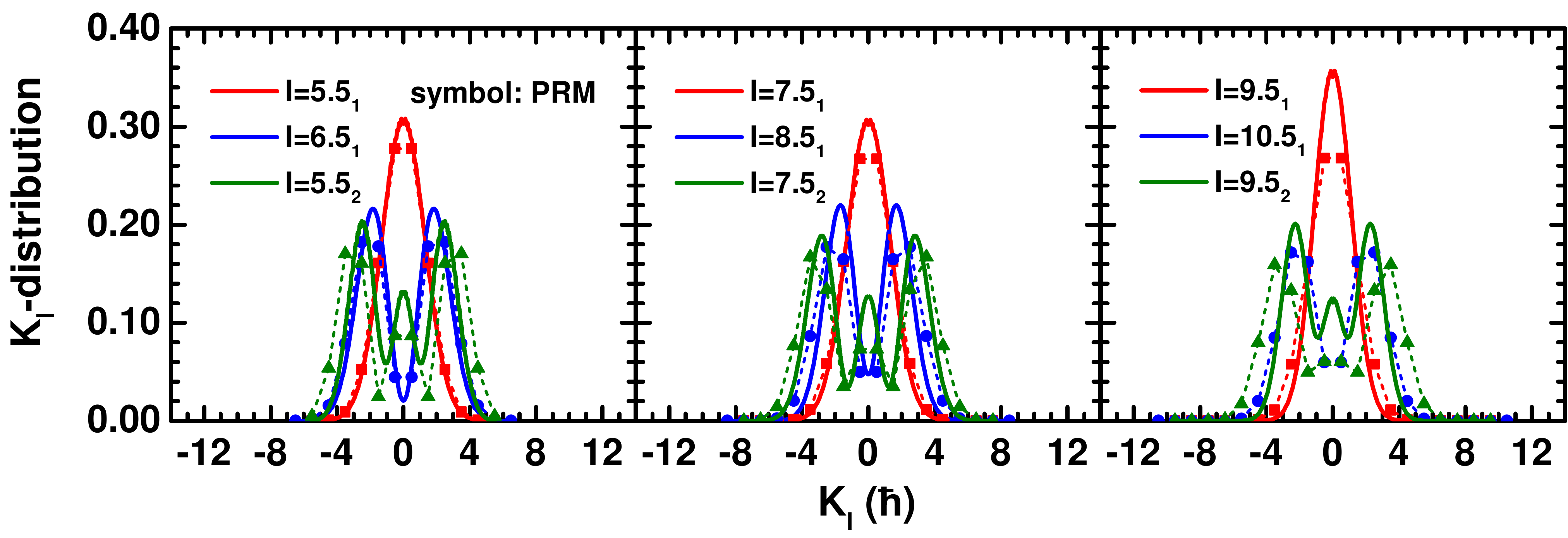}\\
\includegraphics[width=13.0 cm]{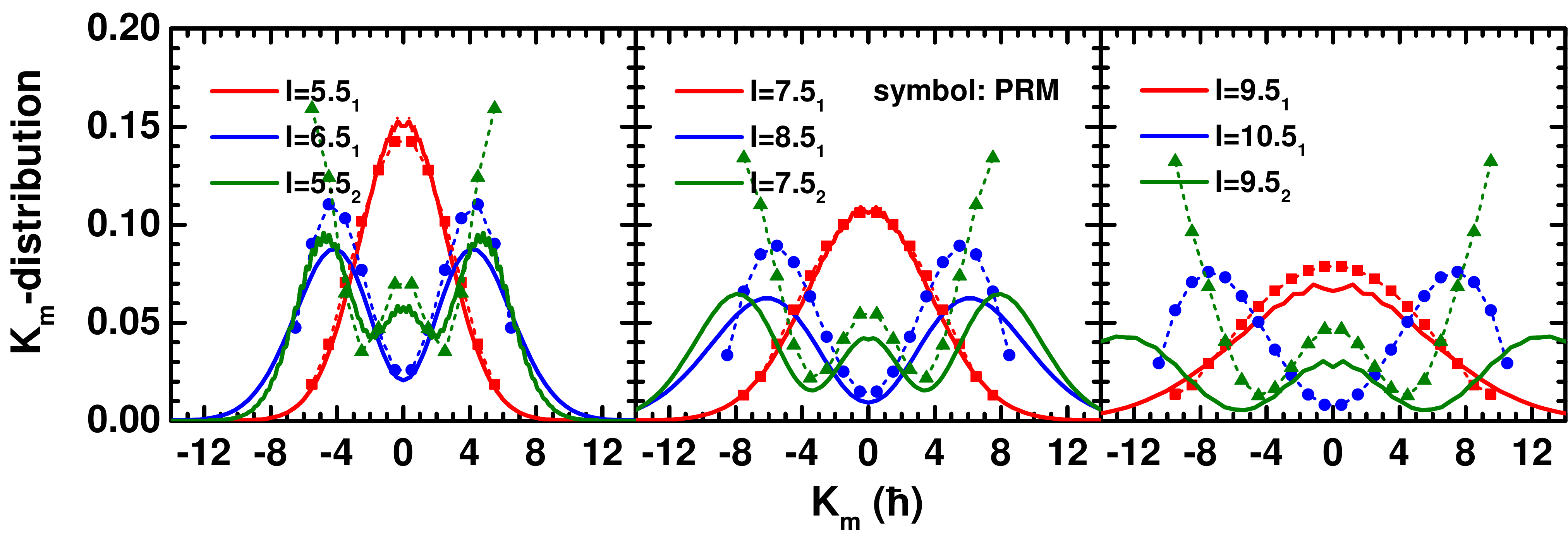}
\caption{\label{f:K-dis_grid} Probability distribution of the total 
  angular momentum projection $K_l$ and $K_m$ for the TW states $n=0$, 
  1, and 2 in the spin region $I\leq 21/2$ obtained for the HO approximation 
  (lines) and PTR. The HO $n=1$ state for  $I= 21/2$ is missing because 
  it is unstable. The tails of the widest HO distribution are cut off.}
\end{figure*}

As reviewed in the Appendix~\ref{sec:AppA}, the authors of 
Ref.~\cite{Tanabe2017PRC} derived a somewhat more accurate expression 
for $H_2$ by means of Holstein-Primakoff boson expansion of the \am 
operators, which includes semiclassical correction terms. After mapping 
$H_2$ on the boson space, it is diagonalized by means of a Bogoliubov 
transformation for bosons. The eigenvalue equation has two real 
solutions: the lower wobbling and the higher cranking solution. They 
are shown in Fig.~\ref{f:Ewob_grid} where they are labeled as HO 
$n=1$, 2, 3 and SP, according to their character. The wobbling mode is 
unstable for $I=21/2$, which is close to the instability of the 
minimum of the classical energy at $\phi_J=0$ in Fig.~\ref{f:ClassAngle}. 
This instability makes the $E_{\textrm{wob}}$ vanish rapidly. 
Also shown are the uncoupled wobbling and cranking modes obtained by 
setting $H_c=0$, which corresponds to $F=G=0$ in Eq.~(\ref{eq:H2HP}). 
The wobbling mode agrees with the HFA for TW in Ref.~\cite{Frauendorf2014PRC}. 
(There are slight numerical differences caused by the mentioned 
semiclassical correction terms.) The FA approximation makes the TW mode 
more stable. As discussed in context of Fig.~\ref{f:ClassAngle} the 
FA moves instability of the minimum of the classical energy at the 
$s$-axis to higher $J$ values.

The slope of the cranking solution for SP is about $1/{\cal J}_s$. This 
corresponds to the difference of the particle alignment 
$j_{\rm{SP}}-j_{\rm{yrast}}$ of about $-1$ between the unfavored and 
favored signature partners, which is expected from the cranking 
Eq.~(\ref{eq:Hcranking}) and consistent with the alignment results 
shown in Fig.~\ref{f:i135Pr}.

Fig.~\ref{f:Ewob_grid} compares the wobbling energies of the exact 
solutions of PTR with the approximate solutions of HP boson expansion. 
For $I=13/2$, 17/2, PTR solutions agree fairly with the $n=1$, 2, 3 HO 
TW phonon excitations. The PTR energy $E_{\textrm{wob}}$ decreases 
with the increasing spin up to $I=29/2$, which is the hallmark of TW. 
After that the TW regime changes into the transition and further into 
the LW regime. The HO approximation gives a collapse of TW at 
$J \sim 21/2$, which is near the \am $J_c$ where the $\phi_J$-minimum  
of the classical and TAC energy become unstable (cf. Figs.~\ref{f:ClassAngle} 
and \ref{f:ANG135PrTAC}). The instability appears earlier than the 
end of the TW regime of exact solutions by PTR.

Fig.~\ref{f:K-dis_grid} compares the $K_l$- and $K_m$-distributions of 
the HO with PTR for $I\leq 21/2$. The HO shows the well known pattern 
generated by the Hermite polynomials of order 0, 1, 2. It is seen both 
in $P(K_l)$ and $P(K_m$) on a different $K$ scale. As discussed in the 
Appendix~\ref{sec:AppA}, the distributions are related as 
$P\left(K_l/\Delta K\right)\approx P\left(K_m \Delta K\right)$, 
where $\Delta K$ is the width of $P(K_l)$, because $\hat J_{l,m}$ 
represent, respectively, the position and momentum operators of 
the TW oscillations. The narrow $K_l$-distributions of HO and PTR 
agree well with each other, although the HO ones are somewhat more 
narrow. The $K_m$-distributions deviate from each other substantially. 
The reason is that the $J$ space is finite, $-I\leq K\leq I$. Mapping 
it on the infinite boson space by means of the HP expansion removes 
this restriction. The wave function can relax beyond the limits, 
which lowers its energy and leads to instability eventually.

For the yrast states $11/2_1$, $15/2_1$, and $19/2_1$, PTR and HO 
agree rather well. For the one-phonon states $13/2_1$ and $17/2_1$, 
the HO distributions extend beyond the limit, which correlates with 
the deviation of the PTR energy from the HO value. The PTR distribution 
has the form of the one-phonon HO distribution cut at the maximal value 
$\vert K_m\vert=I$. The cut-off does not substantially modify SCS plots 
for these states in Fig.~\ref{f:TWmaps1}, which look as expected for 
one-phonon states. The SCS map  of the $21/2_1$ state, which is unstable 
for the HO approximation, deviates noticeably from the harmonic 
one-phonon shape. This indicates the proximity of the instability of 
the TW regime, where the classical energy $E_{\rm{class}}(\phi)$ 
substantially deviates from a parabola.     

The PTR distributions for the states $11/2_2$, $15/2_2$, and $19/2_2$ 
deviate strongly from the HO distributions with phonon number $n=2$.
For small $\vert K_m\vert$ both distributions agree. Instead of the outer
bumps of the HO two-phonon distribution, the PTR one is concentrated near 
$\vert K_m \vert =I$, which is reflected by  the maxima near $\phi=\pm\pi/2$ 
in the SCS maps for the states in Fig.~\ref{f:TWmaps1-2}. It is appropriate 
to classify these states as strongly perturbed two-phonon structures, 
which are in the axis-flip regime. 

For the simple TRM,  a detailed study of the relation between the exact
solutions and the HO oscillators approximation based on the HP expansion 
is given in Ref.~\cite{W.X.Shi2015CPC}, which may be instructive concerning 
the foregoing discussion. 

\begin{figure}[ht]
\includegraphics[trim= 0 5cm 0 0, clip,width=\columnwidth]{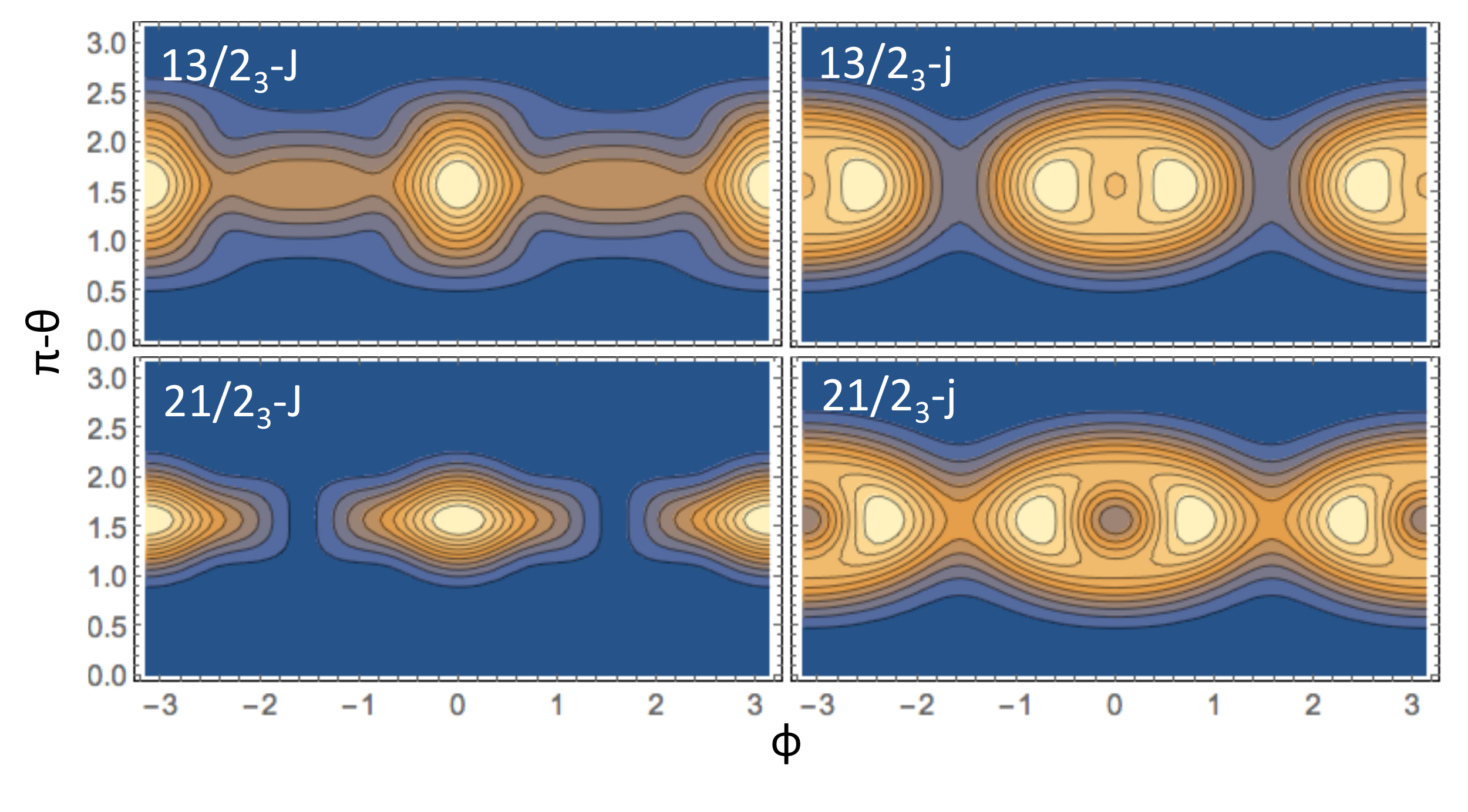}
\caption{\label{f:TWmapsSP} SCS probability distributions 
  $P(\theta \phi)_{I\nu}$ of the total angular momentum $\bm{J}$ (left 
  panel) and the particle angular momentum $\bm{j}$ (right panel) 
  for the states $13/2_3$ and $21/2_3$, which are classified as the 
  signature partner of the $11/2_1$ and $19/2_1$ yrast states.}
\end{figure}

Fig.~\ref{f:TWmapsSP} shows the SCS maps of the states $13/2_3$ and $21/2_3$, 
which we classify as the signature-partner cranking excitation. The probability of 
the total \am $\bm{J}$ has a maximum at the $s$-axis. The particle \am $\bm{j}$
executes a precession cone about the $s$-axis. This is expected for the 
unfavored signature partner, which is interpreted as rotation about the 
$s$-axis with one unit less of particle \am along this axis. The 
$\bm{j}$-precession is compensated by the counter-precession of the 
core \am$\bm{R}$, such that $\bm{J}$ remains aligned with the $s$-axis.

For the $13/2_3$ state, the $\bm{J}$-map has small bumps at $\phi=\pm \pi$,
which disappear for the $17/2_3$ and  $21/2_3$ states. We assign the 
bumps to a weak mixing with the third wobbling states $13/2_2$, $17/2_2$, 
and $21/2_2$, which have a density maximum there (see Fig.~\ref{f:TWmaps1-2}).
The mixing goes away with increasing $I$ because the wobbling and SP states 
move away from each other (see Figs.~\ref{f:E135Pr} and \ref{f:Ewob163Lu}). 
This mixing between wobbling and cranking modes is the reason why the 
distribution of the third wobbling state $13/2_2$ in Fig.~\ref{f:TWmaps1-2}
looks different from the ones at higher $I$. The SCS maps are sensitive to
mixing between the wobbling and SP modes.

\begin{figure}[ht]
\begin{center}
 \includegraphics[width=1.0\columnwidth]{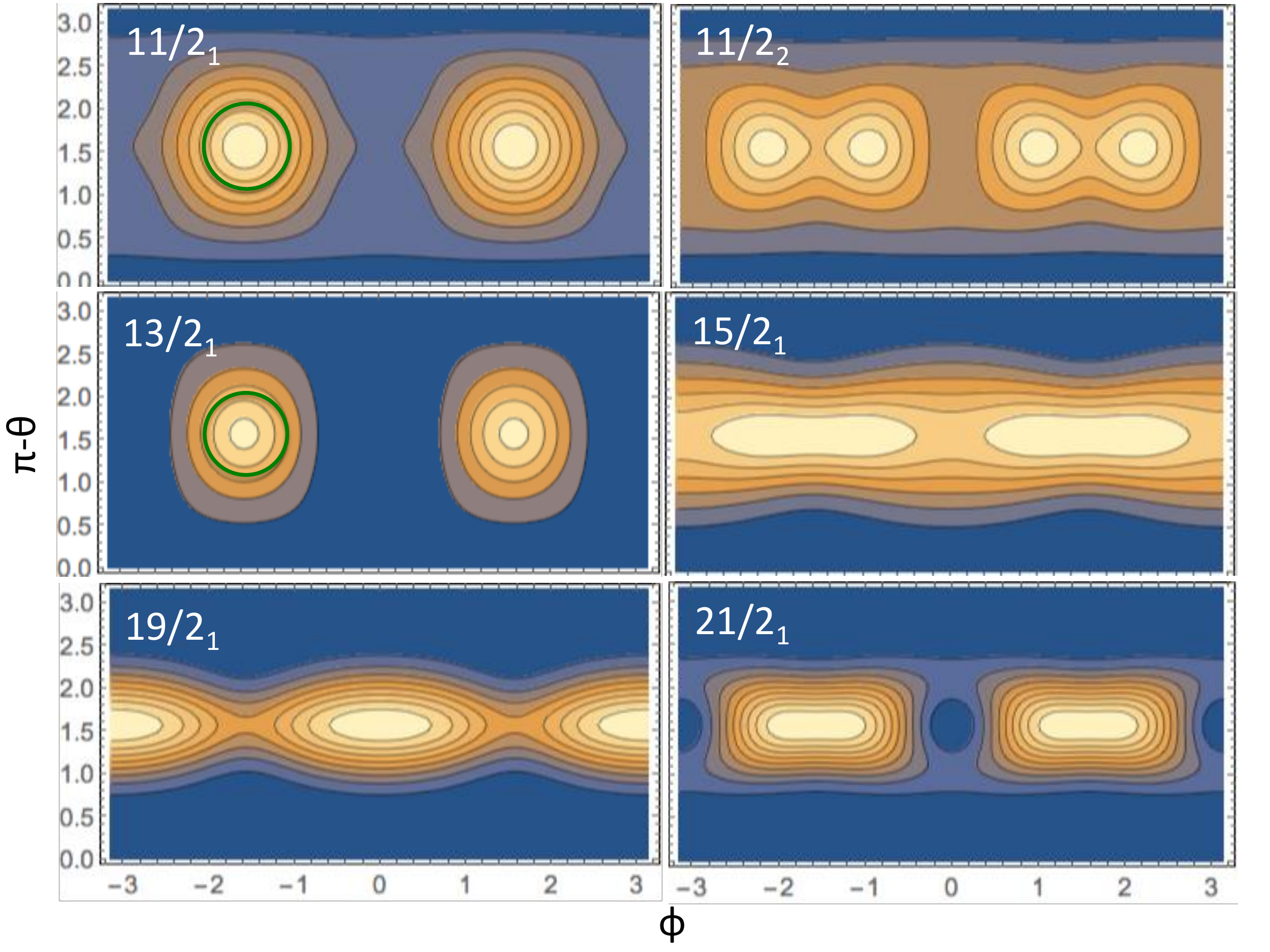}
\end{center}
\caption{\label{f:TWmapsR1} SCS probability distributions 
 $P(\theta \phi)_{I\nu}$ of the core angular momentum $\bm{R}$ 
 for some of TW states $I\nu$ of the PTR states shown in 
 Fig.~\ref{f:E135Pr}. See caption of Fig.~\ref{f:TRmaps1} 
 for details.}
\end{figure}

\begin{figure*}[!bpth]
 \includegraphics[height=7.0 cm]{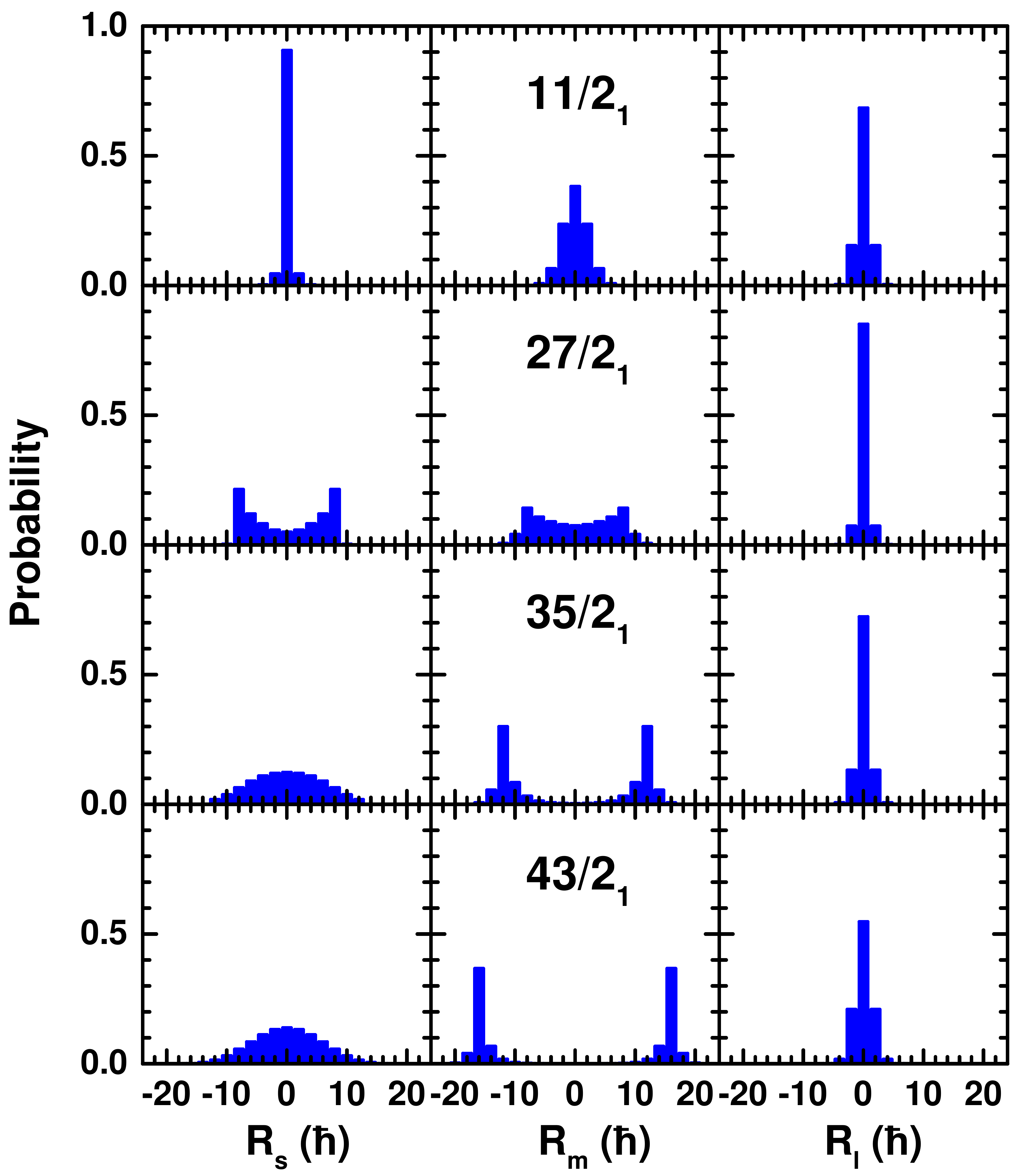}~~
 \includegraphics[height=7.0 cm]{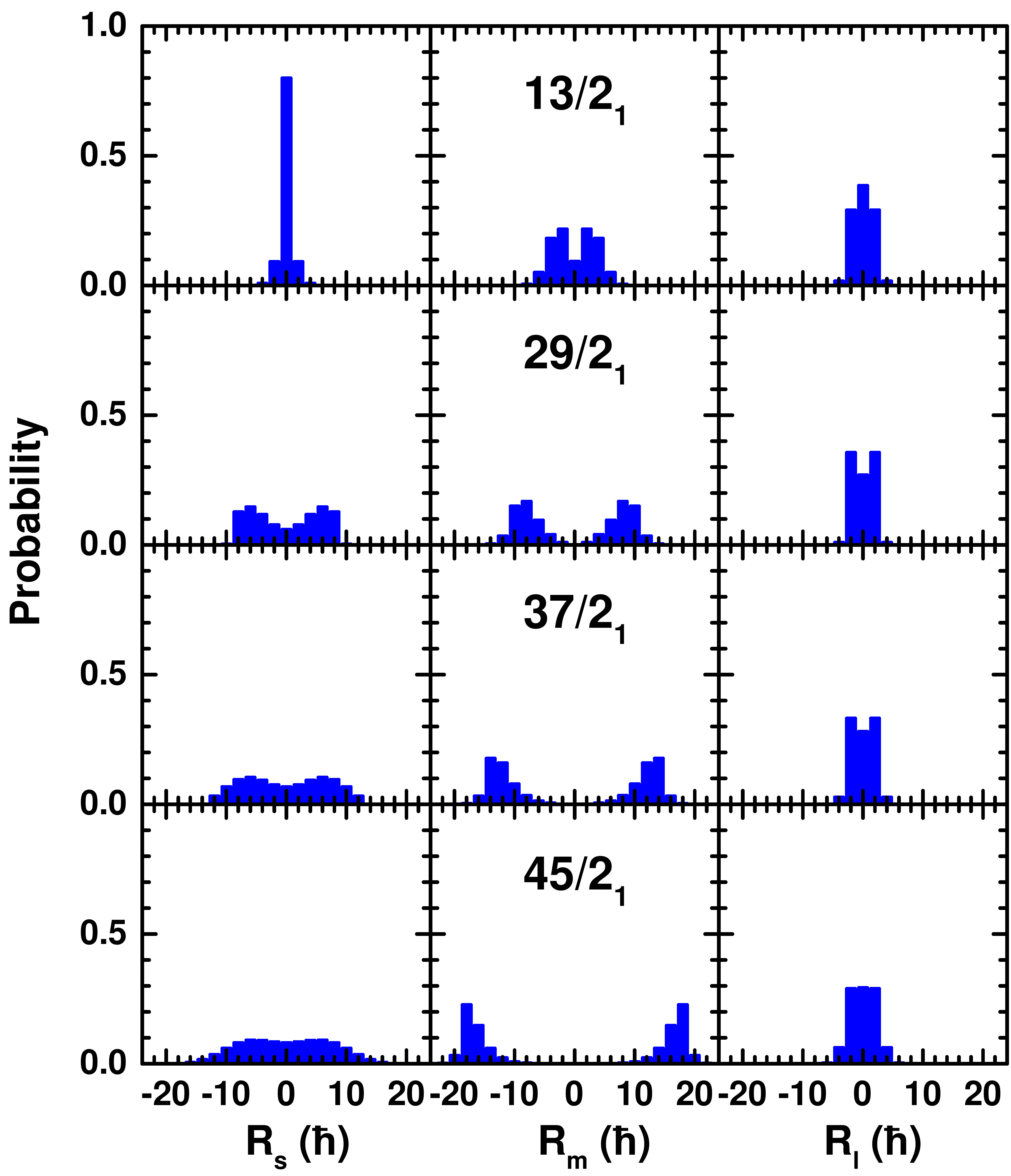}~~
 \includegraphics[height=7.0 cm]{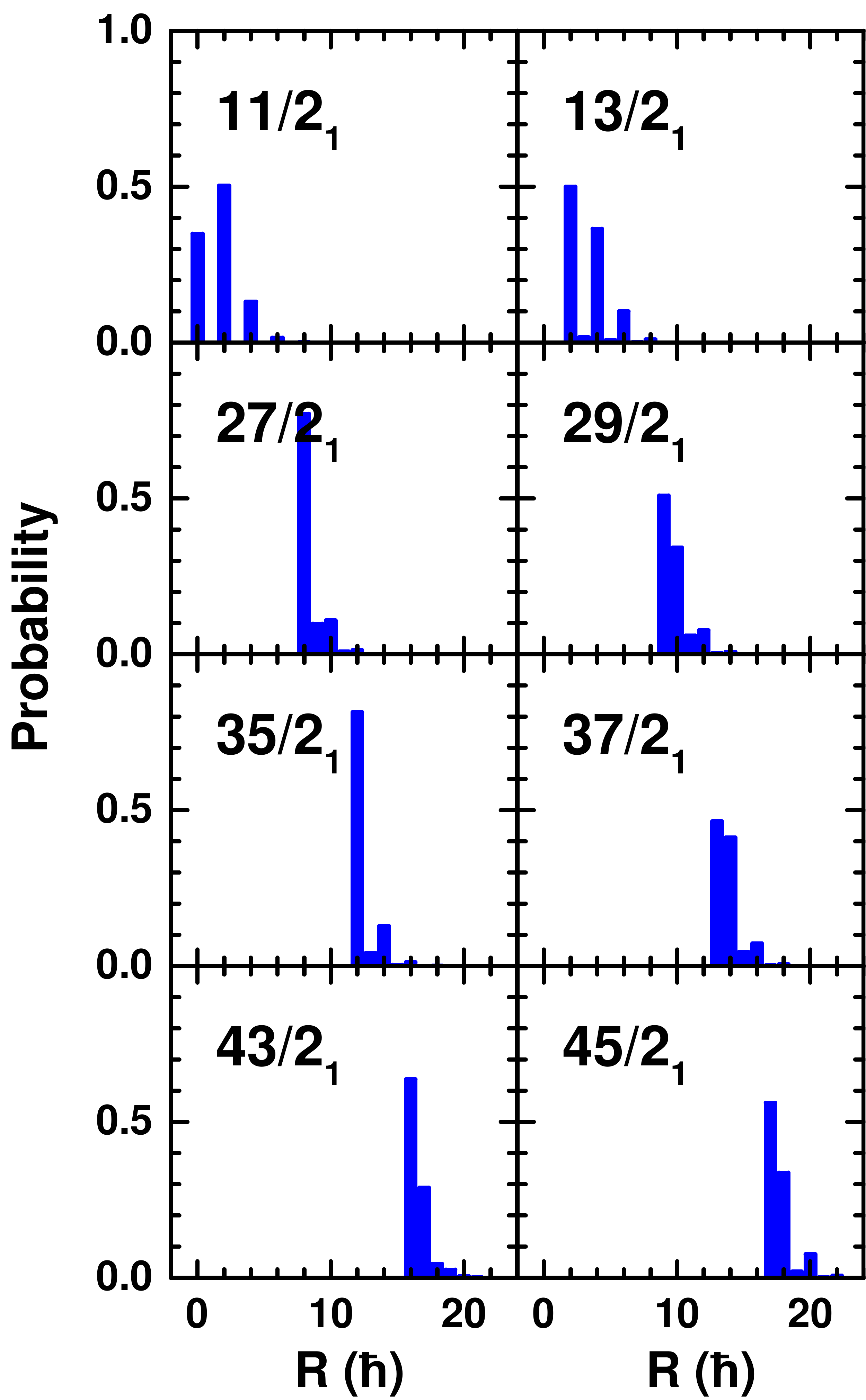}
 \caption{\label{f:KRplots}  Probability distributions of the 
 projection of rotor angular momentum $P_{KR}$ on the three axes 
 for some yrast states (left panel) and wobbling states (middle panel) 
 and the total rotor \am $P_R$ (right panel) of some PTR states 
 shown in Fig.~\ref{f:E135Pr}.}
\end{figure*}

Fig.~\ref{f:TWmapsR1} shows the SCS probability distributions
$P(\theta \phi)_{I\nu}$ of the core angular momentum $\bm{R}$. The 
peaks of the $P(\theta \phi)_{I\nu}$ are consistent with the root 
mean square expectation values of the angular momentum components 
shown in Fig.~\ref{f:ANG135Pr}. If $R_m$ is larger than $R_s$, 
the peaks are located at $\phi=\pm \pi/2$, while if $R_m$ 
is smaller than $R_s$, the peaks are located at $\phi=0$, $\pm \pi$.
For the yrast state $11/2_1$ the small rotor \am of $\bar{R}\approx 1.8$ 
(taken from Figs.~\ref{f:ANG135Pr} or \ref{f:KRplots}) is 
distributed around $\phi=\pm\pi/2$. It counteracts the zero-point 
oscillations of $\bm{j}$ such that the smaller zero-point oscillations 
of $\bm{J}$ result. For the wobbling state $13/2_1$ the rotor \am of  
$\bar{R}\approx 3.7$ is also distributed around $\phi=\pm \pi/2$. 
It generates the maxima of the probability distribution at the two 
turning points $\theta=\pi/2$ in the $\bm{J}$-SCS map shown in 
Fig.~\ref{f:TWmaps1}. In contrast to the yrast states  there is a 
phase difference of between the left and right blobs. For the yrast 
states $15/2_1$ and $19/2_1$ the center of the probability moves toward 
the $s$-axis, along which $\bm{R}$ is aligned in the classical 
picture. Yet the fluctuations remain large. For the wobbling states
$17/2_1$ (not shown) and $21/2_1$ there are  maxima near $\phi=\pm \pi/2$ 
and minima at $\phi=0$, $\pi$, which reflect the phase change. 
Combining these $\bm{R}$-distributions with the pertaining 
$\bm{j}$-distributions in Fig.~\ref{f:TWmapsjlow} results in 
the $\bm{J}$-distribution in Fig.~\ref{f:TWmaps1}.

Fig.~\ref{f:KRplots} shows the probability distributions of the 
projections of rotor angular momentum $P_{KR}$ on the three axes 
for some yrast states (left panel) and wobbling states (middle panel). 
The right panel displays the fraction of the total rotor \am $P_R$ 
in the same PTR states.

The distributions of the $R_s$ component of the yrast states with signature 
$11/2_1+2n$ reflect the $I$-dependence of $\langle \hat R_s^2\rangle$ in 
Fig.~\ref{f:ANG135Pr}. First it widens, and above $J_c$ it does not change 
much. The $R_m$-distribution of the yrast states continuously shift to 
larger values of $\langle \hat R_m^2\rangle$. The $R_m$-distributions remain 
narrowly centered around zero. For the sequence of the TW wobbling 
states with the signature $13/2_1+2n$ the distributions of the $\bm{R}$ 
components develop in an analogous way, which is expected from 
Fig.~\ref{f:ANG135Pr}. The only difference is that the $R_m$-distributions 
of the yrast states have a maxima at $\phi=0,~\pi$, where the ones of 
the wobbling states have minima. The difference is caused by the 
different symmetry of the wave functions with opposite signature, 
which was discussed in the context of the $\bm{R}$-SCS maps. 
 
The right panel of Fig.~\ref{f:KRplots} show the probabilities of the 
different rotor angular momenta $R$. The distributions are restricted by 
the conservation of the total \am, i.e., $|I-j| \leq R \leq I+j$. 
Except for $I=11/2$ the lowest possible value $R=I-j$ has by far the 
highest probability in the yrast sequence. That is, the energy gain by 
reducing the rotor \am overcomes the energy loss by reorienting $\bm{j}$. 
It should be kept in mind that the energy dependence on the orientation 
of $\bm{j}$ and $\bm{R}$ is weaker than for the classical energy expression. 
The reason is the large uncertainty in the orientation of $\bm{j}$, 
which has been discussed in the context of Fig.~\ref{f:Effective_angle}.
The probability of $R=I-j$ is maximal for the wobbling states as well, 
while the fraction of $R=I-j+1$ is substantial. From the $R$-plots, 
one sees that $R$ is an approximate quantum number in the yrast 
band ($I> 15/2$), but not in the wobbling band. The admixture of the
states with $R=I-j$ and $R=I-j+1$ in the wobbling band generates larger 
angle between $\bm{j}$ and $\bm{R}$ in the wobbling cone.

The HO approximation is a useful tool to identify the two fundamental 
modes of the PTR model. The collective TW mode represents the periodic 
motion of $\bm{J}$ with respect to the $s$-axis in the body-fixed frame or, 
equivalently, the same periodic counter-motion of the $s$-axis in the 
laboratory frame. The single particle SP mode represents a de-alignment 
of $\bm{j}$ from the $s$-axis, which is accompanied by a counter motion 
of $\bm{R}$ such that $\bm{J}$ does not wobble. Both modes are coupled. 
The coupling is weak enough that the resulting normal modes can be 
clearly classified as TW and SP. The HO model approximates the full 
PTR reasonably well for zero- and one-phonon states with $I$ values 
below $J_c$. The deviations (unharmonicities) are substantial. 
However they do not change the topological character of the motion, 
which we use to classify the modes. That is, the classification of 
the collective mode as ``transverse wobbling" is not restricted to 
the HFA as claimed by the authors of Ref.~\cite{Lawrie2020PRC} and
by means of which it was introduced in Ref.~\cite{Frauendorf2014PRC}. 
It is of topological nature: TW means that the total \am vector $\bm{J}$ 
revolves the $s$-axis, which is transverse to the $m$-axis with the 
largest \momid. Although not explicitly stated in Ref.~\cite{Frauendorf2014PRC} 
and subsequent discussions~\cite{Frauendorf2018PS, Matta2015PRL, 
Streck2018PRC, Timar2019PRL, Sensharma2019PLB, Q.B.Chen2019PRC_v1, 
Nandi2020PRL}, TW classification was understood in this more general 
topological sense when applied to the quantal PTR calculations.

\subsection{Transverse wobbling versus alternative 
interpretations}\label{s:TWvsAlt}

The Triaxial Strongly Deformed (TSD) bands in the Lu isotopes, which 
represent the first experimental evidence for wobbling mode in 
nuclei~\cite{Odegaard2001PRL}, were studied by several authors in the 
framework of the PTR. The authors of Ref.~\cite{Frauendorf2014PRC} 
classified the bands TSD1 and TSD2 in $^{163}$Lu as the zero- and 
one-TW states. Other authors  arrived at different interpretations 
because they started with different assumption about the moments of 
inertia and used approximation schemes. In this section we judge these 
interpretations from our perspective.

\begin{figure}[ht]
    \includegraphics[width=0.8\linewidth]{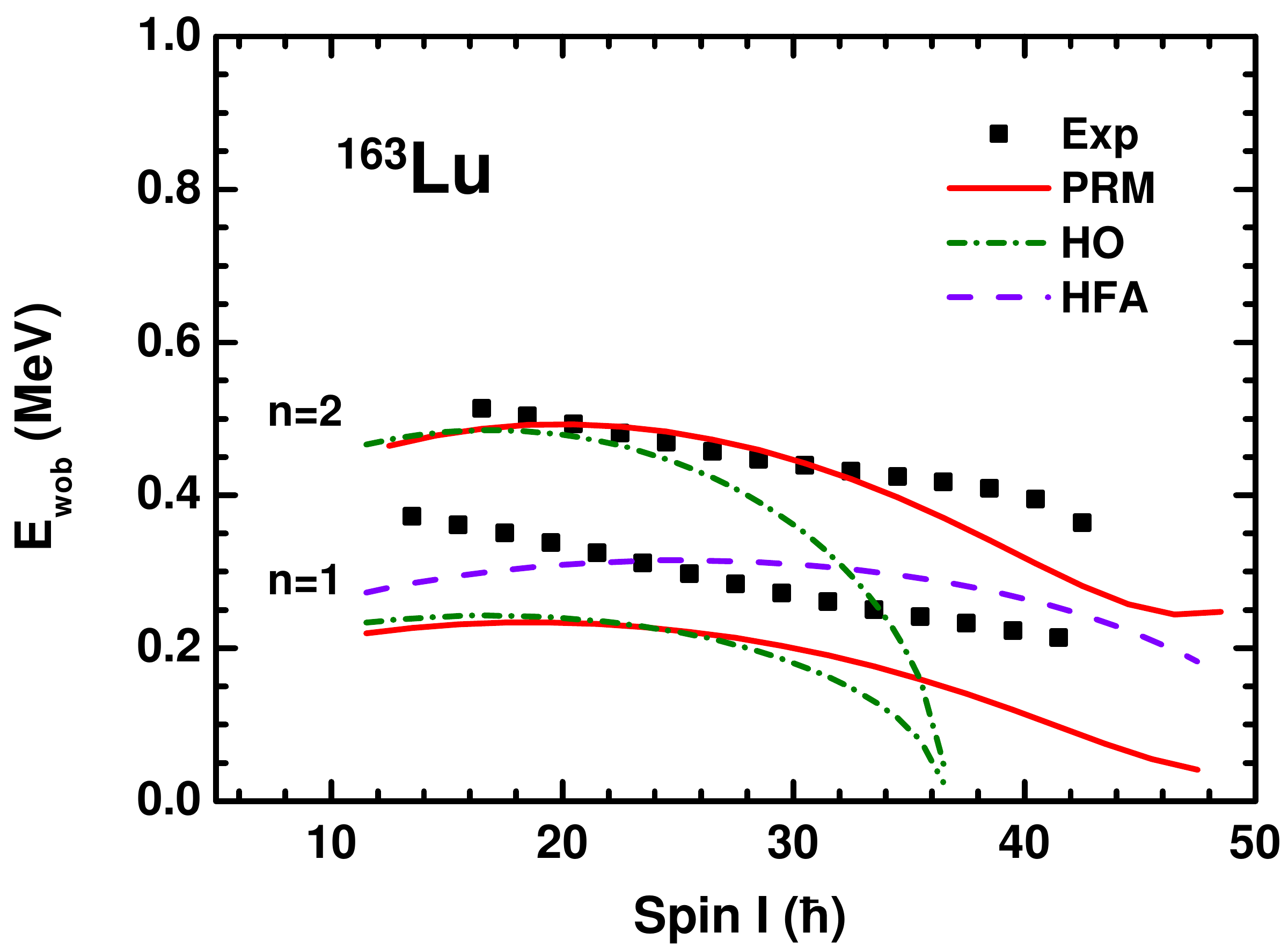}
    \caption{Excitation energies of the wobbling band with phonon numbers 
    $n=1$ and $2$ with respect to the yrast band in $^{163}$Lu 
    calculated by PTR, HP boson expansion (labeled as HO), and HFA.} 
    \label{f:Ewob163Lu}
\end{figure}

\begin{figure*}[ht]
  \includegraphics[width=\linewidth]{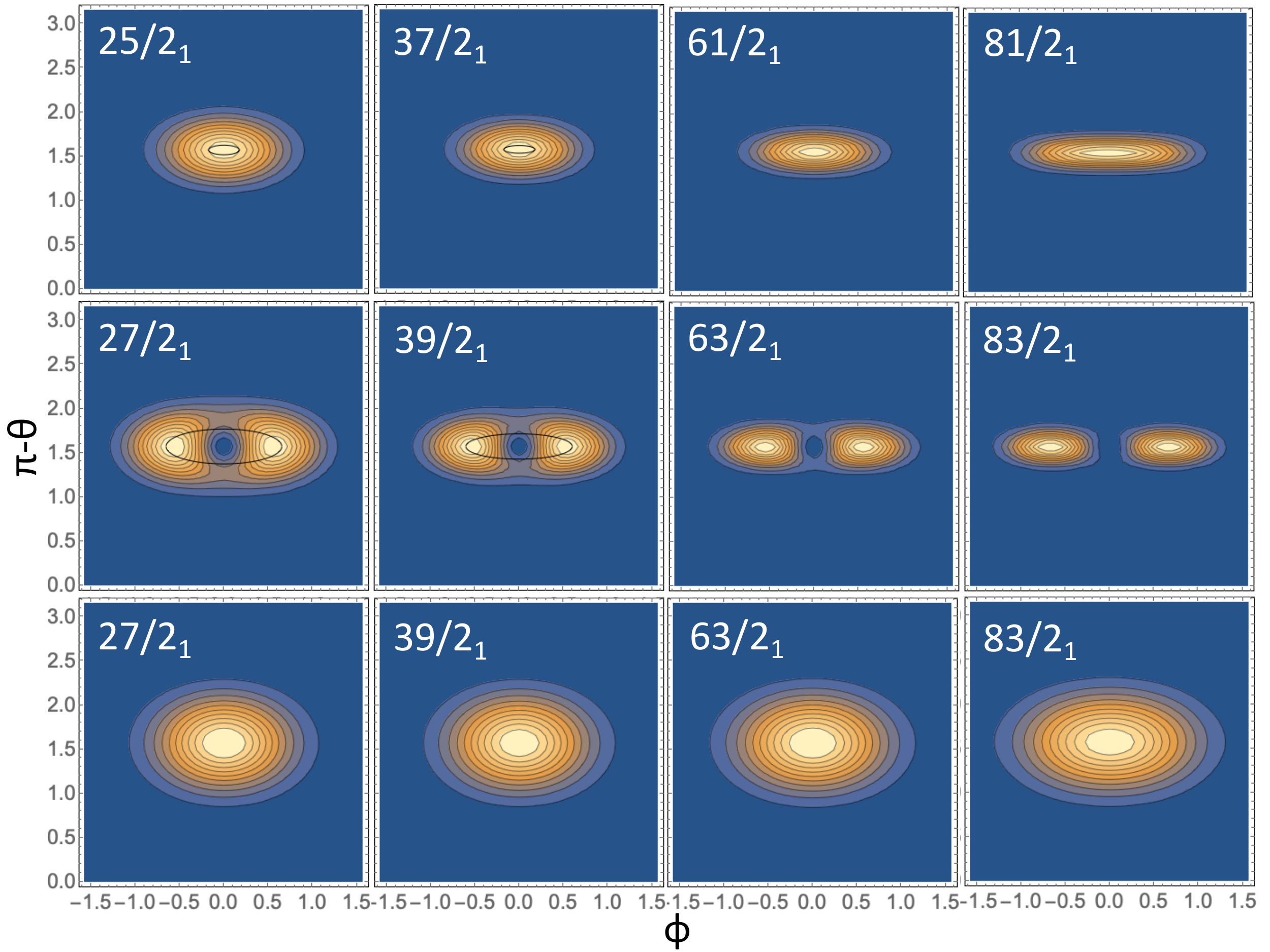}
   \caption{SCS probability distributions $P(\theta \phi)_{I\nu}$ 
   of the total angular momentum $\bm{J}$ (upper and middle rows) 
   and particle \am $\bm{j}$ of the $i_{13/2}$ proton (lower row) 
   for some yrast and one-phonon states in $^{163}$Lu. The classical 
   orbits orbitals calculated by means of the HFA approximation are 
   shown by black ovals.}
   \label{f:TWmaps163Lu}
 \end{figure*}  
 
 \begin{figure}[ht]
    \centering
    \includegraphics[width=0.9\columnwidth, trim=5.5cm 11cm 0 0cm, clip]{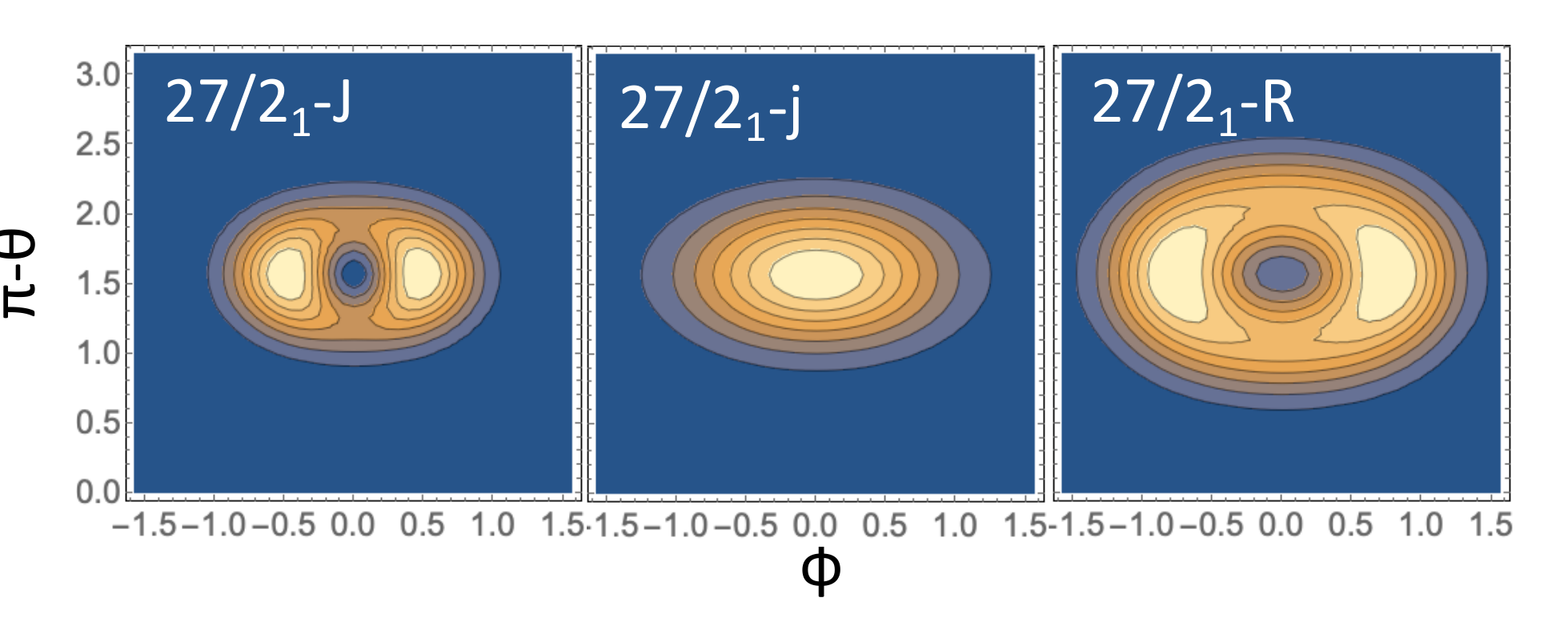}
    \caption{SCS probability distributions $P(\theta \phi)_{I\nu}$ 
    of the total angular momentum $\bm{J}$, the particle angular 
    momentum $\bm{j}$, and the core angular momentum $\bm{R}$ for the 
    state $27/2_1$ of $^{163}$Lu calculated by the PTR assuming the 
    rigid body ratios between the \momis as given in Ref.~\cite{Raduta2020PRC}. 
    See caption of Fig.~\ref{f:TRmaps1} for details.}
    \label{f:TWmaps_rigid}
\end{figure}

In addition to $^{135}$Pr, the studies to be discussed focused
on $^{161-165}$Lu. For this reason, we first present our perspective 
on $^{163}$Lu. We repeated the PTR calculation of Ref.~\cite{Frauendorf2014PRC}, 
where the parameters are listed in Table I ($^{163}$Lu - fit). 
Fig.~\ref{f:Ewob163Lu} compares the PTR wobbling energies $E_{\textrm{wob}}$ 
with the ones obtained from the experimental bands TSD1, TSD2, and TSD3, 
the HO approximation, and the HFA approximation. As discussed in the 
previous section for $^{135}$Pr, the HO mode becomes unstable at $J_c=36.5$ 
near the instability of classical rotation about the $s$-axis while the 
TW mode of the PTR remains stable up to $J\approx 45$. The HFA 
approximation is stable up to $J\approx 50$. Its 
deviation from the PTR curve quantifies the coupling between the 
pure TW and cranking modes. The SP state is admixed with amplitudes 
$-0.06$, $-0.08$, $-0.10$ into the states $I=27/$, 39/2, 63/2, 
respectively. The HO model is only reliable below $J=30$. 

The SCS maps in Fig.~\ref{f:TWmaps163Lu} clearly demonstrate the 
persistence of the TW regime over almost the whole range of \am shown. 
The yrast states with signature $13/2+2n$ represent rotation about 
the $s$-axis. The yrare states with signature $15/2+2n$ show the 
rims which characterize the one-phonon TW excitation. Their energy 
relative to the yrast sequence is shown in Fig.~\ref{f:Ewob163Lu}. 
The coupling between the TW and the SP modes is seen in the lower 
row of Fig.~\ref{f:TWmaps163Lu}. The $\bm{j}$-map becomes more 
elongated with increasing $I$ because the wobbling of $\bm{J}$ 
pulls $\bm{j}$  more and more away from the $s$-axis. This correlated 
motion increases the probability to stay near the two $\phi=\pm \pi/2$ 
turning points and reduces the probability near the $\phi=0$, $\pi$. 
The TW mode changes from the HO form at $I=27/2$ to flipping between 
the turning points at $I=63/2$. The TW regime ends at around $I=81/2$. 

\begin{figure*}[ht]
    \includegraphics[width=\linewidth]{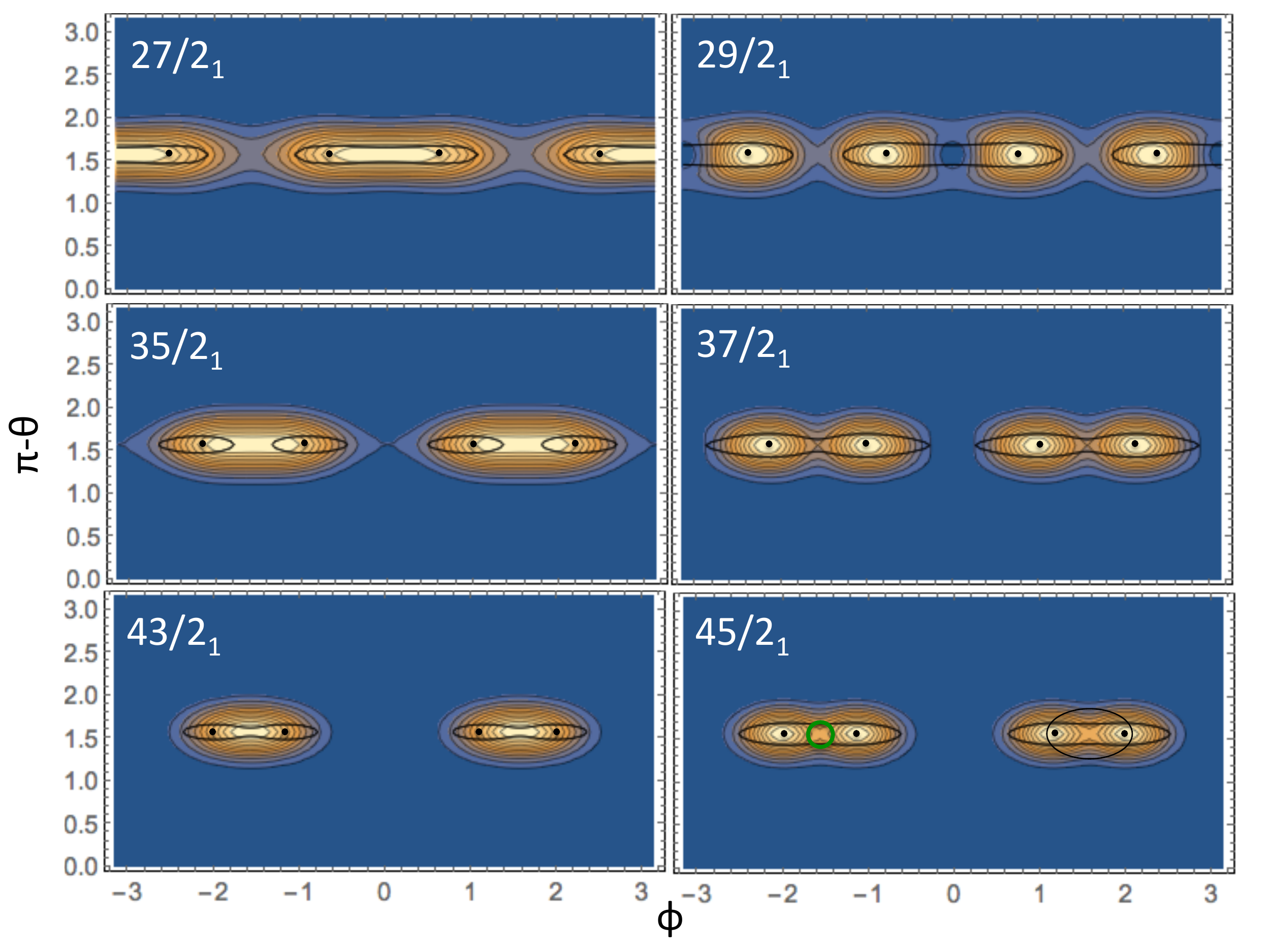}
    \caption{\label{f:TWmaps2} SCS probability distributions 
    $P(\theta \phi)_{I\nu}$ of the total angular momentum $\bm{J}$ 
    for some of the TW states $I\nu$ of the PTR states shown in 
    Fig.~\ref{f:E135Pr}. See caption of Fig.~\ref{f:TRmaps1} for 
    details. The dots are localized at the minimum of the classical 
    energy $\phi_J$ shown in Fig.~\ref{f:ClassAngle}. The curves show 
    a contour calculated by minimizing the energy for given angles $\theta_j$ 
    and $\phi_j$ with respect to the orientation of  $\bm{j}$. The energy 
    of the contour is the PTR value shown in Fig.~\ref{f:E135Pr}. The contour 
    is the classical orbit in adiabatic approximation. The ellipse shows the 
    path along which the phase is shown in Fig.~\ref{f:TWphase452}.} 
\end{figure*}

\begin{figure}[ht]
    \centering
    \includegraphics[width=8.5cm,trim= 5.4cm 0 0 0, clip]{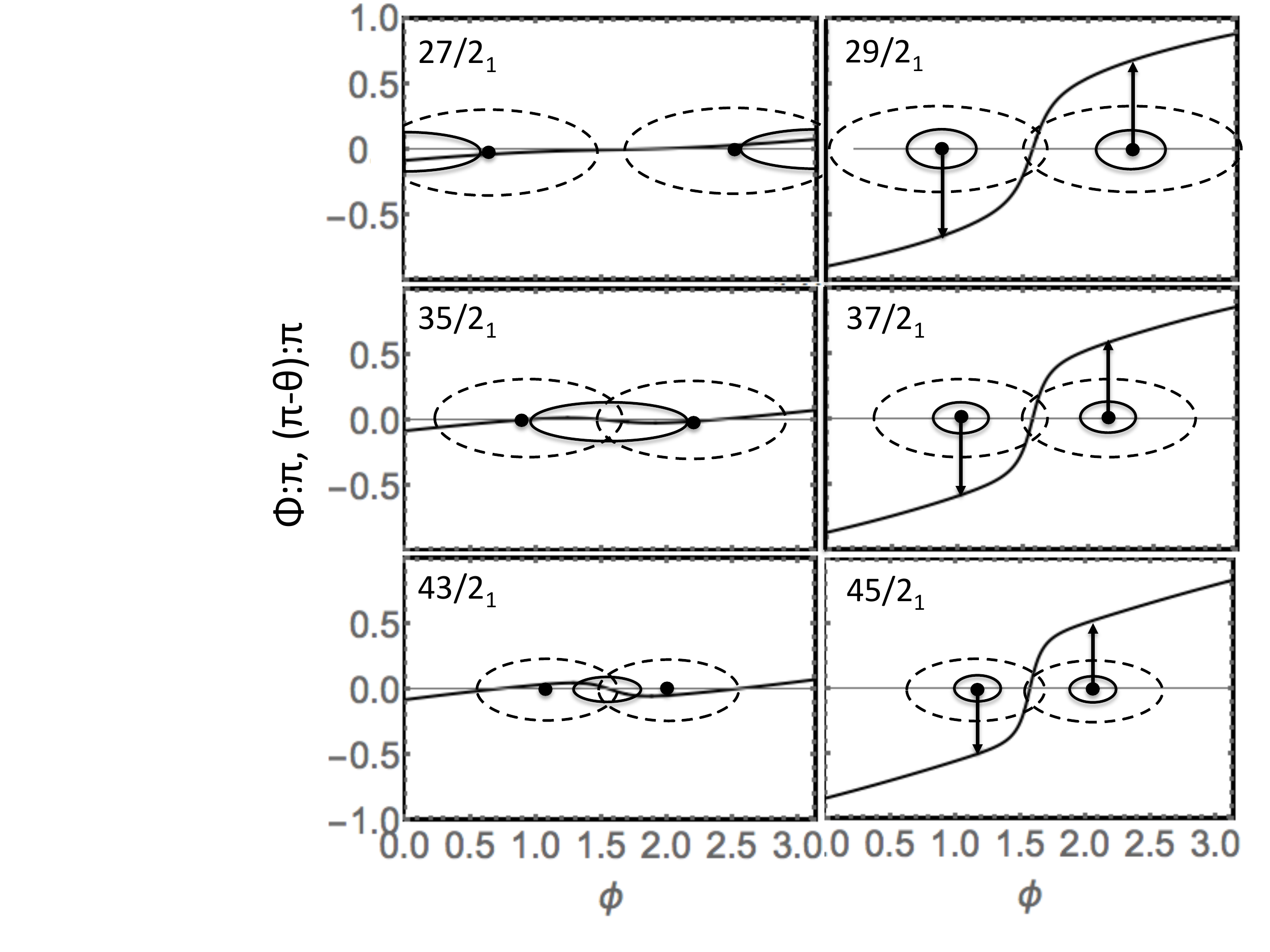}
    \caption{Phase difference $\Phi(\theta=\pi/2+0.05,\phi,
    \theta_0=\pi/2+0.05, \phi_0=\pi/2)$ calculated along a path 
    parallel and slightly above the $\phi$-axis. The full drawn 
    ellipses enclose the region of the largest probability of 
    the PTR states. The dashed ellipses enclose the region of substantial 
    probability for a state that represents uniform rotation about 
    the axes which connect the origin with the dots. The dots are 
    located at the angles $\pm \phi_J$ in Fig.~\ref{f:ClassAngle}.}
    \label{f:TWphase273543}
\end{figure}

\begin{figure}
    \centering
    \includegraphics[width=\columnwidth, trim= 1cm 8cm 1cm 0, clip]{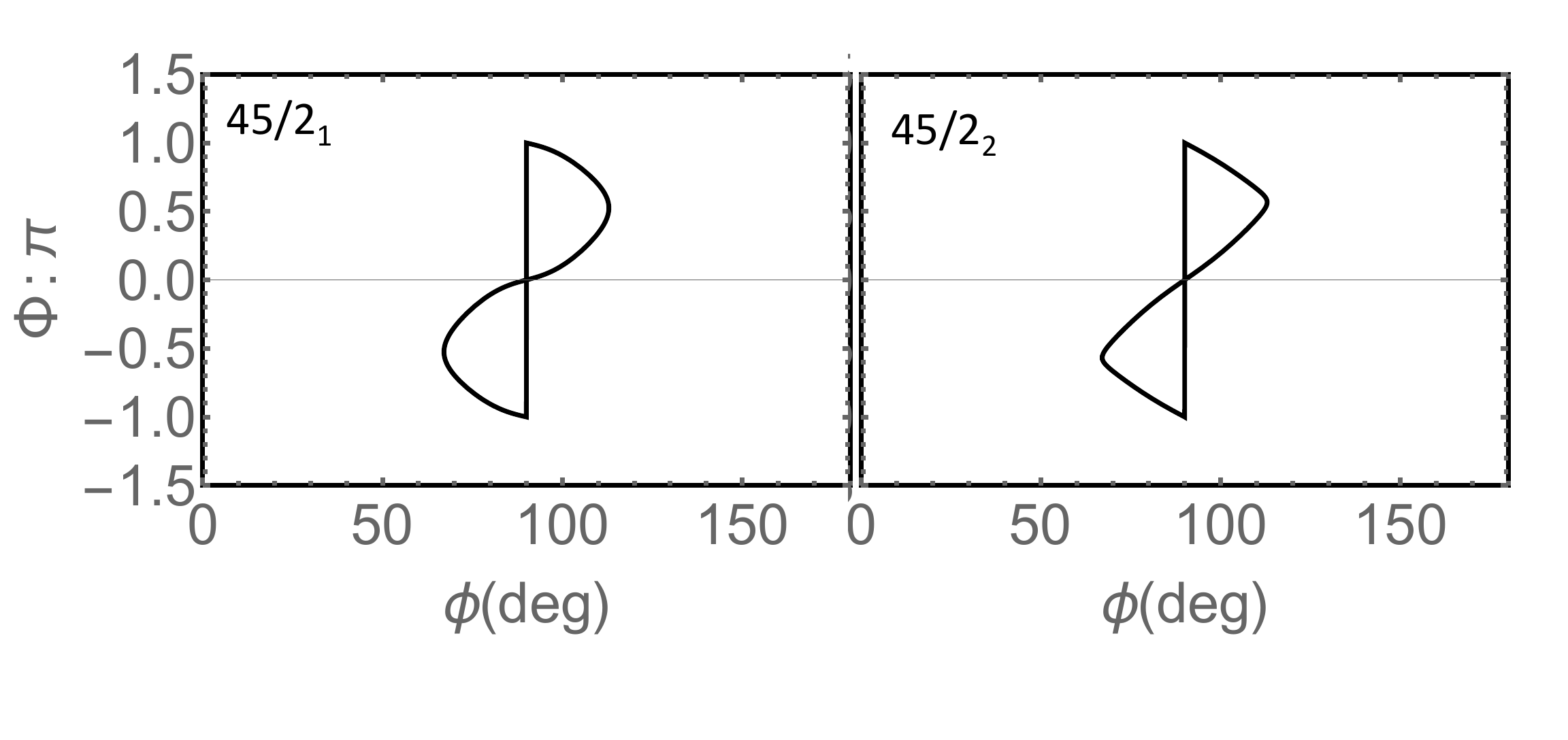}
    \caption{Phase difference $\Phi(\theta\phi, \theta_0\phi_0=\pi/2)$ for 
    the states $45/2_1$ (left) and  $45/2_2$ (right) along the elliptical path shown in 
    Figs.~\ref{f:TWmaps2} and ~\ref{f:TWmaps2-2} relative to the point $\phi_0=\pi/2$ 
    at the upper branch.}
    \label{f:TWphase452}
\end{figure}

\begin{figure*}[ht]
    \includegraphics[width=\linewidth]{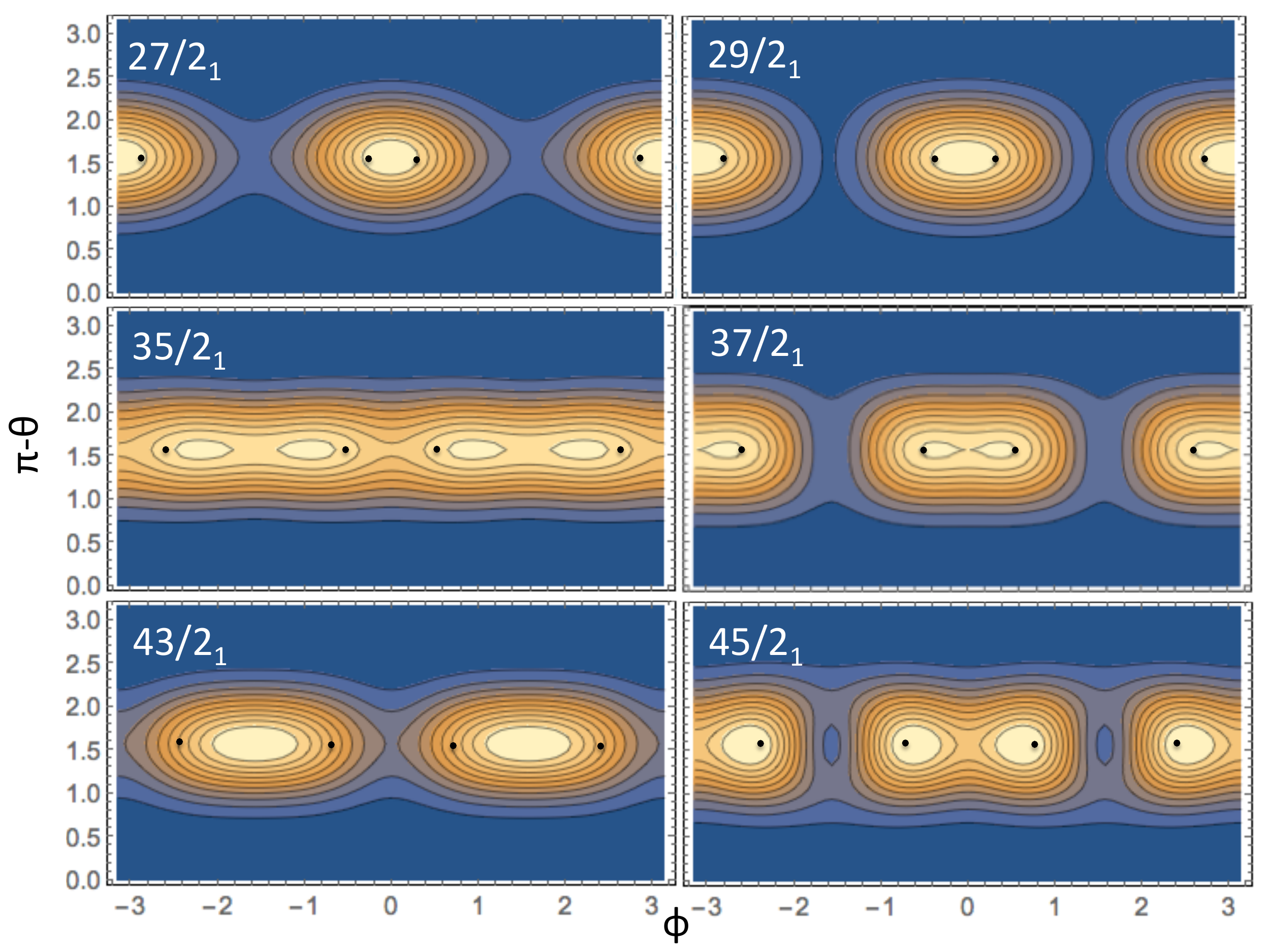}
    \caption{\label{f:TWmapsjhigh} SCS probability distributions 
    $P(\theta \phi)_{I\nu}$ of the particle angular momentum $\bm{j}$ 
    of some states $I\nu$ of the PTR states shown in Fig.~\ref{f:E135Pr}. 
    The dots are located at the values of $\phi_j$ of the minimum 
    of the classical energy shown in Fig.~\ref{f:ClassAngle}.}
\end{figure*}

Next we want to clarify the terminology. Bohr and Mottelson 
introduced wobbling in Nuclear Structure II~\cite{Bohr1975} p.190 ff.
They describe the mode as ``...the precessional motion of the axes with 
respect to the direction of $I$; for small amplitudes this motion has 
the character of a harmonic vibration..." (p.191). The quote indicates 
that they had in mind that the mode may have an anharmonic character 
as well, as it is common to speak about anharmonic vibrations 
of a certain type. The name wobbling is quite appropriate, it denotes 
the motion of the \am with respect to the body-fixed axes, which 
coincides (except the time direction) with the motion of the axes 
of the density distribution in the laboratory system, where the 
\am vector stands still. ``Wobbling" describes the staggering motion 
of a thrown baseball or the swaying motion of the earth axis. 
The authors of Ref.~\cite{Frauendorf2014PRC} defined the termini 
``transverse wobbling" (TW) and ``longitudinal wobbling" (LW) 
in the same topological sense. They discussed the harmonic limits 
of the modes combined with the frozen alignment (FA) approximation 
in order to obtain the analytical HFA expressions for a quick qualitative 
estimate, which allowed them to explain the underlying physics in a 
transparent way. The quantitative studies of TW in $^{135}$Pr and 
$^{163}$Lu were carried out in the frame work of the PTR model 
without any approximation, where using the name TW in the topological 
sense was self-understood. The authors of Ref.~\cite{Lawrie2020PRC}
misunderstood the use of the name TW described in Refs.~\cite{Frauendorf2014PRC, 
Matta2015PRL, Timar2019PRL, Sensharma2019PLB} by falsely restricting 
it to the HFA limit of the mode and called the general PTR states with 
precessional nature ``tilted precession'' (TiP) states. We think one 
should not replace established terminology without a good reason. We 
consider it confusing to have different names for one and the same 
mode: TW for the harmonic limit and TiP when there are anharmonicites 
present. Speaking about harmonic or anharmonic wobbling is the 
appropriate terminology in our view.

The authors of Ref.~\cite{Lawrie2020PRC} introduced the TiP concept starting 
from the triaxial rotor with two equal moments of inertia, which has a 
fixed component of the total \am vector $\bm{J}$ perpendicular to one 
of the principal axes. The TiP interpretation was applied to the general 
quasiparticle+triaxial rotor (QTR) system. The TiP bands represent QTR 
energies, where the particle is assumed to be in the state $k_s=11/2$, 
which is the FA approximation of Ref.~\cite{Frauendorf2014PRC}.
Fig.~10 of Ref.~\cite{Lawrie2020PRC} compares the TiP approximation with 
the HFA approximation of Ref.~\cite{Frauendorf2014PRC} for TW. FA is 
a good approximation up to $I=23/2$, which can be seen in 
Figs.~\ref{f:TWmapsjlow} and \ref{f:ANG135Pr} and the $k_s$ plots 
(not shown), which are strongly concentrated near $k_s=11/2$. In 
this TW regime Fig.~10 of Ref.~\cite{Lawrie2020PRC} corresponds 
to our Fig.~\ref{f:Ewob_grid}, in the context of which we have 
discussed the appearance of anharmonicities with increasing \am 
and excitation energy. To claim a fundamental difference between 
TiP and TW appears inappropriate from our point of view. Above 
$I=23/2$ (see  Figs.~\ref{f:ClassAngle} and \ref{f:TWmapsjhigh}) 
the FA is no longer good. The lowest TiP energies in Fig.~10 of 
Ref.~\cite{Lawrie2020PRC} discontinuously merge into two 
degenerate bands, which is in stark contrast to the exact QTR 
energies. We will discuss this regime in Sec.~\ref{sec:TWtoLW}.

The authors of Ref.~\cite{Tanabe2017PRC} claim that the TW mode 
is unstable. Their claim is based on the stability criterion 
(\ref{eq:Astab}). As discussed in Sec.~\ref{sec:TW-SP}, the HO 
model predicts the instability of the TW mode too early. Moreover, 
the authors' assumption of irrotational-flow \momis destabilizes the 
TW mode further, such that for the relevant $I$ values the TW mode 
is unstable. In Ref.~\cite{Frauendorf2018PRC} it is shown that the 
microscopic ratios between the three \momis deviate from 
the irrotational-flow ratios. Thus we think that considering two 
of the ratios as adjustable parameters and determining them by a 
fit of the experimental wobbling energy is a legitimate procedure. 
The \momis determined in this way in Ref.~\cite{Frauendorf2014PRC} 
provide the stable TW regions discussed before.  

The authors of Ref.~\cite{Tanabe2017PRC} assume ad hoc that the 
ratios between the \momis are the same as for a rigid triaxial 
ellipsoid. The triaxiality parameter $\gamma$ is adjusted to the 
ratios between intra- and inter-band $B(E2)$ values. For rigid-body 
ratios the \momi of the $s$-axis is the largest. That is, the 
wobbling mode is longitudinal (LW)  according to the classification 
introduced in Ref.~\cite{Frauendorf2014PRC}. Fig.~\ref{f:TWmaps_rigid} 
shows the SCS maps for this LW scenario, which is stable for all $I$ 
values. As characteristic for LW, the wobbling frequency increases 
with $I$. To account for the decrease seen in experiment, the authors 
introduce an $I$-dependent scaling factor. They justify the scaling 
by the reduction of the pair correlations. 

There is a fundamental problem with the assumption in 
Ref.~\cite{Tanabe2017PRC}, which has been  discussed in 
Ref.~\cite{Frauendorf2018PRC} in detail. The assumption 
of rigid body \momis is in conflict with the quantal 
nature of the triaxial rotor. The indistinguishability of the 
nucleons requires that the \momi of a symmetry axis has to be zero, 
and any relation between the triaxiality parameter $\gamma$ and the 
\momis has to obey this requirement. The irrotational-flow relation 
is consistent with it, and microscopic cranking calculations fulfil 
it as well. For this reason we think that the longitudinal wobbling 
scenario of Ref.~\cite{Tanabe2017PRC} is problematic.

The authors of Ref.~\cite{Raduta2020PRC} map the PTR model on the SCS basis,
which is generated by two complex parameters $z$, $s$ (which are equivalent 
to the angles $\theta_J$, $\phi_J$ and $\theta_j$, $\phi_j$ in Eq.~(\ref{eq:SCS}) 
for the $J$ and $j$ spaces, respectively). They approximate the yrast 
sequence TSD1 by minimizing the energy with respect to $z$, $s$. The 
yrare sequence TSD2 is determined by minimizing the energy with respect 
to $z$, $s$ as well, which is possible because the SCS spaces of opposite 
signature are distinct. They call TSD2 ``signature partner band". In our 
view this is an unfortunate choice because the name is conventionally 
used for single particle excitations in the context of cranking model, 
which has a different structure than TSD2. As explained in 
Sec.~\ref{sec:TW-SP}, we adopt the conventional name. The TSD3 band 
is treated as a small-amplitude vibrational excitation built on TSD2. 
The authors assume rigid-body ratios between the \momisd. This has 
the consequence that the vibration has  longitudinal character, which 
leads the authors to the conclusion that the Lu isotopes are in the 
longitudinal regime. However, as discussed above, there is a
fundamental problem with the assumption of rigid-body ratios between 
the \momis and consequently with the interpretation.

\subsection{Transition to longitudinal wobbling}\label{sec:TWtoLW}

Above the critical \am $J_c=10.5$ the minima of the classical energy 
move away from $\phi_J=0,~\pi$. As seen in Fig.~\ref{f:Eclass}, there 
are four minima with the same energy that are located at $\pm \phi_J$, 
$\pi \pm \phi_J$ with $\phi_J$ shown Fig.~\ref{f:ClassAngle}. That is, 
the rotational axis of the yrast state tilts into the $s$-$m$-plane, 
where four equivalent orientations exist. The panel $I=33/2$ in 
Fig.~\ref{f:Eclass} suggests to interpret the yrast and first wobbling 
states as superpositions of states localized at the four minima, which 
have to obey the D$_2$ symmetry of the PTR model. In other words, 
the rotational axis flips between the four equivalent positions 
with equal probability for each. The interpretation is 
analogous to the H\"uckel model for the $\rm{C}_6\rm{H}_{12}$ molecule. 
The electronic structure is described  as a superposition of the 
electron orbitals localized at the atoms, which is called linear 
combination of atomic orbitals (LCAO) approximation. Based on 
this interpretation, we classify the regime as axis-flip wobbling.

The TAC approximation for the PTR treats the orientation of $\bm{J}$ 
classically. As for the classical energy, there are four solutions at 
the angles $\pm\phi_J$, $\pi\pm\phi_J$ with $\phi_J$ given in 
Fig.~\ref{f:ClassAngle}, which have the same energy. The TAC solutions 
spontaneously break the D$_2$ symmetry of the PTR on the mean field level 
(Tilted Axis Cranking --- TAC solutions), which implies that the intrinsic 
signature is broken as well. In the case of strong symmetry breaking the 
two sequences of opposite signatures merge into one $\Delta I=1$ 
band~\cite{Frauendorf2001RMP, Frauendorf2018PS}. The broken 
symmetry is restored by combining the TAC solutions with equal weight 
and appropriate phase~\cite{Q.B.Chen2014PRC, Q.B.Chen2016PRC_v1}. 
The mixing generates an energy difference between the two signature 
branches, which corresponds to the minimal distance of the $I_1=11/2+2n$ 
and $I_2=13/2+2n$ sequences in Figs.~\ref{f:E135Pr} and \ref{f:Ewob_grid}. 

Fig.~\ref{f:TWmaps2} shows the SCS maps for $\bm{J}$ above the critical 
\am $J_c$. The figure indicates the locations of the minima of the 
classical energy as dots. As seen in Fig.~\ref{f:ClassAngle}, the 
corresponding $\phi_J$ values are not far from those obtained in the 
TAC approximation for the PTR, which are somewhat closer to $\pi/2$. 
That is, the dots indicate the centers of the TAC mean field distributions. 
The SCS map of the state $27/2_1$ can be interpreted as the even 
superposition of the TAC states that are located left and right of 
$\phi_J=0$. As illustrated in Fig.~\ref{f:TWphase273543}, the densities 
overlap. For the state $27/2_1$ the phase along the path $\theta=\pi/2+0.05$ 
is almost constant, which is also the case for other values of $\theta$. 
Therefore the wave functions of the two TAC states combine constructively, 
which enhances the density in the overlap region. The phase change between 
the dots symmetric to $\pm\pi/2$ is small as well. That is, the signature 
of the linear combination is $11/2+2n$. 

For the state $29/2_1$ the overlap of the TAC states symmetric to 
$\pm\pi/2$ is large. The phase along the path $\theta=\pi/2+0.05$ 
changes by about $\pi$ between the dots. The most rapid change appears 
in the overlap region. The phase change is similar rapid on  
paths with different $\theta$ values, which leads to destructive 
interference between the two TAC states and a reduction of 
the density in the overlap region. The phase change between the
dots at $\pm \phi_J$ via 0 or $\pi\mp\phi_J$ via $\pi$ is small, which 
gives the linear combination the required signature of $13/2+2n$. 

There is an additional mechanism at work. In case of constructive 
interference, the increase of the probability in the overlap 
region pulls the vector $\bm{j}$ there, because this increases the 
Coriolis coupling term $-(A_s j_s J_s + A_m j_m J_m)$. 
The energy gain means that the two TAC states ``attract" each other. 
The attraction is seen for the state $35/2_1$ in Fig.~\ref{f:TWmapsjhigh}, 
which shows the SCS maps for $\bm{j}$. The dots indicate the location of 
the classical angles of $\bm{j}$ in the $s$-$m$-plane shown in 
Fig.~\ref{f:ClassAngle}. The maxima of the probability density are 
closer to $\pm \pi/2$ than the dots, which enhances the range of 
the overlap and the density therein. As a consequence, the 
probability density for $\bm{J}$ stretches between the dots 
with no minima at $\phi=0$, $\pi$.

In case of destructive interference, the reduction of the density 
leads to a loss of energy, which means that two states ``repel" each 
other. The repulsion is seen in Fig.~\ref{f:TWmapsjhigh} for the state 
$37/2_1$. The density maxima are further away from $\pm \pi/2$ than 
the dots. The distribution of $\bm{J}$ in Fig.~\ref{f:TWmaps2} shows 
four maxima at the dots. That is, for the $37/2_1$ wobbling state 
the nucleus flips between uniform rotation about the axes 
tilted by the angles $\pm\phi_J$, $\pi\mp\phi_J$ into the 
$s$-$m$-plane. 

The same mechanism acts for the states $35/2_1$ and $37/2_1$ of opposite 
signature. Because the angle $\vert \phi_J \vert $ is larger, 
the two TAC states symmetric to $\pm\pi/2$ overlap stronger and 
dominate the interference. 

With increasing \am the two tilted axes approach each other toward 
the $m$-axis. As seen in Figs.~\ref{f:TWmaps2} and \ref{f:TWphase273543} 
the state $I=43/2$ has a probability distribution which corresponds 
to rotation about the $m$-axis similar to the $8_1$ state in 
Fig.~\ref{f:TRmaps1}. The classical energy minima at $\phi_J$ are 
well outside the central density maximum. The same holds for 
the $\phi_j$ and the SCS map for $\bm{j}$ in Fig.~\ref{f:TWmapsjhigh}.
The attraction between the two TAC states caused by the
Coriolis term is so strong that they merge. Their zero point 
fluctuations completely wash out the small barrier of the 
classical energy between them.  

The maxima of the probability distribution of the $45/2_1$ state 
are located near the minima of the classical energy at $\phi_J$. 
Like discussed for the $29/2_1$ state, the reason is the destructive 
interference between the TAC states. The vector $\bm{J}$ wobbles 
with respect to the $m$-axis, which corresponds to wobbling of the 
$m$-axis with respect to the space-fixed $\bm{J}$ axis in the laboratory 
system. The opposite action of the Coriolis term makes the energy 
difference between the states of opposite signature increasing with $I$. 
These are the signatures of longitudinal wobbling (LW). 

Fig.~\ref{f:TWphase452} shows that the phase increment on a path 
enclosing the probability maxima is $2\pi$, which is expected for a 
one-phonon excitation. For the state $43/2_1$ there is no phase 
increment along the same path.

\begin{figure*}[ht]
    \includegraphics[width=\linewidth]{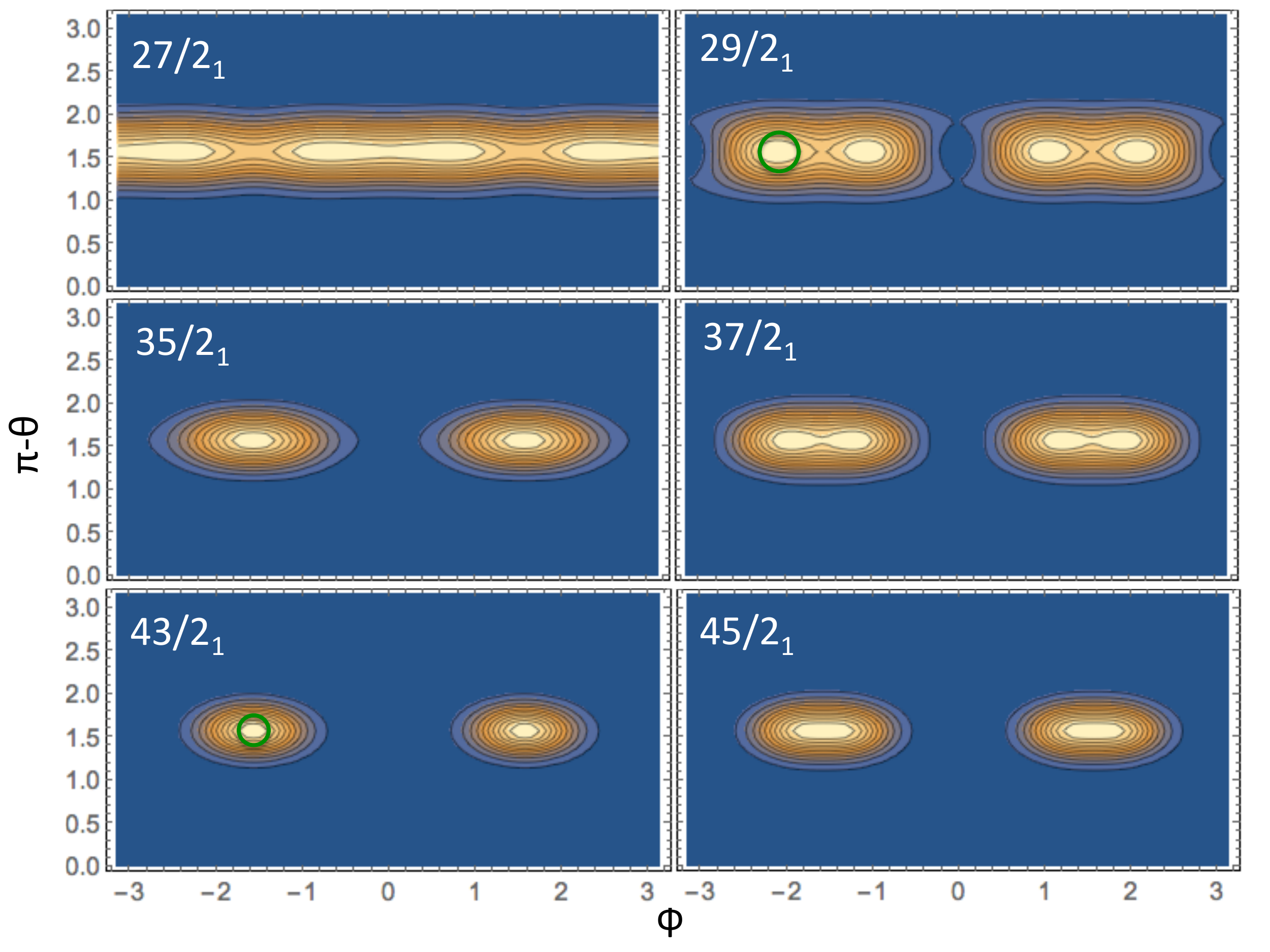}
    \caption{\label{f:TWmapsR2} SCS probability distributions 
    $P(\theta \phi)_{I\nu}$ of the core angular momentum $\bm{R}$ for 
    some of the states $I\nu$ of the PTR states shown in Fig.~\ref{f:E135Pr} 
    above the critical \am $J_c$. See caption of Fig.~\ref{f:TRmaps1} for 
    details.}
\end{figure*}

Fig.~\ref{f:TWmapsR2} shows the $\bm{R}$-SCS maps for the same states 
as the Fig.~\ref{f:TWmaps2} for $\bm{J}$ and Fig.~\ref{f:TWmapsjhigh} 
for $\bm{j}$. For the states $43/2_1$ and $45/2_1$, the $\bm{R}$-distributions 
have maxima at $\phi=\pm \pi/2$, i.e., $\bm{R}$ fluctuates about 
the $m$-axis without generating a phase gain. The $\bm{j}$-distribution 
for $43/2_1$ has maxima at $\phi=\pm \pi/2$, i.e., $\bm{j}$ fluctuates 
about the $m$-axis without generating a phase gain. Their sum $\bm{J}$ 
fluctuates the same way without generating a phase gain, which is 
characteristic for the zero-point motion of the zero-phonon state. The 
$\bm{j}$-distribution for $45/2_1$ has maxima near $\phi=\pm\phi_j$, 
$\pi \mp \phi_j$ with $\phi_j$ shown in Fig.~\ref{f:ClassAngle}. 
There is a phase gain of about $\pi$ between them, which is indicated 
by the minima at $\pm\pi/2$, $\pi\mp\pi/2$. The phase gain transfers 
to the sum $\bm{J}$, which adds to the phase gain of $2\pi$ for the 
circumferential path. 

The discussion of figures leads to another perspective on the LW regime. 
Without the rotor-particle coupling, $\bm{R}$ has a minimum and $\bm{j}$ 
a maximum at the $m$-axis. The coupling tries to align $\bm{j}$ 
with $\bm{R}$, which results in the classical energy $E(\phi_J)$ 
with a maximum at $90^\circ$ and two minima at 68$^\circ$ 
and $180^\circ-68^\circ$ separated by a shallow barrier (cf. 
Fig.~\ref{f:ClassAngle}). The zero- and one-phonon states 
are the lowest even and odd states within this double-well potential. 

The authors of Ref.~\cite{Frauendorf2014PRC} introduced the LW mode 
based on the assumption that the odd particle is tightly aligned with
the axis of the largest \momid. The \am vector $\bm{J}$ executes a 
harmonic precession cone about the axis. This structure differs to 
some extend from the LW limit in the current study. The $\bm{J}$-SCS 
map of the $45/2_1$ state in Fig.~\ref{f:TWmaps2} represents more 
of a flipping between the two orientations of the rotational axis $\bm{J}$ 
that are marked by the dots. It reflects the flipping of $\bm{j}$ 
between the two orientations symmetric to $\pi/2$ tilted by the angle 
$\pi/2-\phi_j$ from the $m$-axis into the $s$-$m$-plane. The 
distribution of $\bm{R}$ is an elongated blob centered with the $m$-axis 
similar to the zero-phonon state $43/2_1$. Deviations from a harmonic 
precession cone (see the state $9_1$ in Fig.~\ref{f:TRmaps1}) are expected
from the presence of the maxima at $\pm\pi/2$ in the classical energy 
seen in Fig.~\ref{f:Eclass}. Above the critical \am $J_c$ the 
wobbling mode consists of flipping between the two tilted axes, where 
the tilt angle $\phi_J$ gradually increases from 0 (TW regime) 
toward $\pi/2$ (LW regime).

\begin{figure*}[ht]
    \includegraphics[width=\linewidth,trim=0 0cm 0 0cm, clip]{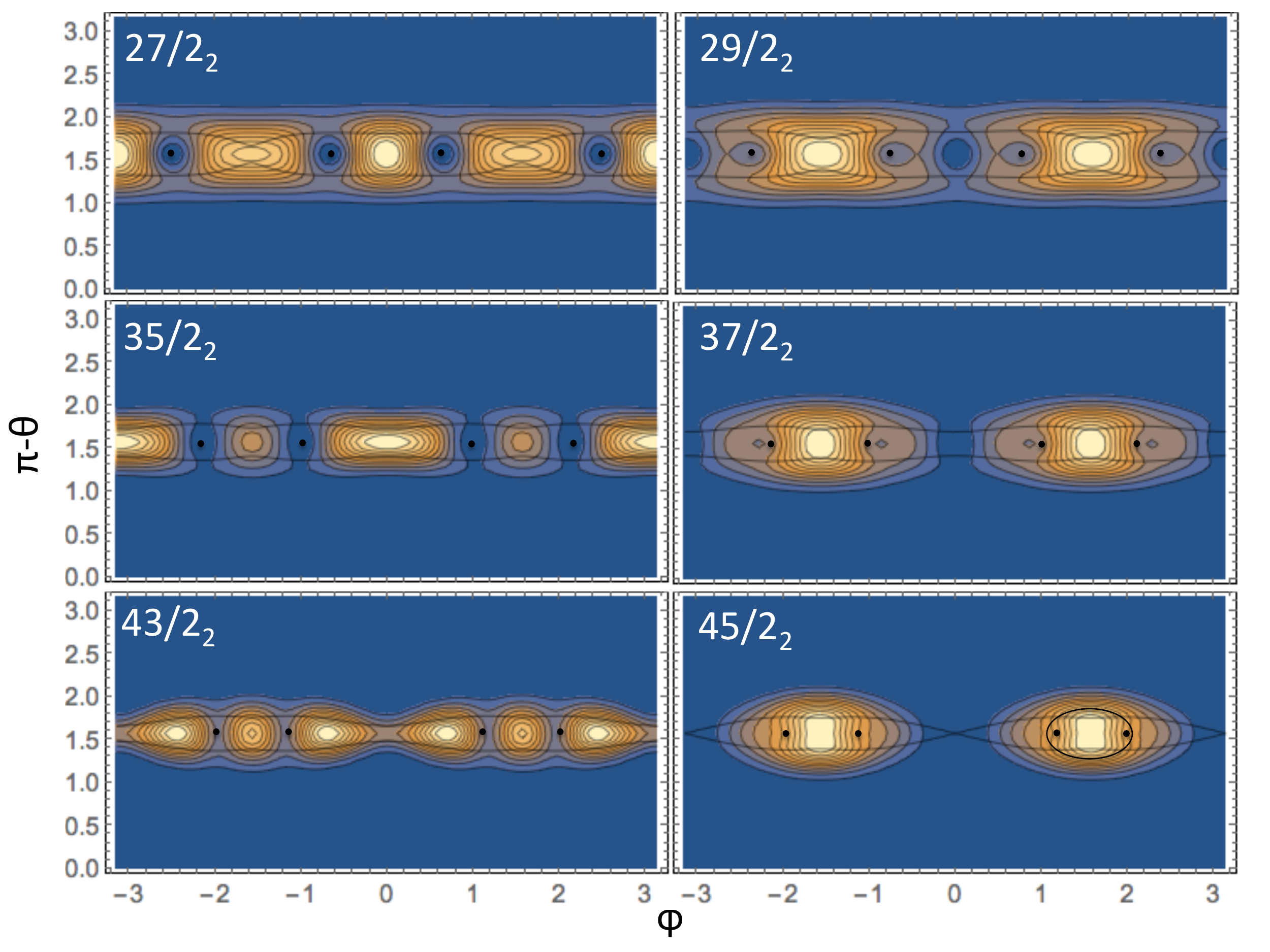}
    \caption{SCS probability distributions $P(\theta \phi)_{I\nu}$ 
    of the total angular momentum $\bm{J}$ for some of the TW states 
    $I\nu$ of the PTR states shown in Fig.~\ref{f:E135Pr}. See caption 
    of Fig.~\ref{f:TRmaps1} for details. The dots are localized at the 
    minimum of the classical energy $\phi_J$ shown in Fig.~\ref{f:ClassAngle}. 
    The curves show a contour calculated by minimizing the energy for 
    given angles $\theta_j$ and $\phi_j$ with respect to the orientation of 
    $\bm{j}$. The energy of the contour is the PTR value shown in 
    Fig.~\ref{f:E135Pr}. The contour is the classical orbit in adiabatic 
    approximation.} 
    \label{f:TWmaps2-2} 
\end{figure*}

\begin{figure}[!ht]
    \includegraphics[width=.47\columnwidth]{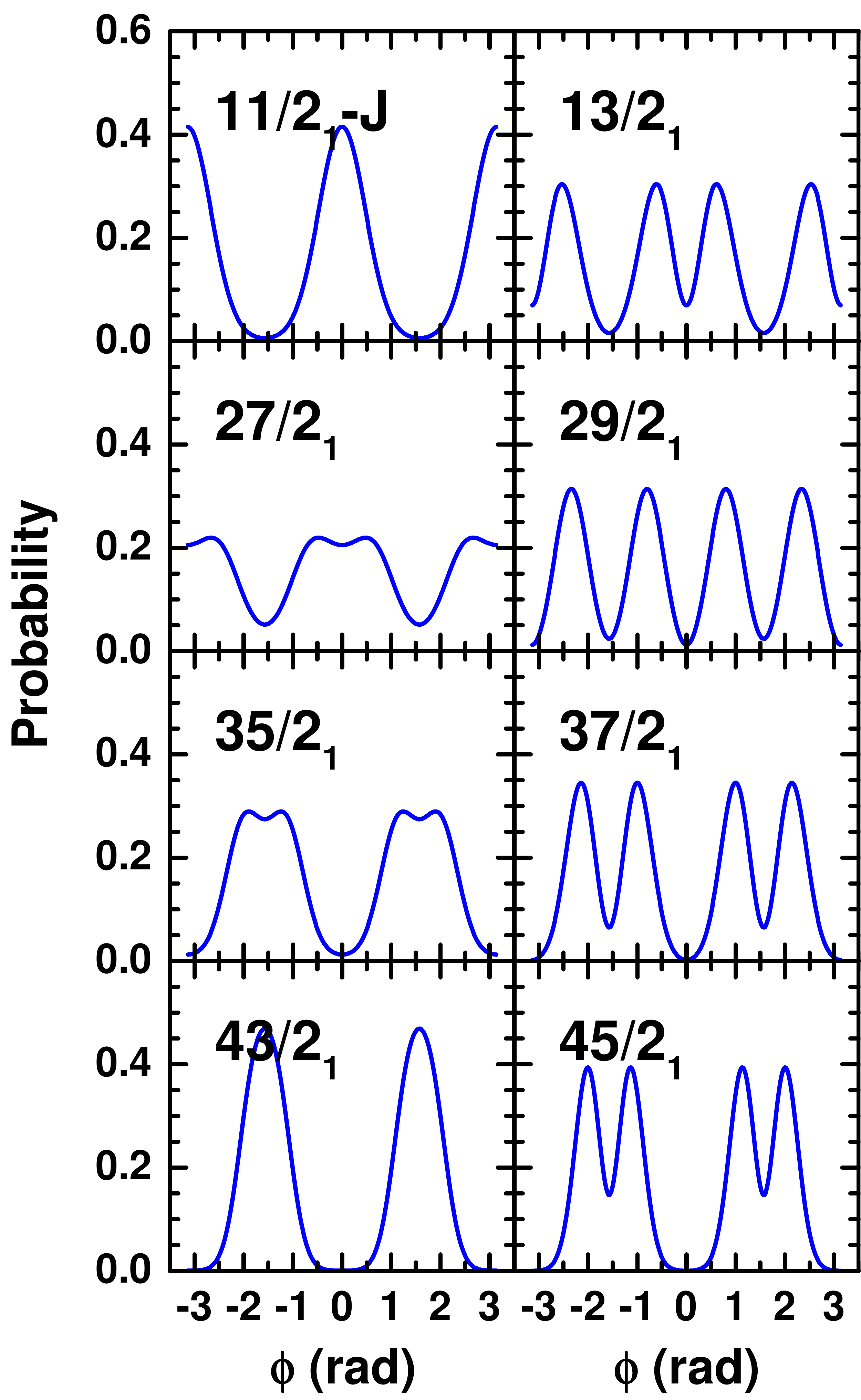}
    \includegraphics[width=.47\columnwidth]{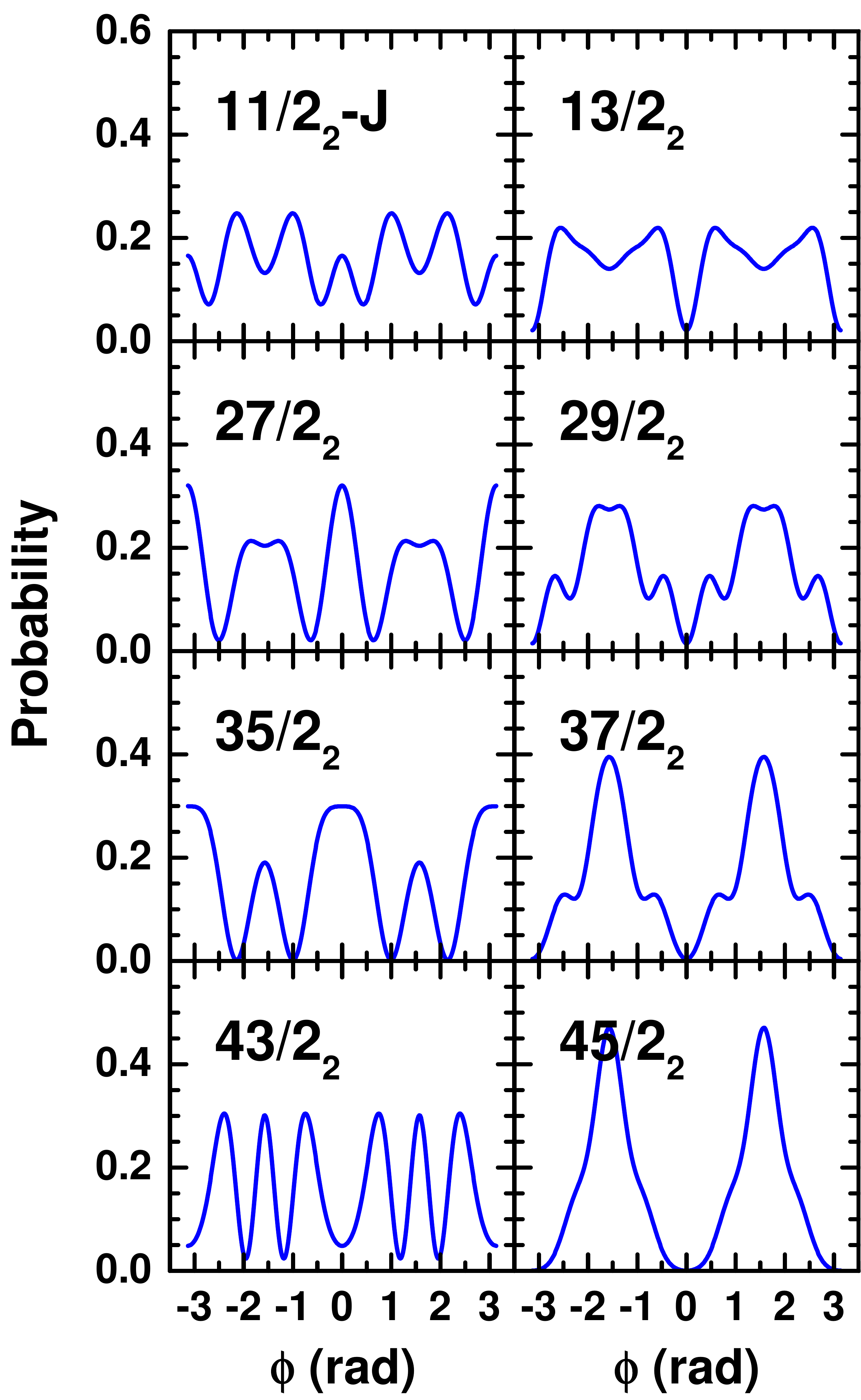}
    \caption{\label{f:density_phi_1st_2nd_I} The probability density of 
    the squeezed $J$-states SSS (\ref{eq:PSSS}) for some yrast and 
    wobbling states.}
\end{figure}

\begin{figure}[!ht]
    \includegraphics[width=.47\columnwidth]{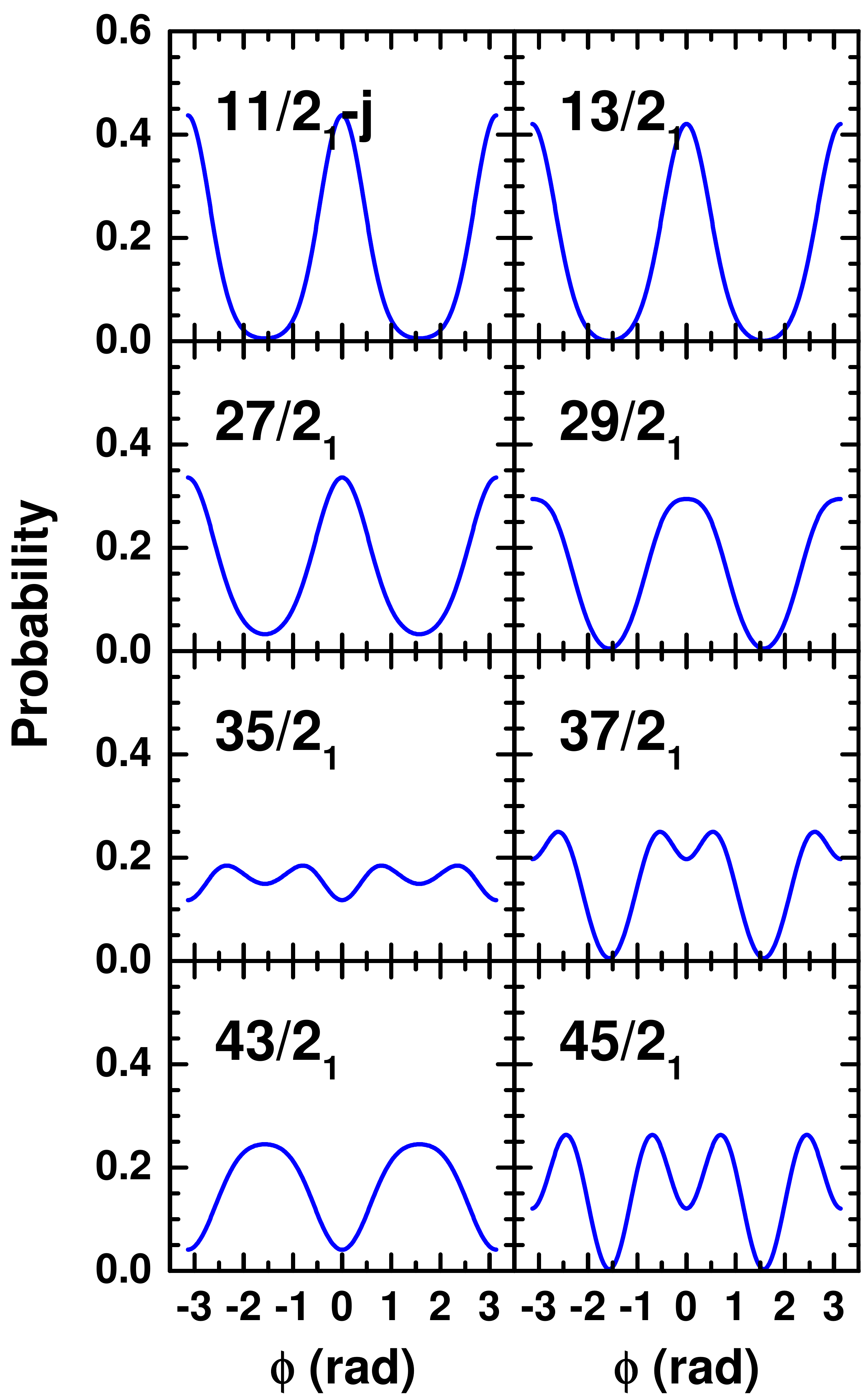}
    \includegraphics[width=.47\columnwidth]{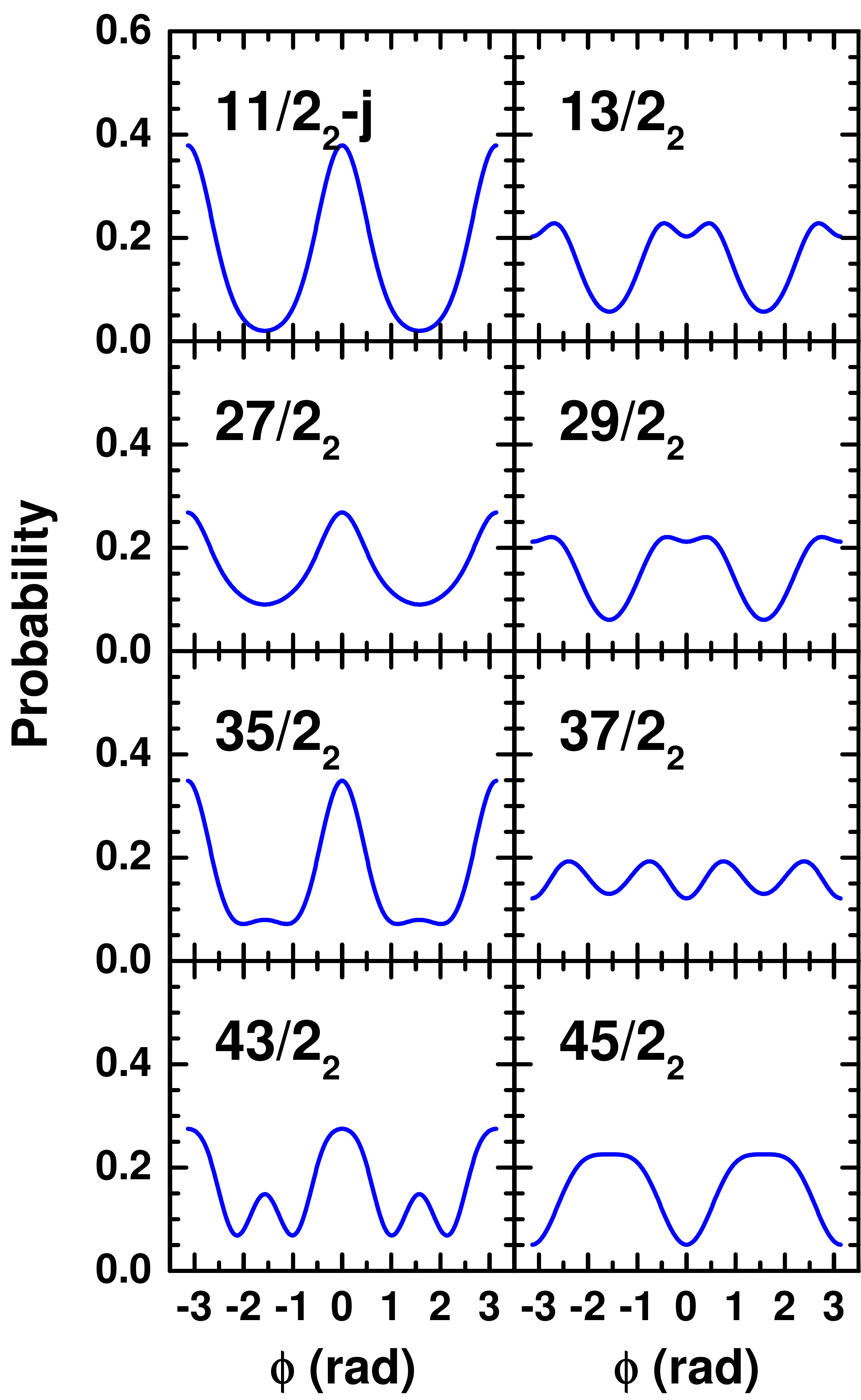}
    \caption{\label{f:density_phi_1st_2nd_j} The probability density of 
    the squeezed $j$-states SSS (\ref{eq:PSSS}) for some yrast and wobbling 
    states.}
\end{figure}

From Fig.~\ref{f:Eclass} one expects that the turning points of the classical 
orbits with two- and three-phonon character change from $\phi=\pm\pi/2$ 
(panel $I=13/2$) to $0$, $\pi$ (panel $I=43/2$) with a region of no turning 
points in between (panel $I=33/2$). The developments of the two-phonon-like 
states with $13/2_2$, $17/2_2$, ..., $43/2_2$ are demonstrated by the 
$\bm{J}$-maps in Figs.~\ref{f:TWmaps1-2} and \ref{f:TWmaps2-2} as well as 
in the SSS plots in Fig.~\ref{f:density_phi_1st_2nd_I}. The probability 
shows six maxima. Four of them correspond to the classical turning points. 
The appearance of two additional maxima is a quantum effect, which characterizes 
the two-phonon nature of the states (see discussions of the $n=2$ structures 
for the TW regime above). The mode has axis-flip character with additional 
quantum bumps. Alternatively, one may understand the pattern as the oscillating 
wave function of the collective $\phi_J$ degree of freedom, which is generated 
by an adiabatic TAC potential similar to the classical one in Fig.~\ref{f:Eclass}. 
This perspective is particularly well visualized by the SSS plots 
for the $n=0$, 1, and 2 states in the corresponding panels of 
Fig.~\ref{f:density_phi_1st_2nd_I}.  

The $n=3$ states $17/2_2$, $21/2_2$, ...., and $45/2_2$ have an axis-flip 
structure with two probability maxima at $\phi_J=\pm\pi/2$. As seen in 
Figs.~\ref{f:TWmaps1-2}, \ref{f:TWmaps2-2}, and \ref{f:density_phi_1st_2nd_I},
with increasing $I$ the width of the probability maxima shrinks 
and they change from the axis-flip structure with the four extrusions 
to near-elliptical shapes. For the state $45/2_2$ they look like the 
$n=0$ shapes with only slight modifications. The difference is seen 
in the $K_l$-plots in Figs.~\ref{f:K-dis_grid} and \ref{f:TWK}. 
The $n=0$ distribution has a maximum at $\vert K_l\vert=1/2$, which 
indicates a symmetric structure, whereas the $n=3$ distribution is 
zero there, which indicates an anti-symmetric structure. The maxima 
are shifted to $\vert K_l\vert$=9/2, which is reflected by the 
elongation of the central density in $\theta$-direction in 
Fig.~\ref{f:TWmaps2-2}.

Fig.~\ref{f:density_phi_1st_2nd_j} shows the response of the proton 
\am in form of SSS plots of the probability distribution for the angle 
$\phi_j$. On can see the change from being centered with the $s$-axis 
in the TW regime to following the $\phi_J$ distributions to a certain 
extension at higher $I$. The figure complements Figs.~\ref{f:TWmapsjlow} 
and \ref{f:TWmapsjhigh}, which have been discussed before. 

\subsection{Transition density maps}

Eq.~(\ref{eq:TDCS}) for calculating transition density plots for triaxial 
rotor in Sec.~\ref{sec:TDMTR} can be easily extended to the case of odd-mass 
nuclei. One just needs to carry out the summation over the odd-particle 
degree of freedom when calculating the transition density matrix in 
Eq.~(\ref{eq:TDM}), i.e.,
\begin{align}
 \rho_{I \to I^\prime}(K, K^\prime)
 =\sum_{k} C_{I^\prime K^\prime k}^*C_{IKk}, \quad I^\prime=I-1.
\end{align}

\begin{figure*}[!ht]
    \includegraphics[width=1.0\linewidth]{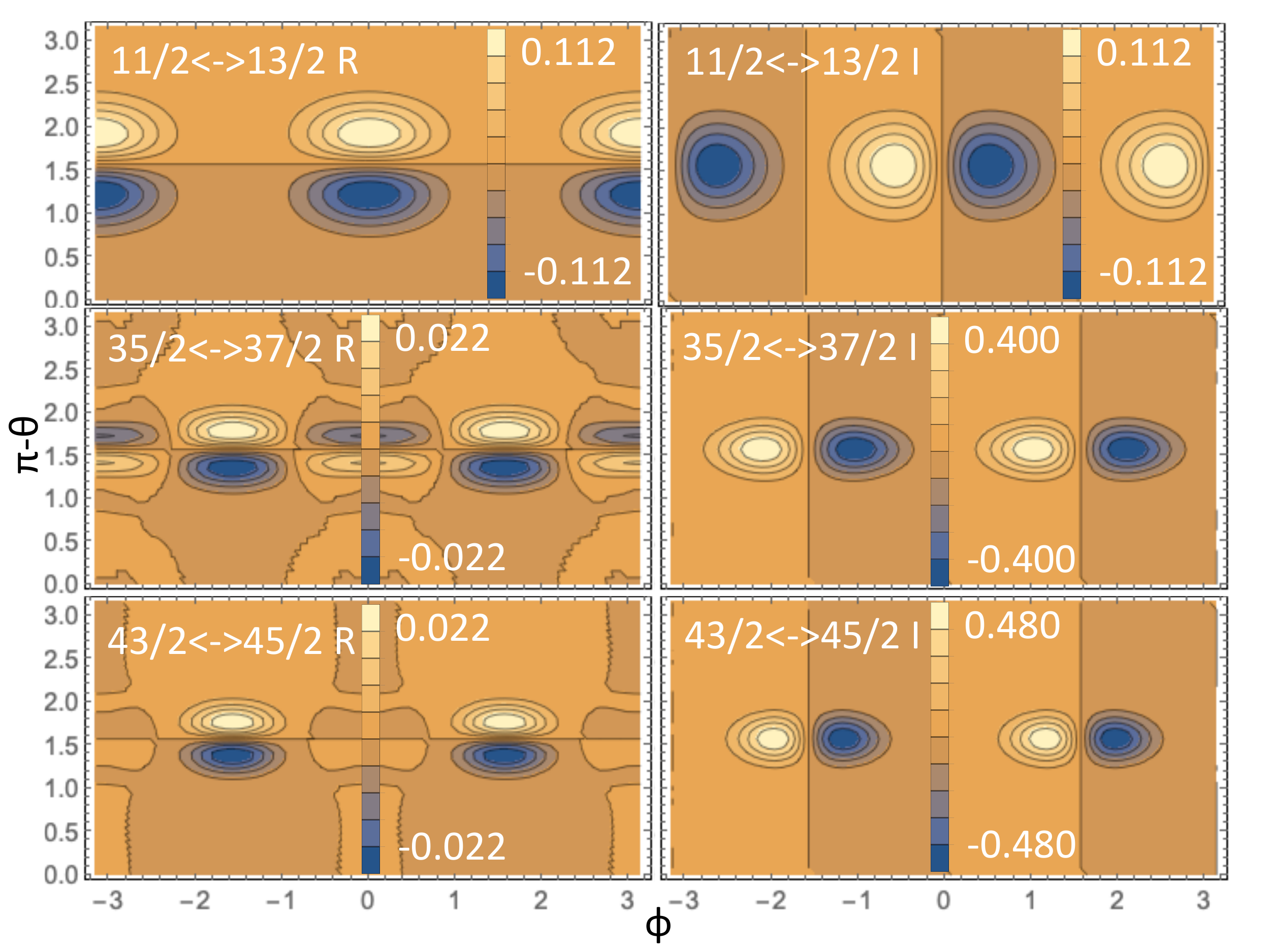}
    \caption{\label{f:TrYtoTWRI} Real part (left panels) and imaginary 
    part (right panels) of the transition density $P(\theta\phi)$ for 
    transitions between the yrast and wobbling states.}
\end{figure*}

\begin{figure*}[!ht]
    \includegraphics[width=1.0\linewidth,trim=0 5.5cm 0 0cm, clip]{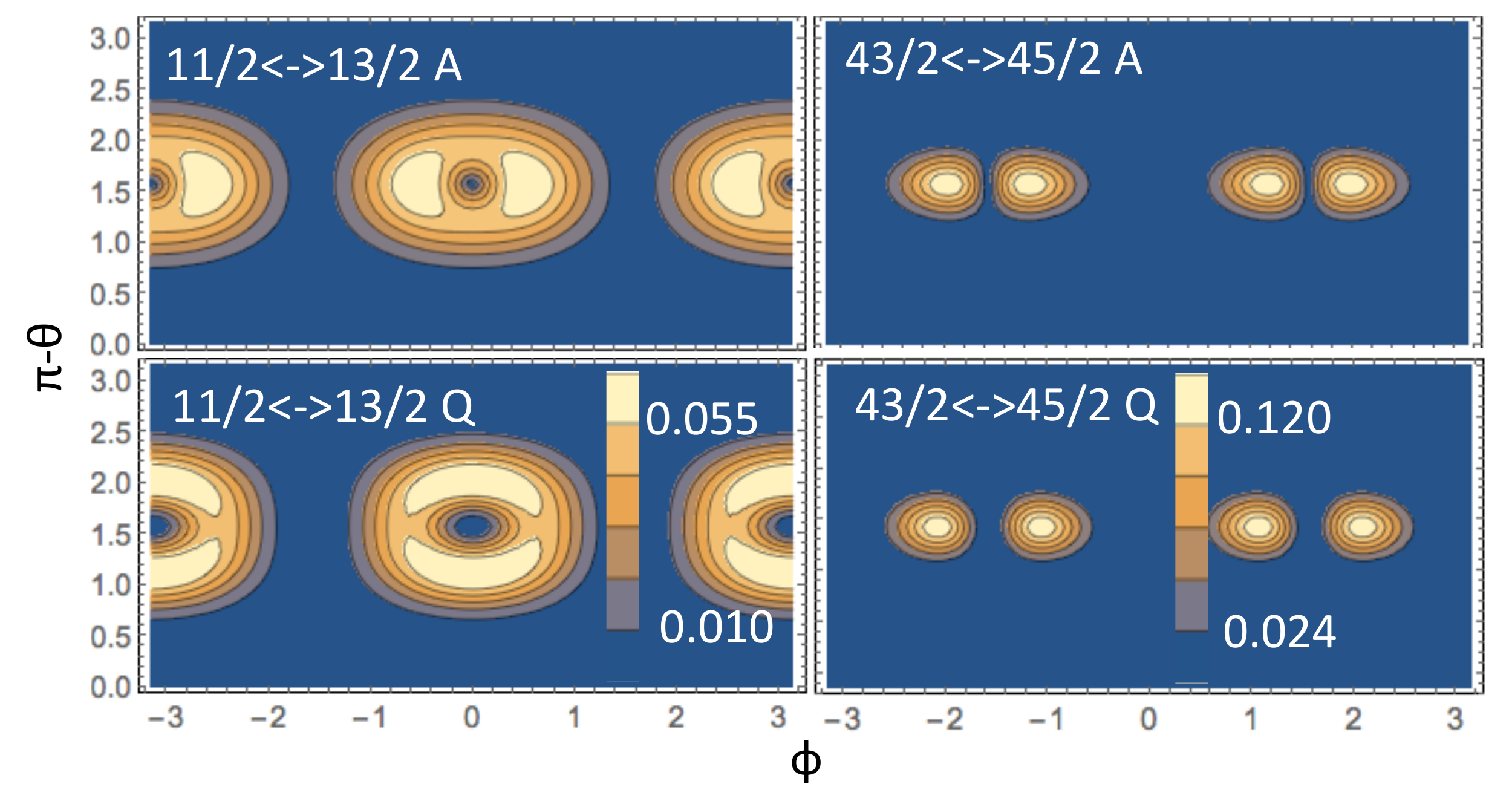}
    \caption{\label{f:TrYtoTWAQ} Absolute value of the transition density 
    $P(\theta\phi)$ (upper panels) and real part of the radiation density 
    $P(\theta\phi) Q_{2-1}(\phi \theta 0)$ given by Eq.~(\ref{eq:Q1int}) 
    (lower panels) for transitions between the yrast and wobbling states.}
\end{figure*}

Fig.~\ref{f:TrYtoTWRI} shows the SCS maps of real and imaginary parts 
of the transition density $P(\theta\phi)$ for transitions between selected  
wobbling and yrast states. Fig.~\ref{f:TrYtoTWAQ} displays the pertaining 
absolute values of $P(\theta\phi)$ and the real parts of 
$P(\theta\phi) Q_{2-1}(\phi\theta0)$, which are given by Eq.~(\ref{eq:Q1int}). 
As discussed for the TR in Sec.~\ref{sec:TDMTR}, the latter represent 
density of the source of the quadrupole radiation which generates the 
transitions between the states. The pertaining imaginary parts
are not shown because they are anti-symmetric. Their integral 
over $\theta$, $\phi$ is zero, such that it does not contribute to
the transition matrix element, which is real in the chosen basis. 
See Fig.~\ref{f:TR9to8} for the TR and its discussion in Sec.~\ref{sec:TDMTR}.  
 
In Fig.~\ref{f:TrYtoTWRI} the two-dimensional wobbling motion of $\bm{J}$ is 
decomposed into a $\theta$-oscillation represented by the real part and 
a $\phi$-oscillation represented by the imaginary part. The absolute values 
of the distributions are shown in Fig.~\ref{f:TrYtoTWAQ}. They look similar 
to an overlay of the pertaining SCS maps of the states connected by the 
transitions in Figs.~\ref{f:TWmaps1} and \ref{f:TWmaps2}. With increasing $I$, 
the real part is progressively suppressed. The transition $13/2_1\rightarrow 11/2_1$ 
in the TW regime has comparable real and imaginary parts. They generate 
the  rim of the absolute value of the transition density, which classically corresponds
to the elliptical precession cone of $\bm{J}$. Both oscillations contribute the transition matrix 
element. For the transition $43/2_1\rightarrow 41/2_1$ in the LW regime the 
$\theta$-oscillation is suppressed. As discussed in Sec.~\ref{sec:TWtoLW}, 
the LW mode is a near one-dimensional $\phi$-oscillation.

\begin{figure}[!ht]
    \includegraphics[width=\columnwidth,trim=5cm  6cm 7cm 0cm clip]{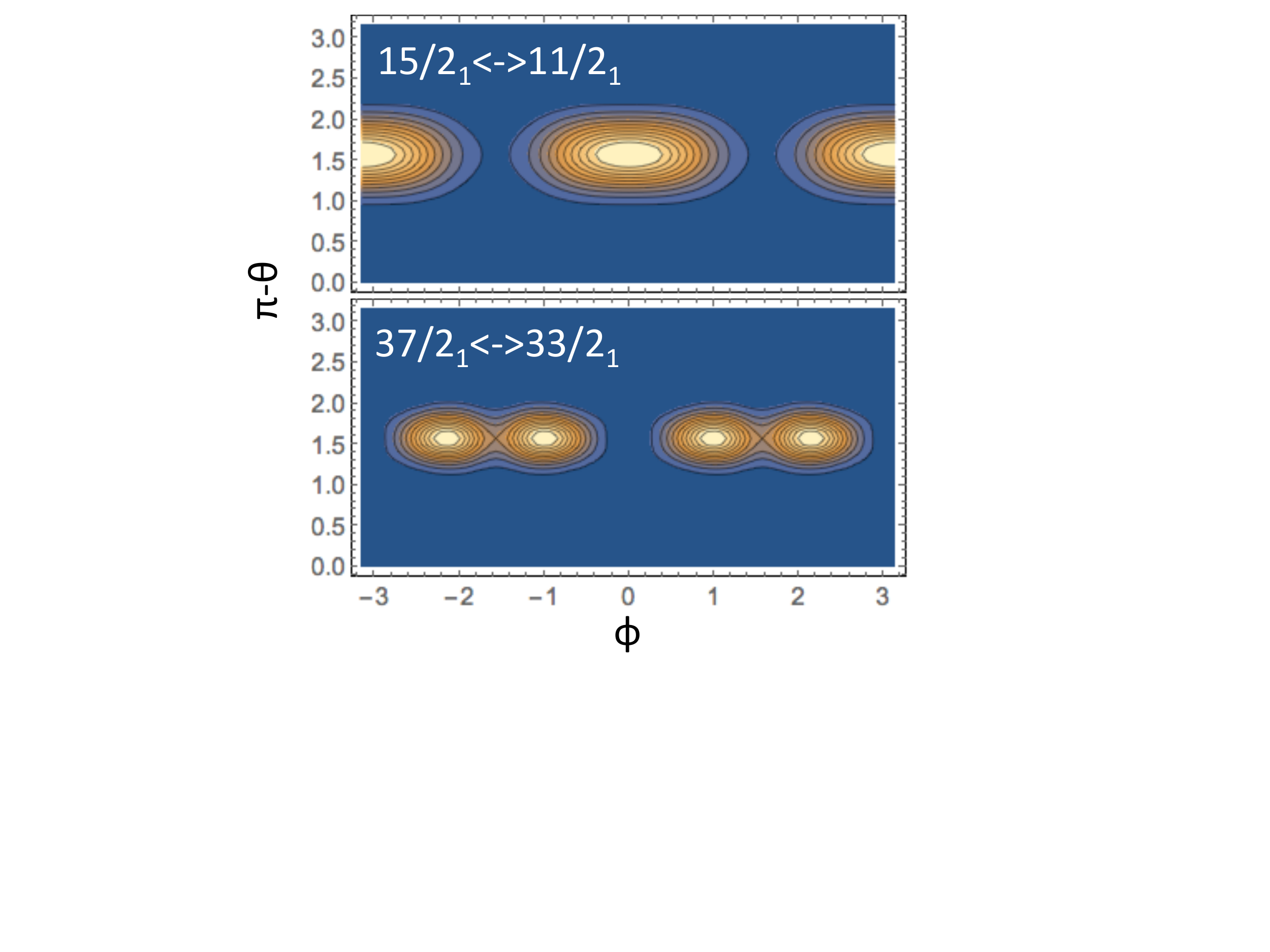}
    \caption{\label{f:TDM_2} Real parts of the radiation density  
    $P(\theta\phi) Q_{2-2}(\phi \theta0)$ defined in analogy to 
    by Eq.~(\ref{eq:Q1int}) for stretched transitions between adjacent 
    yrast and adjacent TW states.}
\end{figure}

One can also define the $P(\theta\phi) Q_{2-2}(\phi\theta0)$ for $E2$
stretched transitions. Its real part is shown in Fig.~\ref{f:TDM_2} 
for two yrast states. The distribution is similar to the $\bm{J}$-maps of 
the states in Figs.~\ref{f:TWmaps1} and \ref{f:TWmaps2}. The same holds for 
the $\Delta I=2$ transitions between the TW states. 
%


\section{Summary and Conclusions}

The classical perspective provides a revealing insight into the physics 
of several high-$j$ particles coupled to a triaxial rotor core. We 
studied in detail the interpretation of the one-particle-plus-triaxial 
rotor system with the intend to evaluate the various methods visualizing the 
structure. For the typical \am range of nuclear experiments $\sim 20 \hbar$ 
it turns out appropriate to keep absolute \am value equal to its quantized 
value and interpret its orientation semiclassically. The non-commutativity 
of the angular momentum components limits the correspondence with the classical 
particle-rotor system. Starting from the finite dimensional Hilbert space spanned 
by the quantized projection of one \am component we compared various methods 
to visualize the structure. We applied the methods to the triaxial rotor without 
the odd particle in order to demonstrate their capabilities for elucidating 
the structure of a simple system with only one kind of \amd.
The description was based on the density matrix, which allowed us 
to straightforwardly generalize the methods to the particle-plus-triaxial 
rotor model (PTR) by studying the reduced density matrices for the 
total, particle, and the rotor angular momenta.

Plots of the root of the expectation values of the square of the \am 
components have been used in the literature for a long time. They provide 
restricted information about the distribution of the \am over the axes. 
The $K$-plots used before as well show the probability distribution of the 
\am projection on one of the axes. They provide additional information 
about the dynamics via their modulation. As a complement we introduced 
the $\phi$-plots which display the probability for the orientation of \am 
vector with respect to a chosen axis. The number of possible angles is 
equal to the dimension of the finite Hilbert space of the \am projections. 
The discreteness of orientation is alien to classical systems. In order 
to establish the closest possible correspondence we introduced the 
over-complete, non-orthogonal Spin Squeezed State (SSS) basis. The 
continuous SSS plots of the pertaining probability density are instructive 
because they look quite like the familiar probability density of 
one-dimensional wave functions. Further, we introduced the continuous set 
the over-complete, non-orthogonal Spin Coherent States (SCS) spanned on 
the Hilbert space, which represent the orientation of the \am vector. The 
SCS representation was studied in great detail because it makes the closest 
contact with the classical pendant of the quantal PTR model. The 
two-dimensional SCS maps provide the probability density distribution as 
function of the polar and azimuthal angles of the \am vector with respect 
to the principal axes of the rotor. They distill, as far as possible, 
the classical orbit which corresponds to the quantum state of interest. 
The development of the phase along a path on the angular momentum sphere 
was also extracted within in the SCS representation. The phase increment 
along a closed path characterizes a state according to the semiclassical 
quantization rules. In addition, SCS maps of the transition density plots 
were proposed to illustrate the  motion of the charge density which 
generates the collective $\Delta I=1,~2$ $E2$-transitions between 
the rotational states.

For the triaxial rotor our analysis established a close correspondence between 
classical and quantal descriptions. The SCS maps trace the classical orbits 
in form of fuzzy rims. For fixed \amd, the phase increment over the closed 
orbits increases according to the semiclassical quantization rules in steps 
of $2\pi$ with the energy of the states. In accordance with the classical rotor,
the structure develops from a harmonic wobbling mode with respect to the 
medium axis with the largest \momi at the lowest energy to harmonic wobbling 
with respect to the long axis with the smallest \momi at the highest energy. 
The wobbling becomes increasingly anharmonic with energy. The mode geometry 
is seen as oval rims that enclose the respective axis points in the SCS maps. 
Near the transition from one to the other topology the mode consists in 
flipping between rotation about the positive and negative directions of 
the short axis seen as two blobs at the points where the short axis penetrates 
the \am sphere. The name \textit{axis-flip wobbling} is suggested for the mode, 
which emerges when classical triaxial objects are spun about the unstable 
axis with the intermediate \momid.

The quantal results obtained by the PTR have been analyzed in detail, 
using $^{135}$Pr and $^{163}$Lu as examples, which have been thoroughly 
studied experimentally. The classical yrast energy defines a critical \amd, 
below which the nucleus rotates about the short axis, and above which 
the nucleus rotates about an axis, which is tilted into short-medium plane 
and moves toward the medium axis with increasing \amd. Below the critical 
\am the transverse wobbling mode appears, where ``transverse" indicates 
that the \am vector revolves the short axis, which is perpendicular to the 
medium axis with the largest \momid. The SCS maps for the total \am visualize 
the topological character most clearly. The states of the lowest $\Delta I=2$ 
sequence appear as blobs centred at the short axis, which indicates their 
zero-phonon nature. The states of the next higher $\Delta I=2$ sequence (of 
opposite signature) appear as oval rims centred at the short axis, which 
indicates their one-phonon nature. The energy difference between the bands, 
which is the wobbling frequency, decreases with the \amd, which is the hallmark 
of transverse wobbling. The states of the third $\Delta I=2$ sequence, appear 
as oval rims centred at the short axis as well, which indicates their 
two-phonon nature. The states of the fourth $\Delta I=2$ sequence appear 
as blobs centred at the medium axis, which indicates axis-flip wobbling 
and signals the instability of transverse wobbling. The geometry of the 
wobbling mode can also be inferred from the $K$-plots, which represent
projections of the probability distributions on the principal axes. The 
analysis of the particle \am shows that it remains well aligned with the 
short axis below the critical \amd. 
 
Above the critical \am the collective wobbling mode has axis-flip character. 
The states of the lowest $\Delta I=2$ sequence can be interpreted as the 
even superposition of two states that represent uniform rotation about
the two equivalent tilted axes, which are related by a reflection through
the short-long or medium-long principal planes. The SCS maps show two blobs
centered to the axes and an overlap region connecting them. The states of 
the next higher $\Delta I=2$ sequence are interpreted as the odd superposition 
of the two states. The SCS maps show the two blobs centered at the axes 
without a connecting overlap region. With increasing \am the two tilted axes 
approach the medium axis from both sides. The overlap of the symmetric yrast 
states strongly increases such that the probability distribution in the SCS 
map becomes an elongated blob centered at the medium axis. For the 
anti-symmetric yrare states the two blobs come close to the medium axis 
where the minimum at the axis remains. As the \am vector executes an oscillatory 
motion with respect to the medium axis, the mode is interpreted as longitudinal 
wobbling, where the symmetric states represent the zero-phonon and the 
anti-symmetric the one-phonon states. The energy difference between the bands, 
which is the wobbling frequency, increases with the \amd, which is the hallmark 
of longitudinal wobbling. Above the critical \am the particle \am follow the 
total \am to some extend. If an adiabatic treatment of the response in the 
framework of the cranking model will lead to a sufficiently accurate effective 
collective Hamiltonian remains to be investigated.

In addition to the wobbling modes, the PTR model contains the cranking modes, 
which are associated with the orientation of the \am of the odd particle. The 
lowest excitation of this type, the signature partner mode, can be clearly 
discriminated from the wobbling modes below the critical \amd. The SCS maps 
of the signature partner states show the precession rims around the short 
axis for the particle \am and for the total \am the blobs at the short axis, 
which indicates uniform rotation.

We demonstrated that the small amplitude approximation in the framework of the 
truncated boson expansion gives a too early instability of the transverse wobbling 
mode. Alternative interpretations of the PTR results were critically reviewed 
from our perspective. The change of terminology by other authors was clarified. 
The assumption of other authors that the ratios between the three \momis 
are given by the expressions for a rigid body contradicts fundamental properties 
of an ensemble of indistinguishable particles.

It was demonstrated that the SCS maps are a powerful tool to visualize the angular 
momentum geometry of rotating nuclei. Their application to the case of several 
particles coupled to a triaxial rotor as well as to the Triaxial Projected 
Shell Model will be studied in forthcoming papers.      
\section*{Acknowledgements}

Supports by the US DoE Grant DE-FG02-95ER40934 as well as the Deutsche
Forschungsgemeinschaft (DFG) and the National Natural Science Foundation
of China (NSFC) through funds provided to the Sino-German CRC 110
``Symmetries and the Emergence of Structure in QCD'' (DFG Grant
No. TRR110 and NSFC Grant No. 11621131001) are acknowledged.          

\begin{appendix}

\section{Harmonic approximation}\label{sec:AppA}

In Ref.~\cite{Tanabe2017PRC}, the Holstein-Primakoff (HP) boson expansion 
was used as a starting point in order to derive the two-oscillator 
approximation of the PTR, to test the stability of the TW regime and 
discern the physical content of the exact solution. For the odd-$A$ case  
two kinds of bosons are introduced: $a$ for the total angular momentum
$\bm{J}$ and $b$ for the single-particle angular momentum $\bm{j}$. They 
identify the nature of each mode. The quantum number $n_\alpha$ 
counts number of the wobbling phonons of $\bm{J}$ and $n_\beta$ the number 
precession excitations of $\bm{j}$ (we use the name cranking excitations). 

For the transverse wobbling motion, the total energy becomes the
lowest when both angular momentum vectors $\bm{J}$ and $\bm{j}$ are 
closely aligned along the 1-axis ($s$-axis) direction, implies that 
the components $J_1$ and $j_1$ should be chosen as the diagonal terms 
in the HP boson representation. The angular momentum components 
are expressed in terms of the boson operators as
\begin{align}
 \label{eq:eq1}
 \hat{J}_+&=\hat{J}_-^\dag=\hat{J}_2+i\hat{J}_3=-a^\dag\sqrt{2I-n_a},\\
 \hat{J}_1&=I-n_a,\\
 \hat{j}_+&=\hat{j}_-^\dag=\hat{j}_2+i\hat{j}_3=\sqrt{2j-n_b}b,\\
 \label{eq:eq2}
 \hat{j}_1&=j-n_b,
\end{align}
with the boson number operators $n_a=a^\dag a$ and $n_b=b^\dag b$. 
The authors of Ref.~\cite{Tanabe2017PRC} expand the square
roots up to the order of $n_a/(2I)$ and $n_b/(2j)$ as
\begin{align}
 \sqrt{2I-n_a}\approx \sqrt{2I}\Big(1-\frac{n_a}{4I}\Big),\\
 \sqrt{2j-n_b}\approx \sqrt{2j}\Big(1-\frac{n_b}{4j}\Big).
\end{align}
Plugging this into PTR Hamiltonian they derive an approximate 
Hamiltonian of fourth order in the boson number. Here we are 
only interested in the terms up to the quadratic 
\begin{align}
 H_{02}= H_0+H_2,
\end{align}
where $H_0$ denotes a constant which collects all the terms
independent of boson operators,
\begin{align}
 H_0
  &=A_1 (I-j)^2+\frac{1}{2}A_{231}\notag\\
  &-\frac{\kappa\cos(\gamma+\pi/3)}{2j}(2j-1)(2j+3),
\end{align}
and $H_2$ the terms bilinear in the boson operators,
\begin{align}\label{eq:H2HP}
 H_2
 &=A(a^\dag a + a a^\dag)+B(a^\dag a^\dag + aa)
  +2F\Big(a^\dag b^\dag + ab\Big)\notag\\
 &+C(b^\dag b + b b^\dag)+D(b^\dag b^\dag +bb)
  +2G\Big(a^\dag b+ab^\dag\Big).
\end{align}
In the above formula, the corresponding coefficients are
\begin{align}\label{eq:H2coef}
 A&=\frac{1}{2}\Big(I-\frac{1}{2}\Big)A_{231}+j A_1,\\
 B&=\frac{1}{2}\Big(I-\frac{1}{4}\Big)A_{23},\\
 C&=\frac{1}{2}\Big(j-\frac{1}{2}\Big)a_{231}+ IA_1,\\
 D&=\frac{1}{2}\Big(j-\frac{1}{4}\Big)a_{23},\\
 F&=\frac{1}{2}(A_2+A_3)\sqrt{Ij}, \\
 G&=\frac{1}{2}A_{23}\sqrt{Ij},\\
 A_{231}&=A_2+A_3-2A_1,\\
 A_{23}&=A_2-A_3,\\
 a_{23}&=A_{23}-2\sqrt{3}\kappa\sin(\gamma+\pi/3)/j,\\
 a_{231}&=A_{231}+6\kappa\cos(\gamma+\pi/3)/j,\\
 A_k&=1/(2\mathcal{J}_k).
\end{align}
The Hamiltonian $H_2$ is diagonalized by means of a Bogoliubov 
transformation, which leads to the eigenvalue equation for the 
frequencies of the eigen modes,
\begin{align}\label{eq:Aev}
 \omega^4-b\omega^2+c=0,
\end{align}
with
\begin{align}
 b&=A^2-B^2+C^2-D^2+2(G^2-F^2),\\
 c&=(A^2-B^2)(C^2-D^2)+(G^2-F^2)^2\notag\\
 &\quad+4FG(AD+BC)\notag\\
 &\quad-2(AC+BD)(F^2+G^2).
\end{align}

This equation has two positive solutions,
\begin{align}\label{eq:Aomega}
 2\omega_{\pm}^2=b\pm \sqrt{b^2-4c},
\end{align}
the lower wobbling and the higher cranking solution,
which are shown in Fig.~\ref{f:Ewob_grid}. Both solutions only
exist when the inequalities
\begin{align}\label{eq:Astab}
 b^2-4c\geq 0, \quad b>0, \quad c>0
\end{align}
hold, which compose the stability conditions for the modes.
For our example the TW mode is stable for $c>0$, which 
results in a critical \am of $I_c=10.5$.  
 
The mapping of the finite $J$ space  on the infinite boson space
means that  $K$ and $k$ become continuous variables. 
Considering $Q=K/\sqrt{I}$ and $q=k/\sqrt{j}$ as coordinates,  
the  pertaining momenta are $P=-i\sqrt{I}d/dK$ and $p=-i\sqrt{j}d/dk$.  
In accordance with Eqs.~(\ref{eq:eq1})-(\ref{eq:eq2}), one 
can express the boson operators in terms of these two generalized coordinates
and momenta by means of the standard relations 
\begin{align}
 a^\dag &=\frac{i}{\sqrt{2}}\Big(\sqrt{I}\frac{d}{dK}-\frac{K}{\sqrt{I}}\Big),\\
 a &=\frac{i}{\sqrt{2}}\Big(\sqrt{I}\frac{d}{dK}+\frac{K}{\sqrt{I}}\Big),\\
 b^\dag &=\frac{i}{\sqrt{2}}\Big(\sqrt{j}\frac{d}{dk}-\frac{k}{\sqrt{j}}\Big),\\
 b &=\frac{i}{\sqrt{2}}\Big(\sqrt{j}\frac{d}{dk}+\frac{k}{\sqrt{j}}\Big).
\end{align}
The Hamiltonian $H_2$  (\ref{eq:H2HP}) is rewritten as
\begin{align}\label{eq:eq3}
 H_2&=-(A+B)I\frac{d^2}{dK^2}-(C+D)j\frac{d^2}{dk^2}\notag\\
    &-(2G+2F)\sqrt{Ij}\frac{d}{dK}\frac{d}{dk}\notag\\
    &+(A-B)\frac{K^2}{I}+(C-D)\frac{k^2}{j}\notag\\
    &+(2G-2F)\frac{Kk}{\sqrt{Ij}},
\end{align}
which re-expresses the two-oscillator approximation as a set of
two coupled differential equations, which describe the 
two oscillations in the $K$ and $k$ coordinates and their couplings. 
The solution is the wave function $\psi(K,k)$. The probability density
\begin{equation}
 P(K)=\int_{-\infty}^\infty \vert \psi(K,k)\vert^2 dk        
\end{equation}
can be compared with the $P(K)$ of the $K$-plots generated 
from the PTR states. 

The eigenvalue problem established by the Hamiltonian (\ref{eq:eq3}) is
solved by discretizing the differential equations.  We assume that there 
are $2n_K+1$ and $2n_k+1$ grid points $K_\mu$ and $k_\nu$ in coordinate 
space distributed symmetrically with respect to the origin,
\begin{align}
 K_\mu &=\mu \Delta_K, \quad \mu=-n_K, -n_K+1, ..., n_K,\\
 k_\nu &=\nu \Delta_k, \quad \nu=-n_k, -n_k+1, ..., n_k.
\end{align}
Taking the requirement of the D$_2$ symmetry of a triaxial nucleus into 
account, $\mu$ and $\nu$ take the following values: $\mu-\nu>0$ and 
$\mu-\nu$ are even numbers; and if $\mu-\nu=0$, $\mu \geq 0$. In 
addition, the grid points ($K_\mu$, $k_\nu$) and ($K_{-\mu}$, $k_{-\nu}$) 
are combined with the phase factor $(-1)^{I-j}$.

The derivatives at $(K_\mu, k_\nu)$ are approximated  (using central 
difference format) as
\begin{align}\label{eq:H2dis}
 &\frac{d\psi(K_\mu, k_\nu)}{dK}=\frac{\psi(K_{\mu+1}, k_\nu)
  -\psi(K_{\mu-1}, k_\nu)}{2\Delta_K},\\
 &\frac{d\psi(K_\mu, k_\nu)}{dk}=\frac{\psi(K_{\mu}, k_{\nu+1})
  -\psi(K_{\mu}, k_{\nu-1})}{2\Delta_k},\\
 &\frac{d^2\psi(K_\mu, k_\nu)}{dK^2}\notag\\
 &\quad =\frac{\psi(K_{\mu+2}, k_\nu)
  -2\psi(K_\mu, k_\nu)+\psi(K_{\mu-2}, k_\nu)}{4\Delta_K^2},\\
 &\frac{d^2\psi(K_\mu, k_\nu)}{dk^2}\notag\\
 &\quad =\frac{\psi(K_{\mu}, k_{\nu+2})
  -2\psi(K_\mu, k_\nu)+\psi(K_{\mu}, k_{\nu-2})}{4\Delta_k^2},\\
 &\frac{d^2\psi(K_\mu, k_\nu)}{dK dk}\notag\\
 &\quad =\frac{1}{4\Delta_K \Delta_k}
 [\psi(K_{\mu+1}, k_{\nu+1}) -\psi(K_{\mu-1}, k_{\nu+1})\notag\\
 &\quad \quad -\psi(K_{\mu+1}, k_{\nu-1})+\psi(K_{\mu-1}, k_{\nu-1})].
\end{align}

To solve the eigenvalue problem we consider the grid points 
as a set of orthonormal states,
\begin{align}
 \langle K_{\mu^\prime}, k_{\nu^\prime}\vert K_{\mu}, k_{\nu}\rangle
  =\delta_{\mu^\prime \mu} \delta_{\nu^\prime \nu},\\
\psi(K_\mu, k_\nu)=\langle\psi\vert K_{\mu}, k_{\nu}\rangle.
\end{align}
The discrete set of differential equation (\ref{eq:H2dis}) becomes a matrix 
problem on the $[K,k]$ space, which we solve by standard numerical 
diagonalization. In the calculations, we take steps $\Delta_K=\Delta_k=0.0625$, 
and $n_K=I/\Delta_K$, $n_k=j/\Delta_k$. We have checked that if $n_K$ and
$n_k$ are doubled, the 
wobbling energy and $K$-plots do not change much. As shown in 
Fig.~\ref{f:Ewob_grid}, the eigenvalues agree with the energies 
of the harmonic spectrum generated from the solutions of 
Eq.~(\ref{eq:Aomega}). 

Eqs.~(\ref{eq:eq3}) and (\ref{eq:H2HP}) correspond to the choice of 
the axes 1, 2, 3 as $s$, $m$, $l$. So the diagonalization of the 
Hamiltonian provides the $K$-plot for the $l$-axis. To obtain the 
$K$-plot for the $m$-axis, we have to choose the $m$-axis as the 
$3$-axis, $l$-axis as the $2$-axis, and $s$-axis as the $1$-axis. 
This can be done by changing the triaxial deformation parameter 
$\gamma=-26^\circ$ to $\gamma=246^\circ$.

Fig.~\ref{f:K-dis_grid} shows the $K_l$- and $K_m$-plots. They look nearly 
the same when scaled appropriately in $K$. This is understood as follows.
Eq.~(\ref{eq:eq3}) implies $J_3=\sqrt{I}Q$ and $J_2=\sqrt{I}P$. That is, 
the $K_m$-plot represents the probability distribution of the momentum 
$P$. For a harmonic oscillator the distributions the wave functions in 
$Q$, $q$ and $P$, $p$ are related by a simple scale transformation. The 
probability distributions of $Q/\Delta Q$ and $P \Delta Q$ are identical, 
where $\Delta Q$ is the oscillator length. For coupled oscillators the 
scaling property holds for the normal coordinates. As the coupling between 
the wobbling and the cranking modes is not very strong in the TW regime 
one has $P\left(K_l/\Delta K\right)\approx P\left(K_m \Delta K\right)$, 
where $\Delta K$ is the width of $P(K_l)$.

\end{appendix}


\end{document}